\RequirePackage{ifpdf}
\documentclass[12pt,letterpaper]{JHEP3}
\usepackage{cite}
\usepackage{epsfig}
\usepackage{amsmath}
\usepackage{amsfonts}
\usepackage{amssymb}
\usepackage{graphicx}
\usepackage{epstopdf}
\usepackage{subfigure}
\usepackage{morefloats}
\usepackage{url}
\usepackage{color,soul}

\graphicspath{{Figures/}} 

\newcommand{\Beq}{\begin{eqnarray}}
\newcommand{\Eeq}{\end{eqnarray}}

\newcommand{\eqn}[1]{Eqn. (\ref{#1})}
\newcommand{\nn}{\nonumber \\}
\newcommand{\grchombo}{\mathtt{GRChombo}}
\def\Lean{\texttt{Lean}}

\newcommand{\tgamma}{\tilde{\gamma}}
\newcommand{\tGamma}{\tilde{\Gamma}}
\newcommand{\tA}{\tilde{A}}
\newcommand{\tD}{\tilde{D}}

\let\normalcolor\relax 
\allowdisplaybreaks

\title{$\grchombo$: Numerical Relativity with Adaptive Mesh Refinement}

\author{Katy Clough${}^{1}$, Pau Figueras${}^{2}$, Hal Finkel${}^{3}$, Markus Kunesch${}^{2}$, Eugene A. Lim${}^{1}$, Saran Tunyasuvunakool${}^{2}$ \\
${}^{1}$ Theoretical Particle Physics and Cosmology Group, King's College London, London, United Kingdom \\
${}^{2}$ Department of Applied Mathematics and Theoretical Physics (DAMTP), University of Cambridge, Cambridge, United Kingdom\\
${}^{3}$ Argonne National Laboratory, Argonne, IL, USA
}

\abstract
{
In this work, we introduce $\grchombo$: a new numerical relativity code which incorporates full adaptive mesh refinement (AMR) using block structured Berger-Rigoutsos grid generation. The code supports non-trivial ``many-boxes-in-many-boxes'' mesh hierarchies and massive parallelism through the Message Passing Interface (MPI). $\grchombo$ evolves the Einstein equation using the standard BSSN formalism, with an option to turn on CCZ4 constraint damping if required. The AMR capability permits the study of a range of new physics which has previously been computationally infeasible in a full $3+1$ setting, whilst also significantly simplifying the process of setting up the mesh for these problems. We show that $\grchombo$ can stably and accurately evolve standard spacetimes such as binary black hole mergers and scalar collapses into black holes, demonstrate the performance characteristics of our code, and discuss various physics problems which stand to benefit from the AMR technique.
} 

\begin{document}
\section{Introduction}

Almost a hundred years after Einstein wrote down the equations of General Relativity \cite{Einstein1916}, solutions of the Einstein equation remain notoriously difficult to find beyond those which exhibit significant symmetries. Even for these highly symmetric solutions, basic questions remain unanswered. A famous example is the question of the non-perturbative stability of the Kerr solution -- more than 50 years after its discovery, it is not known whether the exterior Kerr solution is stable. The main difficulty of solving the Einstein equation is its non-linearity, which defies perturbative approaches. 

One of the main approaches in our hunt for solutions is the use of numerical methods. Numerical methods have been used to solve the Einstein equation for many decades, but the past decade has seen tremendous advances. A particular watershed moment was the breakthrough in evolving the inspiral mergers of two black holes \cite{Pretorius:2005gq,Baker:2005vv,Campanelli:2005dd} in 2005, a crucial milestone in the growth of numerical relativity as a discipline and as a tool. The other driver of this development is an explosion in the availability of large and powerful supercomputing clusters and the maturity of parallel processing technology such as the Message Passing Interface (MPI) and OpenMP, which open up new computational approaches to solving the Einstein equation.

We anticipate that this development will continue to accelerate, partly driven by upcoming observational projects. The gravitational wave detector LIGO is expected to start Advanced LIGO science runs in late 2015, and there are hopes that the sensitivity might be good enough to achieve a first detection of gravitational waves from binaries. In the longer term, the European Space Agency (ESA) has designated the space-based eLISA detector an L3 launch slot (expected launch date around 2034), and the LISA Pathfinder spacecraft has a firm launch date of late 2015. 

Beyond searching for gravitational waves and black holes, numerical relativity is now beginning to find uses in the investigation of other areas of fundamental physics. For example, standard GR codes are now being adapted to study modified gravity \cite{Berti:2015itd}, cosmology \cite{Wainwright:2014pta,Johnson:2011wt} and even string theory motivated scenarios  \cite{Cardoso:2012qm,Chesler:2013lia,Cardoso:2014uka,Choptuik:2015mma}. In particular, there is an increasing focus on solving GR coupled to matter equations in the strong-field regime: cosmic string evolution with GR, realistic black hole systems with accretion disks, non-perturbative systems in the early universe, etc. This nascent, but growing, interest in using numerical relativity as a mature scientific tool to explore other broad areas of physics is one key motivation of this work. Since it is often difficult to have an intuitive picture of the entire evolution ahead of time, the code must be able to automatically adapt to ensure that all regions of interest always remain adequately resolved.

In the numerical GR community, this requirement is largely met through a moving-box mesh refinement scheme. This type of setup consists of hierarchies of boxes nested around some specified centres, and the workflow typically requires the user to specify the exact size of these boxes beforehand. These boxes are then moved around, either along a prespecified trajectory guided by prior estimates, or by automatically tracking certain quantities or features in the solution as it evolves. Boxes which come within a certain distances of each other may also be allowed to merge. A number of moving-box mesh refinement codes have been made public over the recent years, many of which are built on top of the well-known $\mathtt{CACTUS}$ framework \cite{Goodale2002a,Loffler:2011ay}. One such implementation is the McLachlan/Kranc code \cite{Brown:2008sb,Kranc:web}, which uses finite difference discretisation and the Baumgarte-Shapiro-Shibata-Nakamura (BSSN) evolution scheme  \cite{Baumgarte:1998te,Shibata:1995we}. Similarly, the \texttt{LEAN} code \cite{Sperhake:2006cy,Zilhao:2010sr}, which  uses the \texttt{CACTUS} framework, and \texttt{BAM} and AMSS-NCKU \cite{Marronetti:2007ya, PhysRevD.82.024005} also implement the BSSN formulation of the Einstein equations.  There is also $\mathtt{GRHydro}$ which implements general-relativistic magnetohydrodynamics (MHD) for the Einstein Toolkit \cite{EinsteinToolkit:web}, building yet another layer of physics on top of evolution codes such as McLachlan/Kranc. There are also non-$\mathtt{CACTUS}$ codes such as $\mathtt{SPeC}$ \cite{Pfeiffer:2002wt} and \texttt{bamps} \cite{Hilditch:2015aba}, which implement the generalised harmonic formulation of the Einstein equations using a pseudospectral method. In addition to these public codes, there is a plethora of closed-source codes.

The moving-box mesh refinement technique has found great success in astrophysically motivated problems such as two-body collision/inspiral. Outside of this realm, however, the setup can quickly become impractical, especially where one expects new length scales of interest to emerge dynamically over the course of the evolution. This can occur generically in highly nonlinear regimes, either by interaction between GR and various matter models, or by gravitational self-interaction itself which can exhibit complicated unstable behaviour in higher dimensions. In such situations, it is necessary to develop a code which has the flexibility to create refinement regions of arbitrary shapes and sizes, anywhere in the computational domain as may be required. This can be achieved by using a fully adaptive mesh refinement (AMR) technique, whose feature is generally characterised by the ability to monitor a chosen quantity at each time step and insert higher resolution sub-regions where this quantity fails to lie within some chosen bounds. Of course, the efficacy of such codes depend crucially on a sensible choice of these criteria, however when implemented correctly they can be an extremely powerful tool. The advantage here is twofold: AMR ensures that small emergent features remain well-resolved at all times, but also that only those regions which require this extra resolution gets refined, thus allowing more problems to fit within a given memory footprint. To the best of our knowledge, \texttt{PAMR/AMRD} \cite{PAMR} and \texttt{HAD} \cite{Neilsen:2007ua} are the only two codes with full adaptive mesh refinement (AMR) capabilities in numerical GR. 

In this work, we introduce $\grchombo$, a new multi-purpose numerical relativity code. $\grchombo$ is built on top of the $\mathtt{Chombo}$ \cite{Chombo} framework. $\mathtt{Chombo}$ is a set of tools developed by Lawrence Berkeley National Laboratory for implementing block-structured AMR for solving partial differential equations. $\grchombo$ features \footnote{Since the \texttt{Chombo} core is dimension-independent for up to six spatial dimensions, $\grchombo$ could potentially be extended to simulate fully higher dimensional spacetimes without any symmetry assumptions, subject to computational resource availability.} include the following.

\begin{itemize}
\item{\emph{BSSN formalism with moving puncture}: $\grchombo$ evolves the Einstein equation in the BSSN formalism. An option to turn on the CCZ4 constraint damping modification\cite{Alic:2011gg,Bona:2003fj} is also available. Singularities of black holes are managed using the moving puncture gauge conditions \cite{Campanelli:2005dd, Baker:2005vv}.}
\item{\emph{Adaptive Mesh Refinement}: $\mathtt{Chombo}$ provides full adaptive mesh refinement with non-trivial nesting topologies via the Berger-Rigoutsos block-structured adaptive mesh algorithm \cite{BergerRigoutsis91}. The user only needs to specify regridding criteria, and $\mathtt{Chombo}$ does the rest. Kreiss-Oliger dissipation is used to control errors, from both truncation and the interpolation associated with regridding. }
\item{\emph{MPI scalability}: $\grchombo$ inherits the parallel infrastructure of $\mathtt{Chombo}$, with ability to scale efficiently to many thousands of CPU-cores per run.}
\item{\emph{Standardized Output and Visualization}: $\grchombo$ uses \texttt{Chombo}'s $\mathtt{HDF5}$ output format, which is supported by many popular visualization tools such as $\mathtt{VisIt}$. In particular, the output files can be used as input files if one chooses to continue a previously stopped run -- i.e. the output files are also checkpoint files.}
\end{itemize}

In this paper, we will detail these capabilities of $\grchombo$ and illustrate how they expand the current field in numerical GR to permit new physics to be explored. The design methodology, scaling properties and performance of $\grchombo$ in a number of standard simulations are included.  

The paper is organized as follows:

In Sec. \ref{sec-GRChombo} we describe the implementation of the code. In particular, in Sec. \ref{sec-eqns} we establish the exact formulation of the equations which were used and our notation conventions, and in Sec. \ref{sec-code} we detail the AMR methodology and other ``coding'' aspects, such as finite differencing, dissipation and load balancing.

In Sec. \ref{sec-newphys} we give examples of several areas of physics which the code is well adapted to explore, and in which it offers advantages over existing codes. 

In Sec. \ref{sec-tests}, we present the results of standard tests, including the Apples with Apples tests \cite{Babiuc:2007}, black holes and black hole mergers, and critical collapse. We test the AMR capabilities of the code, its robustness to regridding errors, and its scaling and convergence properties.  

We discuss our results and future directions in Sec. \ref{sec-dis}. 

Videos of several tests conducted in this paper, and examples of some new problems being tackled using the code, can be viewed via our website at http://grchombo.github.io.

\section{$\grchombo$} \label{sec-GRChombo}

In this section, we will describe our numerical implementation of the Einstein equation in $\grchombo$.

\subsection{$\grchombo$ equations and notation conventions} \label{sec-eqns}
 
The purpose of this subsection is to clearly state the equations of motion that we have implemented and our conventions for completeness. Since these are standard in the field, the experienced reader may want to skip this subsection.

Many numerical relativity codes implement the so called BSSN formulation of the Einstein equation \cite{Nakamura:1987zz,Shibata:1995we,Baumgarte:1998te}. This formulation expresses the Einstein equation in a strongly hyperbolic form, and together with the $``1+\log"$ slicing \cite{Bona:1994dr} and the ``gamma-driver" gauge conditions \cite{Alcubierre:2002kk}, has allowed the stable simulation of dynamical spacetimes of interest, including black hole binaries. 

More recently, other refined formulations of the Einstein equation based on the Z4 system \cite{Bona:2003fj,Gundlach:2005eh} have been proposed, most notably the Z4c formulation \cite{Bernuzzi:2009ex} and the CCZ4 formulation \cite{Alic:2011gg}.\footnote{Both the BSSN and CCZ4 equations have been written in a fully covariant form \cite{Brown:2009dd,Baumgarte:2012xy,Sanchis-Gual:2014nha}. These covariant formulations can be advantageous in certain cases, and we plan to implement them in the future.} In the Z4 system, both the Hamiltonian and the momentum constraint are promoted to dynamical variables and hence constraint violating modes can propagate and eventually exit the computational domain. This may potentially result in a more stable evolution. In addition, the Z4 system can be augmented with damping terms so that constraint violating modes can be exponentially suppressed. In practical terms, the changes required between the CCZ4 equations and the standard BSSN equations are minimal and in \texttt{GRChombo} we have implemented both. 

In this work, we follow the indexing convention of \cite{ShapiroBook}. The signature is $(-+++)$, and low-counting Latin indices $a,b,\dots$ are abstract tensor indices while Greek indices $\mu,\nu,\dots$ denote spacetime component indices and run from $0,1,2,3$. Spatial component indices are labeled by high-counting Latin indices $i,j,\dots$ which runs from $1,2,3$.  Unless otherwise stated, we set $G=1$ and $c=1$.

\subsubsection{Evolution equations}

The Z4 system with constraint damping is \cite{Gundlach:2005eh} 
\begin{equation}
R_{ab} +\nabla_a\,Z_b + \nabla_b\,Z_a-\kappa_1\left[n_a\,Z_b + n_b\,Z_a-(1+\kappa_2)\,g_{ab}\,n^c\,Z_c\right]=8\,\pi\left(T_{ab} -\frac{1}{2}\,g_{ab}\,T\right)
\label{eq:Z4eqs}
\end{equation}
where $R_{ab}$ is the Ricci tensor associated with the metric $g$ on the spacetime manifold ${\mathcal M}$, and $\nabla$ is the corresponding metric compatible covariant derivative. $T_{ab}$ is the stress-energy tensor of the matter and $T\equiv g_{ab}\,T^{ab}$ is its trace. If we set $Z^a=0$, the Z4 equations \eqn{eq:Z4eqs} reduce to the standard (trace-reversed) Einstein equation. Here $\kappa_1$ and $\kappa_2$ are parameters which control the damping. 

In the \texttt{GRChombo} code we use the standard $3+1$ ADM decomposition of the spacetime metric, 
\begin{equation}
ds^2=-\alpha^2\,dt^2+\gamma_{ij}(dx^i + \beta^i\,dt)(dx^j + \beta^j\,dt)\,,
\end{equation}
so that $\gamma_{ij}$ is the induced metric on the spatial slices and 
\begin{equation}
n^\mu = \frac{1}{\alpha}\left(\partial_t^\mu - \beta^i\,\partial_i^\mu\right)\,,
\end{equation}
is the corresponding timelike unit normal. The extrinsic curvature is defined as
\begin{equation}
K_{ij} = -\frac{1}{2}\,(\pounds_n\gamma)_{ij}\,,
\end{equation}
where $\pounds$ denotes the Lie derivative. As is customary, we decompose the induced metric as $\gamma_{ij}=\frac{1}{\chi^2}\,\tilde\gamma_{ij}$ so that $\det\tilde\gamma_{ij}=1$ and $\chi = \left(\det\gamma_{ij}\right)^{-\frac{1}{6}}$. Similarly, the extrinsic curvature is decomposed into its trace, $K=\gamma^{ij}\,K_{ij}$, and its traceless part so that
\begin{equation}
K_{ij}=\frac{1}{\chi^2}\left(\tilde A_{ij} + \frac{1}{3}\,K\,\tilde\gamma_{ij}\right)\,,
\end{equation}
with $\tilde\gamma^{ij}\,\tilde A_{ij}=0$. In the Z4 system, one further defines $\Theta$ as the projection of the Z4 four-vector along the normal timelike direction, $\Theta \equiv -n_\mu\,Z^\mu$. Finally, the spacelike components of the four-vector, $Z_i$, are included in a variable $\hat\Gamma^i$ defined as
\begin{equation}
\hat\Gamma^i \equiv \tilde\Gamma^i + 2\,\tilde\gamma^{ij}\,Z_j\,,
\end{equation}
where $\tilde\Gamma^i=\tilde\gamma^{jk}\,\tilde\Gamma^i_{~jk}$ and $\tilde\Gamma^i_{~jk}$ are the Christoffel symbols associated to the conformal metric $\tilde\gamma_{ij}$,
\begin{equation}
\tilde \Gamma^i_{~jk} = \frac{1}{2}\,\tilde\gamma^{il}\left(\partial_j\tilde\gamma_{kl} + \partial_k\tilde\gamma_{jl} - \partial_l\tilde\gamma_{jk}\right)\,.
\end{equation}
Summarizing, the dynamical variables for the Z4 system are
\begin{equation} 
\{\chi,\,\tilde\gamma_{ij},\,K,\,\tilde A_{ij},\,\Theta,\,\hat\Gamma^i\}\,.
\end{equation}
Setting to zero the Z4 four-vector, $Z^\mu=0$, this system reduces to the standard BSSN system. 

Finally,  we recall the various components of the matter stress tensor in the standard $3+1$ decomposition: 
\begin{equation}
\rho = n_a\,n_b\,T^{ab}\,,\quad S_i = -\gamma_{ia}\,n_b\,T^{ab}\,,\quad S_{ij} = \gamma_{ia}\,\gamma_{jb}\,T^{ab}\,,\quad S = \gamma^{ij}\,S_{ij}\,.
\label{eq:Mattereqns}
\end{equation}

We are now ready to write down the evolution equations for CCZ4 system in the standard $3+1$ decomposition \cite{Alic:2011gg}:
\begin{align}
&\partial_t\chi=\frac{1}{3}\,\alpha\,\chi\, K - \frac{1}{3}\,\chi \,\partial_k \beta^k + \beta^k\,\partial_k \chi\,, \label{eqn:dtchi}\\
&\partial_t\tilde\gamma_{ij} =-2\,\alpha\, \tA_{ij}+\tgamma_{ik}\,\partial_j\beta^k+\tgamma_{jk}\,\partial_i\beta^k-\frac{2}{3}\,\tgamma_{ij}\,\partial_k\beta^k +\beta^k\,\partial_k \tgamma_{ij}\,, \label{eqn:dttgamma} \\
&\partial_t K = -\gamma^{ij}D_i D_j \alpha + \alpha\left(R + 2\,D_iZ^i + K^2 - 2\,K\,\Theta\right) + \beta^i\partial_iK \nonumber\\
&\hspace{1.2cm}-3\,\alpha\,\kappa_1(1+\kappa_2)\Theta+4\pi\,\alpha(S-3\,\rho), \label{eqn:dtK} \\
&\partial_t\tilde A_{ij} = \chi^2\left[-D_iD_j \alpha + \alpha\left( R_{ij} + D_iZ_j + D_jZ_i-8\pi\,\alpha \,S_{ij}\right)\right]^\textrm{TF}  \nn
&\hspace{1.3cm}+ \alpha \tA_{ij}(K-2\,\Theta)-2\,\alpha\,\tA_{il}\,\tA^l{}_j+\tA_{ik}\,\partial_j\beta^k + \tA_{jk}\,\partial_i\beta^k \nn
&\hspace{1.3cm}-\frac{2}{3}\,\tA_{ij}\,\partial_k\beta^k+\beta^k\,\partial_k \tA_{ij}\,, \label{eqn:dtAij}\\
&\partial_t\Theta =\frac{1}{2}\,\alpha\left(R + 2\,D_iZ^i-\tA_{ij}\,\tA^{ij} + \frac{2}{3}\,K^2 - 2\,\Theta\,K\right) \nn
&\hspace{1.2cm} -Z^i\,\partial_i\alpha + \beta^k\,\partial_k\Theta - \alpha\,\kappa_1\,(2+\kappa_2)\Theta - 8\pi\,\alpha\,\rho\,,\\
&\partial_t\hat\Gamma^i=-2\,\tA^{ij}\,\partial_j \alpha +2\,\alpha\left(\tilde\Gamma^i_{jk}\,\tA^{jk}-\frac{2}{3}\,\tilde\gamma^{ij}\partial_j K - 3\,\tA^{ij}\frac{\partial_j \chi}{\chi}\right) \nn
&\hspace{1.3cm} +\beta^k\partial_k \hat\Gamma^{i} +\tilde\gamma^{jk}\partial_j\partial_k \beta^i +\frac{1}{3}\,\tilde\gamma^{ij}\partial_j \partial_k\beta^k \nn
&\hspace{1.3cm} + \frac{2}{3}\,\tilde\Gamma^i\,\partial_k \beta^k -\tilde\Gamma^k\partial_k \beta^i 
+2\,\kappa_3\left(\frac{2}{3}\,\tilde\gamma_{ij}\,Z_j\,\partial_k\beta^k - \tilde\gamma^{jk}\,Z_j\,\partial_k\beta^i\right) \nn
&\hspace{1.3cm} + 2\,\tilde\gamma^{ij}\left(\alpha\,\partial_j\Theta - \Theta\,\partial_j\alpha - \frac{2}{3}\,\alpha\,K\, Z_j\right) 
-2\,\alpha\,\kappa_1\,\tilde\gamma^{ij}\,Z_j - 16\pi\,\alpha\,\tilde\gamma^{ij}\,S_j\,.\label{eqn:dtgamma}
\end{align}
Here $D_i$ is the metric compatible covariant derivative with respect to the physical metric $\gamma_{ij}$ and $[\ldots]^\textrm{TF}$ denotes the trace free part of the expression inside the parenthesis. The three-dimensional Ricci tensor, $R_{ij}$, is split as
\begin{equation}
R_{ij} = \tilde R_{ij} + R^\chi_{ij}\,,
\end{equation}
where
\begin{equation}
\tilde R_{ij} = -\frac{1}{2}\tgamma^{lm}\partial_m\partial_l\tgamma_{ij}+\tGamma^k\tGamma_{(ij)k}+\tgamma^{lm}(2\tGamma^k_{l(i}\tGamma_{j)km}+\tGamma^k_{im}\tGamma_{klj}) \label{eqn:conformalR}
\end{equation}
and
\begin{equation}
R^{\chi}_{ij}=\frac{1}{\chi}(\tD_i\tD_j\chi + \tgamma_{ij}\tD^l\tD_l \chi)-\frac{2}{\chi^2}\tgamma_{ij}\tD^l\chi \tD_l\chi. \label{eqn:Rchi}
\end{equation}
where $\tilde D_i$ is the metric compatible covariant derivative with respect to the conformal metric $\tilde\gamma_{ij}$. Note that the three-dimensional Ricci Scalar is then $R = \gamma^{ij}R_{ij}$.

Equations \eqn{eqn:dtchi}--\eqn{eqn:dtgamma} are the CCZ4 evolution equations as originally presented in \cite{Alic:2011gg}, including the extra damping parameter $\kappa_3$. This parameter controls the coupling of some quadratic terms in the evolution equation for $\hat \Gamma^i$. The choice $\kappa_3=1$ corresponds to the fully covariant CCZ4 system, but as discussed in \cite{Alic:2011gg},  it leads to instabilities in the evolution of spacetimes containing black holes. More recently, \cite{Alic:2013xsa} showed that replacing $\kappa_1\to \kappa_1/\alpha$ in \eqn{eqn:dtchi}--\eqn{eqn:dtgamma} allows to stably evolve black hole spacetimes whilst retaining the full covariance of the CCZ4 system. In $\grchombo$ we have included a parameter that allows us to switch from the original formulation of the CCZ4 system to the more recent one proposed in \cite{Alic:2013xsa}, with the aforementioned redefinition of $\kappa_1$.

Note that in the actual evolution, the values of the three-vector $Z_i$ are computed from the knowledge of the evolved variable $\hat\Gamma^i$ and $\tilde\Gamma^i$, which is computed from the conformal metric, $\tilde\gamma_{ij}$. Finally, we note that the evolution equations \eqn{eqn:dtchi}--\eqn{eqn:dtgamma} reduce to the standard BSSN equations upon setting $\Theta=0$ and $Z^i=0$, and using the Hamiltonian constraint, \eqn{eqn:HamiltonianConst}, in the evolution equation for $K$, \eqn{eqn:dtK}, to eliminate the Ricci scalar $R$.

\subsubsection{Gauge conditions}

To complete the set of evolution equations, we need to choose slicing conditions -- we specify the gauge via driving conditions for the lapse $\alpha$ and shift $\beta^i$ \cite{Alcubierre:2002kk}. The optimal gauge conditions are in general physics dependent, and $\grchombo$ allows the user to code in whichever gauge conditions are well adapted to the application at hand. However, the most commonly used conditions which have been implemented in $\grchombo$ are detailed below.

The \emph{alpha-driver} condition is usually written as a first order differential equation
\begin{equation}
\partial_t \alpha = -\mu_{\alpha_1}\alpha^{\mu_{\alpha_2}}K + \mu_{\alpha_3}\beta^i\partial_i \alpha. \label{eqn:alphadriver}\,
\end{equation}
The commonly used $1+\log$ slicing applicable for black hole inspirals corresponds to $\mu_{\alpha_1}=2$, $\mu_{\alpha_2}=1$ and $\mu_{\alpha_3}=1$. On the other hand, the \emph{maximal slicing} condition, which preserves $K=0$ and $\partial_t K=0$ at all slices, is a second order differential equation
\begin{equation}
D^2\alpha = \alpha[K_{ij}K^{ij}+4\pi(\rho+S)], \label{eqn:maximalslicing}
\end{equation}
which is useful for spherically symmetric collapse problems such as the critical scalar collapse scenarios.

We specify the evolution equation for $\beta^i$ using the \emph{gamma-driver} conditions \cite{Alcubierre:2002kk},
\begin{eqnarray}
\partial_t \beta^i& =& \eta_1 B^i\, \label{eqn:betadriver}\\
\partial_t B^i &=& \mu_{\beta_1}\alpha^{\mu_{\beta_2}}\partial_t \hat\Gamma^i-\eta_2 B^i\, ,\label{eqn:gammadriver}
\end{eqnarray}
where $B^i$ is an auxiliary vector field, while $\eta_1$, $\eta_2$, $\mu_{\beta_1}$ and $\mu_{\beta_2}$ are input parameters. The usual hyperbolic gamma-driver condition uses the parameters $\eta_1=3/4$, $\mu_{\beta_1}=1$, $\mu_{\beta_2}=0$ and $\eta_2=1$. We have also included parameters that allow us to turn on standard advection terms in \eqn{eqn:betadriver}--\eqn{eqn:gammadriver}. 

In our tests in Sec. \ref{sec-vacblack} and Sec. \ref{sec-choptuik}, where black holes are present, we manage the singularities with the so-called \emph{moving punctures method} \cite{Campanelli:2005dd,Baker:2005vv}, which is a combination of the $1+\log$ slicing for $\alpha$ and gamma-driver for $\beta^i$. In addition, we hard code the condition $\alpha>0$ as is usual practice.

\subsubsection{Constraint equations}

$\grchombo$ computes both the Hamiltonian constraint,
\begin{equation}
H = R + K^2-K_{ij}K^{ij}-16\pi \rho , \label{eqn:HamiltonianConst}
\end{equation}
and the momentum constraint,
\begin{equation}
M_i = \gamma^{jk}(\partial_l K_{ij}-\partial_iK_{jl}-\Gamma^m_{~jl}K_{mi}+\Gamma^m_{~ij}K_{lm})-8\pi S_i , \label{eqn:MomentumConst}
\end{equation}
in order to monitor the accuracy of the calculation. 

For the algebraic constraints of BSSN, we do not enforce (by hand) the condition that the conformal metric has a determinant of one, but we do enforce after each timestep that $\tA_{ij}$ is traceless.

\subsubsection{Scalar matter evolution equations} \label{eqn:scalarmatterevolve}

We have included a single minimally coupled scalar field $\phi$ as matter content
\begin{equation}
{L}_{\phi} = \frac{1}{2}\nabla_\mu \phi\nabla^{\mu} \phi + V(\phi) \label{eqn:scalaraction},
\end{equation}
with the equation of motion
\begin{equation}
\nabla_{\mu}\nabla^{\mu}\phi - \frac{dV}{d\phi}=0 \label{eqn:scalarEOM}.
\end{equation}
As is usual, we decompose the second order \eqn{eqn:scalarEOM} into two first order variables $\phi$ and $\Pi_M$
\begin{equation}
\Pi_M \equiv \frac{1}{\alpha}(\partial_t \phi -\beta^i\partial_i \phi) \label{eqn:phiM}.
\end{equation}
We note that our $\Pi_M$ is negative of $\Pi$ in some references, e.g. \cite{ShapiroBook}. \eqn{eqn:scalarEOM} is then decomposed into the following equations
\begin{equation}
\partial_t \phi = \alpha \Pi_M +\beta^i\partial_i \phi \label{eqn:dtphi}
\end{equation}
and
\begin{equation}
\partial_t \Pi_M=\beta^i\partial_i \Pi_M +\gamma^{ij}(\alpha\partial_j\partial_i \phi + \partial_j \phi\partial_i \alpha)+\alpha\left(K\Pi_M-\gamma^{ij}\Gamma^k_{ij}\partial_k \phi+\frac{dV}{d\phi}\right). \label{eqn:dtphiM} 
\end{equation} 
We also use the energy momentum tensor of the scalar field 
\begin{align}
T_{\mu \nu} = \nabla_\mu \phi \nabla_\nu \phi - \tfrac{1}{2} g_{\mu \nu} (\nabla_\lambda \phi \, \nabla^\lambda \phi + 2V)
\end{align} 
to calculate the matter components of the BSSN/CCZ4 system via \eqn{eq:Mattereqns}.

\subsection{$\grchombo$ code implementation} \label{sec-code}

$\grchombo$ is a physics engine built around the publicly-available adaptive-mesh framework $\mathtt{Chombo}$ \cite{Chombo}. $\grchombo$ solves the system of  hyperbolic partial differential equations of the Einstein equation and scalar matter content (see section \ref{sec-eqns}) using a finite difference scheme.

A key feature of $\grchombo$ is its highly flexible adaptive mesh refinement capability -- to be precise, $\grchombo$
uses Berger-Oliger style \cite{bergeroliger,BergerColella} adaptive mesh refinement with Berger-Rigoutsos \cite{BergerRigoutsis91} block-structured grid generation. $\grchombo$ supports full non-trivial mesh topology -- i.e. many-boxes-in-many-boxes. Morton ordering is used to map grid responsibility to neighbouring processors in order to optimize processor number scaling.

\subsubsection{Discretization and Time-stepping} \label{sect:discretization}

We would like to evolve a set of fields in space (the state-vector  $\Phi({\bf x},t) = \{\phi_1,\phi_2,\phi_3,\dots\}$) through time $t$ via the equations of motion
\begin{equation}
\frac{\partial \Phi}{\partial t} = {\cal F}(\Phi) ,
\end{equation}
where ${\cal F}$ is some operator on $\Phi$ which, in the case of the Einstein equation, is non-linear. 

In $\grchombo$, both the space and time coordinates are discretized. Evolution in time is achieved through time-stepping $t\rightarrow t+\Delta t$, where at each time step we compute the fluxes for each grid point individually. Time stepping is implemented using the standard 4th Order Runge-Kutta method, and hence, as usual, we only need to store the values of the state-vector at each time step.

$\Phi$ itself is discretized into a cell-centered grid. Spatial derivatives across grid points are computed using standard 4th order stencils for all spatial derivatives, except for advection terms which are implemented using an upwind stencil. The form of the stencils used exactly follows equations (2.2) through (2.6) of \cite{Zlochower:2005bj}.

\subsubsection{Berger-Rigoutsos Block-structured AMR} \label{sect:BRAMR}

$\grchombo$ uses {\tt Chombo}'s implementation of the Berger-Rigoutsos adaptive-mesh-refinement algorithm \cite{BergerRigoutsis91}, which is one of the standard block-structured AMR schemes.  Block-structured AMR regrids by overlaying variable size boxes, instead of remeshing on a cell-by-cell basis (the ``bottom-up'' approach). The main challenge is to find an efficient algorithm to \emph{partition} the cells which need regridding into rectangular ``blocks''. In this section, we will briefly discuss the algorithm.

For a given grid at some refinement level $l$ where $l=0$ is the base level and $l_{max}$ is some preset maximum refinement level, we first ``tag'' cells for which refining is required. The refinement condition used by $\grchombo$ is discussed later in this section. The primary problem of AMR is to efficiently partition this grid into regions which require adaptive remeshing. In \emph{block-structured} AMR these regions are boxes in 3D or rectangles in 2D. \emph{Efficiency} is measured by the ratio of tagged over untagged cell points in the final partitions. 

In each partition, we compute the \emph{signatures} or traces of the tagging function $f(x,y,z)$ of any given box
\begin{eqnarray}
X(x) &=& \int f(x,y,z) dy dz, \\
Y(y) &=& \int f(x,y,z) dx dz, \\
Z(z) &=& \int f(x,y,z) dy dx,
\end{eqnarray}
where $f(x,y,z) =1$ if it is tagged for refinement and $0$ otherwise. Given these traces, we can further compute the Laplacian of the traces $\partial^2_x X(x)$, $\partial^2_y Y(y)$ and $\partial^2_z Z(z)$. Given the Laplacians, the algorithm can search for all (if any) inflection points \emph{individually for each direction} -- i.e. the locations of zero crossings of the Laplacian, and then pick the one whose $\delta (\partial^2_i X_i)$ is the greatest (corresponding to the line -- or plane in 3D -- separating the largest change in the Laplacian). This point then becomes the line of partition for this particular dimension. Roughly speaking, this line corresponds to an edge between tagged and untagged cells in the orthogonal directions of the signature. Furthermore, if there exists a point $x_i$ with zero signature $X_i(x_i)=0$ (i.e. no cells tagged along the plane orthogonal to the direction), then this ``hole'' is chosen to be the line of partition instead.

After a partitioning, we check whether or not each partition is \emph{efficient}, specifically whether it passes a user-specified threshold or \emph{fill factor}, $\epsilon <1.0$, 
\begin{equation}
\frac{\mathrm{Tagged~Cells}}{\mathrm{Total~Cells}} > \epsilon
\end{equation}
If this is true, then we check if this box is \emph{properly nested}\footnote{Properly nested means that (1) a $l+1$ level cell must be separated from an $l-1$ cell by at least a single $l$ level cell and (2) the physical region corresponding to a $l-1$ level cell must be completely filled by $l$ cells if it is refined, or it is completely unrefined (i.e. there cannot be ``half-refined'' coarse cells).} \cite{BergerColella, bergeroliger} and if so we accept this partition and the partitioning for this particular box stops. If not, then we continue to partition this box recursively until either all boxes are accepted or partitioning no longer can be achieved (either by the lack of any tagged cells or reaching a preset limit on the number of partitions). Furthermore, $\grchombo$ allows one to set the maximum partition size, which if exceeded will force a partitioning of the box. 

Note that a higher value of $\epsilon$  means that the partitioning will be more aggressive which will lead to a higher efficiency in terms of final ratio of tagged to untagged cells -- generating more boxes in the process. However, this is not necessarily always computationally better as partitioning requires computational overhead, which depends on the number and topology of the processors. The ideal fill ratio is often a function of available processors, their topology and of course the physical problem in question.

A partitioned box is then \emph{refined}, i.e. its grids split into a finer mesh using the (user definable) refinement ratio $n^l = \delta x^{l+1}/\delta x^l$, and this process continues recursively until we either have no more tagged cells, or when we reached a preset number of refinement levels $l_{max}$.

Finally we need to specify  a prescription for tagging which cells are required to be refined. $\grchombo$ tags a cell when any (set of) user selected fields $\phi \in \Phi$ pass a chosen threshold $\sigma(\phi)$, which sets a limit on the $L^2$ norm of the change in the value of the field across that cell, i.e. 
\begin{equation}
f(x,y,z)=\left\{
  \begin{array}{cc}
   1 & \mathrm{if}~\sqrt{\sum\limits_{i=1}^{3}  (\phi({\bf x}+\delta x \, \hat{\bf x}_i) - \phi({\bf x}-\delta x \, \hat{\bf x}_i))^2} > \sigma(\phi) \\
    \\
   0 & \mathrm{otherwise}.
    \end{array}
    \right. \label{eqn:tagging} 
\end{equation}

This condition can be augmented, for example by using estimated truncation errors as tagging conditions instead.

Partitioning can be done at every time-step for each refinement level and this is a user preset choice per refinement level. However, the user may wish to select a lower frequency because it might be useful to not partition at every timestep for a given refinement level. One consideration is that it is important to let numerical errors dissipate (e.g. via Kreiss-Oliger dissipation, see Sec. \ref{sect:kreissoliger}) before remeshing. Once a new hierarchy of partitions is determined, we interpolate via linear interpolation from coarse to fine mesh, and average from fine to coarse mesh.

Since the finer mesh has a smaller Courant number, each mesh level's timestep is appropriately reduced via
\begin{equation}
\Delta t^{l+1} = \frac{\Delta t^l}{n^l}.
\end{equation}

$\grchombo$ follows standard Berger-Collela AMR evolution algorithm \cite{BergerColella}. Starting from the coarsest mesh, it advances the coarse mesh 1 time step i.e. $t \rightarrow t+\Delta t^l$. Then it advances the next finest mesh $n^l$ times until the fine mesh ``catches up'' with the coarse mesh time. Once both coarse and fine mesh are at the same time $t$, $\grchombo$ synchronizes them by averaging over the fine cells to the coarse cell values. We add that in a conservative system, this simple synchronization is not conservative and requires proper \emph{refluxing} -- the coarse fluxes are replaced with a time-averaged fine mesh fluxes. This step incurs additional overhead, and is at the moment not implemented by $\grchombo$ as GR equations are not conservative. Nevertheless, we intend to implement conservative refluxing as an option in a future version of $\grchombo$.

\subsubsection{Load Balancing}  \label{sect:loadbalance}

$\grchombo$'s efficiency when running on a large number of distributed-memory nodes is highly dependent on efficient load balancing of the available computational work across those nodes. Load balancing seeks to avoid the situation where most of the nodes are waiting for some small subset of nodes to finish their computational work, and it does this by seeking to distribute the amount of work to be done per time step evenly among all of the nodes. This can be non-trivial when AMR boxes at many different refinement levels are simultaneously being evolved across the system. In addition, even within a single node, multiple OpenMP threads might be running, and the per-node workload needs to be balanced amongst those threads.

For the inter-node load balancing, $\grchombo$ leverages {\tt Chombo}'s load balancing capabilities to distribute the AMR boxes among the available nodes. It does this by building a graph of the boxes to be distributed, adding edges between neighbouring and overlapping boxes. A bin packing / knapsack algorithm is used to balance the computational work among nodes, where the work is assumed to be proportional to the number of grid points, and then an exchange phase is used to minimise the communication cost. Because this load balancing procedure can be costly, we normally run it only every few time steps. In between runs of the load balancing procedure, new boxes generated by AMR refinement stay on the node which holds the parent box.

Within each node, the computational work is divided amongst the available OpenMP threads by iterating over the boxes to process using OpenMP's dynamic scheduling capability. This allows each thread to take the next available box from the queue of unprocessed boxes, instead of deciding ahead of time which boxes each thread will process. This is important because the boxes are varying in size. We generally divide even the coarsest level into multiple boxes so that it can be processed in parallel by multiple threads.

\subsubsection{Kreiss-Oliger Dissipation} \label{sect:kreissoliger}

In a finite difference scheme, instabilities can arise from the appearance of high frequency spurious modes. Furthermore, regridding generates errors an order higher than the typical error of the evolution operator, hence it is doubly crucial that we control these errors. The standard prescription to deal with this is to implement some form of numerical dissipation to damp out these modes. $\grchombo$ implements $N=3$ Kreiss-Oliger \cite{TUS:TUS1547} dissipation. In this scheme, for all evolution variables $u\in \{\tA_{ij},\tgamma_{ij},K,\chi,\Theta,\tGamma^{i}\}$, the evolution equations are modified as follows
\begin{equation}
\partial_t u_m \rightarrow \partial_t u_m + \frac{\sigma}{64 \Delta x}(u_{m+3}-6u_{m+2}+15u_{m+1}-20u_{m}+15u_{m-1}-6u_{m-2}+u_{m-3}),
\end{equation}
where $m\pm n$ labels the grid point $m$, $n$ the total offset from $m$ and $\sigma$ is an adjustable dissipation parameter usually of the order ${\cal O}(10^{-2})$. This 3rd order scheme is accurate as long as the integration order of the finite difference scheme is 5 or less (which it is in our implementation using 4th order Runge-Kutta).

\subsubsection{Boundary Conditions} \label{sect:BC}

$\grchombo$ supports both periodic (in any direction) boundary conditions, as well as any particular boundary conditions the user may want to specify (such as Neumann or Dirichlet types). A particular popular type of boundary condition is the so-called Sommerfield \cite{Alcubierre:2002kk} boundary condition, where out-going radiation is dissipated away. For any field $f$, we impose the condition at the boundary
\begin{equation}
\frac{\partial f}{\partial t} = -\frac{v x_i}{r} \frac{\partial f}{\partial x_i}  -v\frac{f-f_0}{r}
\end{equation}
where $r= \sqrt{x_1^2+x_2^2+x_3^2}$ is the radial distance from the center of the grid, $f_0$ is the desired space-time at the boundary (typically Minkowski space for asymptotically flat spacetimes) and $v$ the velocity of the ``radiation'', which is typically chosen to be 1.

\subsubsection{Initial conditions} \label{sect:IC}

$\grchombo$ supports several ways of entering initial conditions. 
\begin{itemize}
\item{Direct equations -- Initial conditions which are described by known analytic equations, such as the Schwarzchild solution, can be entered directly in equations form.}
\item{Checkpointing -- The {\tt HDF5} format output files from $\grchombo$ doubles as checkpointing files. A run can simply be continued from any previous state as long as its {\tt HDF5} output file is available. }
\item{Entering from data file -- $\grchombo$ allows one to insert data from a file.}
\item{Relaxation -- $\grchombo$ has a rudimentary capability to solve for the initial metric given some initial mass distribution, and assuming a moment of time symmetry and conformal flatness. Given a guess for $\chi$, $\grchombo$ relaxes it to the correct initial metric using a dissipation term which is proportional to a user chosen dissipation coefficient times the Hamiltonian constraint.}
\end{itemize}

The initial conditions used in this paper are mostly analytic or approximate analytic solutions, and so are entered directly into the code.  In the critical collapse, a Mathematica numerical solution as a function of the radius is interpolated onto the initial grid. 

\section{Using $\grchombo$ for new physics} \label{sec-newphys}

The primary advantage of $\grchombo$ over existing publicly available codes is its robust AMR ability. In this section we discuss several physical systems which can be studied with \texttt{GRChombo} but that would be hard (if not impossible) to simulate using codes based on moving-box mesh refinement. Most of the examples discussed in this section are still ``work in progress" by the authors and the results will be presented in separate publications. 

\subsection{Asymmetric scalar field bubbles}
\label{sec-newbubs}

One of the most fascinating and as yet not fully understood aspects of general relativity is the appearance of critical phenomenon in gravitational collapse as first discovered by Choptuik \cite{Choptuik:1992jv}. A comprehensive review can be found in \cite{Gundlach:2007gc}. 

Briefly, if we have an initial configuration, such as a Gaussian shaped bubble of scalar field, and allow this to evolve under the action of gravity, the result will be either the formation of a black hole, or dispersal of the field to infinity depending on the ``strength" of the initial data. Varying any one initial parameter $p$ of the configuration (such as the height of the bubble), one finds that there is a critical point $p^*$ at which the transition between the two end states occurs, and that the mass of the black hole created on the supercritical side follows the scaling relation,

\begin{equation}
M \propto (p - p^*)^{\gamma},
\end{equation}
where the scaling constant $\gamma$ is universal in the sense that it does not depend on the choice of family of initial data. For a massless scalar in a spherically symmetric collapse, $\gamma$ has been numerically determined to be around 0.37.

The other key phenomenon which is observed is that of self-similarity in the solutions, or ``scale-echoing''. Close to the critical point, and in the strong field region, the value of any gauge independent field $\phi$ at a point $x$ and time $T$ exhibits the following scaling relation,

\begin{equation}
\phi (x,t) = \phi(e^{\Delta} x, e^{\Delta} T) ,
\end{equation}
where $\Delta$ is a dimensionless constant with another numerically determined value of 3.44 for a massless scalar field in the spherical case. The time $T$ here is measured ``backwards" - it is the difference between the critical time at which the formation of the black hole occurs and the current time, with time being the proper time measured by a central observer. 

What one sees is therefore that, as the time nears the critical time by a factor of $e^{\Delta}$, the same field profile is seen but on a scale $e^{\Delta}$ smaller. This scale-echoing may be either continuous or discrete. In Choptuik's seminal paper \cite{Choptuik:1992jv}, a 1+1 adaptive mesh code was used to study such behaviour near the critical point. Since then there has been some progress in studying the phenomenon in non spherically symmetric cases, see \cite{Abrahams:1993wa,Choptuik:2003ac,Sorkin:2010tm,Healy:2013xia,Hilditch:2013cba,Hilditch:2015aba}, but progress in making firm conclusions has been slower than expected, due to the extremely high refinements required to study the stages of the collapse, which are magnified three-fold in full $3+1$ codes. 

\begin{figure}[t]
\subfigure{
\includegraphics[width=.425\textwidth]{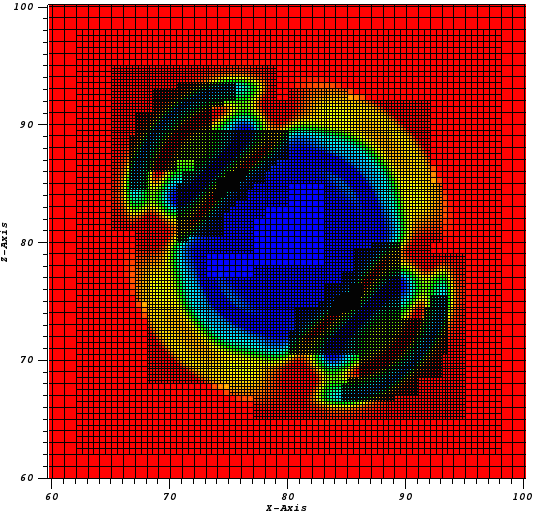}
}
\subfigure{
\includegraphics[width=.45\textwidth]{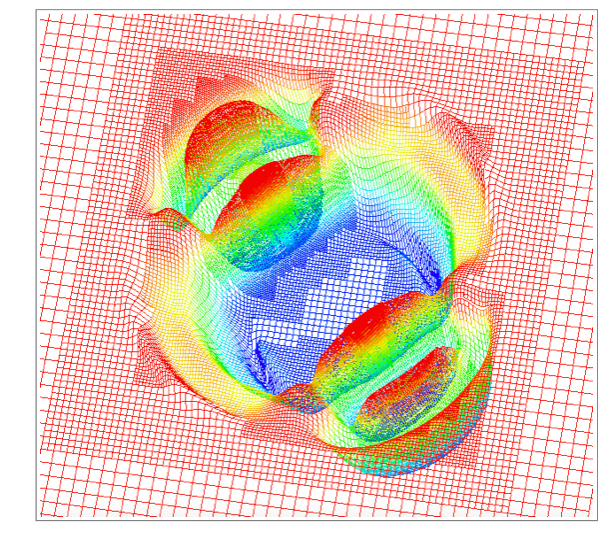}
}
\caption{An asymmetric scalar field bubble collapse: In the left image, a slice through the centre of the field bubble is shown at a particular time during the collapse, with the instantaneous mesh superimposed. This same slice is shown in 3D in the right image to better illustrate how the mesh is adapted to the curvature. The AMR efficiently tracks the scales in the profile. In addition, new levels are added as the profile shrinks, leading to a more efficient computation in terms of memory and computational resources, and requiring less ``human input'' ahead of and during the run.
\label{fig-BubCol}}
\end{figure}

We are currently investigating the problem of asymmetric bubble collapse with $\grchombo$. A snapshot of the evolution of one such example is shown in Figure \ref{fig-BubCol}. The mesh refines so that the field profile is consistently resolved as the critical time is approached. Note that since the profile is highly irregular, with a range of length scales represented in different (disconnected) regions, $\grchombo$ provides a significant advantage over moving-box mesh refinement in terms of computational cost. In addition to the adaptation of the mesh to the local curvature, $\grchombo$ can automatically add in new levels of refinement during the evolution, which allows the scale-echoing behaviour to be probed consistently as the profile shrinks even in the absence of spherical symmetry. The results will be discussed in detail in a separate publication.

\subsection{Higher dimensional black holes/anti-de Sitter}
\label{sec-newother}

In recent years it has been realized that the dynamics of general relativity, even in vacuum, beyond the traditional asymptotically flat four-dimensional setting is much richer than previously anticipated. Some of these new directions involve considering more than four spacetime dimensions and/or new boundary conditions, such anti-de Sitter (AdS) or Kaluza-Klein (KK) asymptotics. Black holes, as primary objects in any theory of gravity, in these new set ups exhibit two important new physical phenomena: Firstly, black holes with topologically non-spherical horizons are possible. The black ring of \cite{Emparan:2001wn} is the first example of an asymptotically flat vacuum black hole with a non-spherical horizon. By now it is clear that this solution is the tip of the iceberg, and many more new types of black holes are known to exist. Secondly, vacuum black holes can be dynamically unstable under gravitational perturbations, as Gregory and Laflamme first showed in the case of black strings in asymptotically KK spacetimes \cite{Gregory:1993vy} . Rapidly spinning asymptotically flat vacuum (and AdS) black holes suffer from these Gregory-Lafllamme-type-of instabilities \cite{Emparan:2003sy,Dias:2009iu,Dias:2010eu}, and new  types of non-axisymmetric instabilities \cite{Shibata:2010wz}.  In fact, anti-de Sitter itself is non-linearly unstable to the formation of black holes \cite{Bizon:2011gg}, and the process is turbulent.  There is a lot of interest in studying the dynamics of gravity, and black holes, in AdS motivated by the gauge/gravity correspondence \cite{Maldacena:1997re}.  See \cite{Emparan:2008eg,Horowitz:2012nnc} for some (relatively recent) reviews with references. 

In a remarkable paper, \cite{Lehner:2010pn} studied the endpoint of the Gregory-Laflamme instability of black strings in five-dimensions. This paper gave convincing evidence that the black string would pinch off in finite asymptotic time, thus providing a potential counter-example of the weak cosmic censorship conjecture in non-asymptotically flat vacuum spacetimes. The evolution the Gregory-Laflamme instability for black strings showed that the horizon develops a fractal structure, with thin necks connecting bulges at different scales. Moreover, the non-linear instability of AdS is of a turbulent nature \cite{Bizon:2011gg}, and in fact AdS black holes also suffer from turbulent type-of-instabilities \cite{Holzegel:2011uu,Carrasco:2012nf,Adams:2013vsa,Yang:2014tla}. A common feature in all these instabilities is that in the fully non-linear regime, new length scales are dynamically generated and a priori one does not know where they will appear. Therefore, if one wants to use numerical GR to determine the endpoints of these instabilities, one needs a numerical method that can automatically resolve these newly generated length scales. Therefore, it seems that full AMR is not an option but a necessity.\footnote{The AMR capabilities of the \texttt{PAMR/AMRD} code were essential in \cite{Lehner:2010pn}.} In addition, with \texttt{GRChombo} one should be able to simulate higher dimensional spacetimes with no symmetry assumptions using the fact that the \texttt{Chombo} core is dimension independent up to six spatial dimensions. This should find applications to the AdS/CFT correspondence, where one may be interested in simulating $(4+1)$-dimensional asymptotically AdS spacetimes in full generality.

\begin{figure}
\begin{center}
\subfigure{
\includegraphics[scale=0.215]{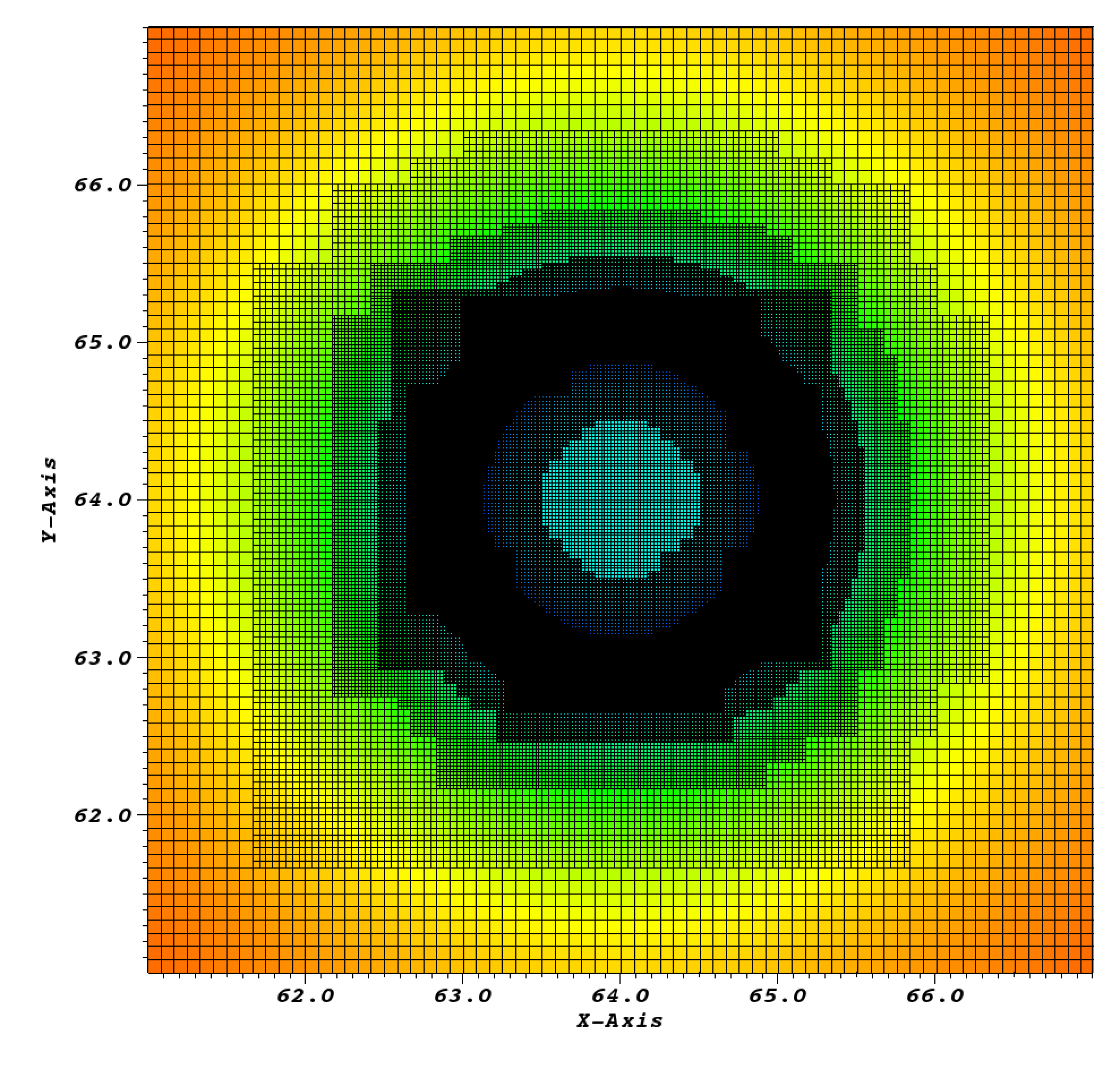}
}
\subfigure{
\includegraphics[scale=0.215]{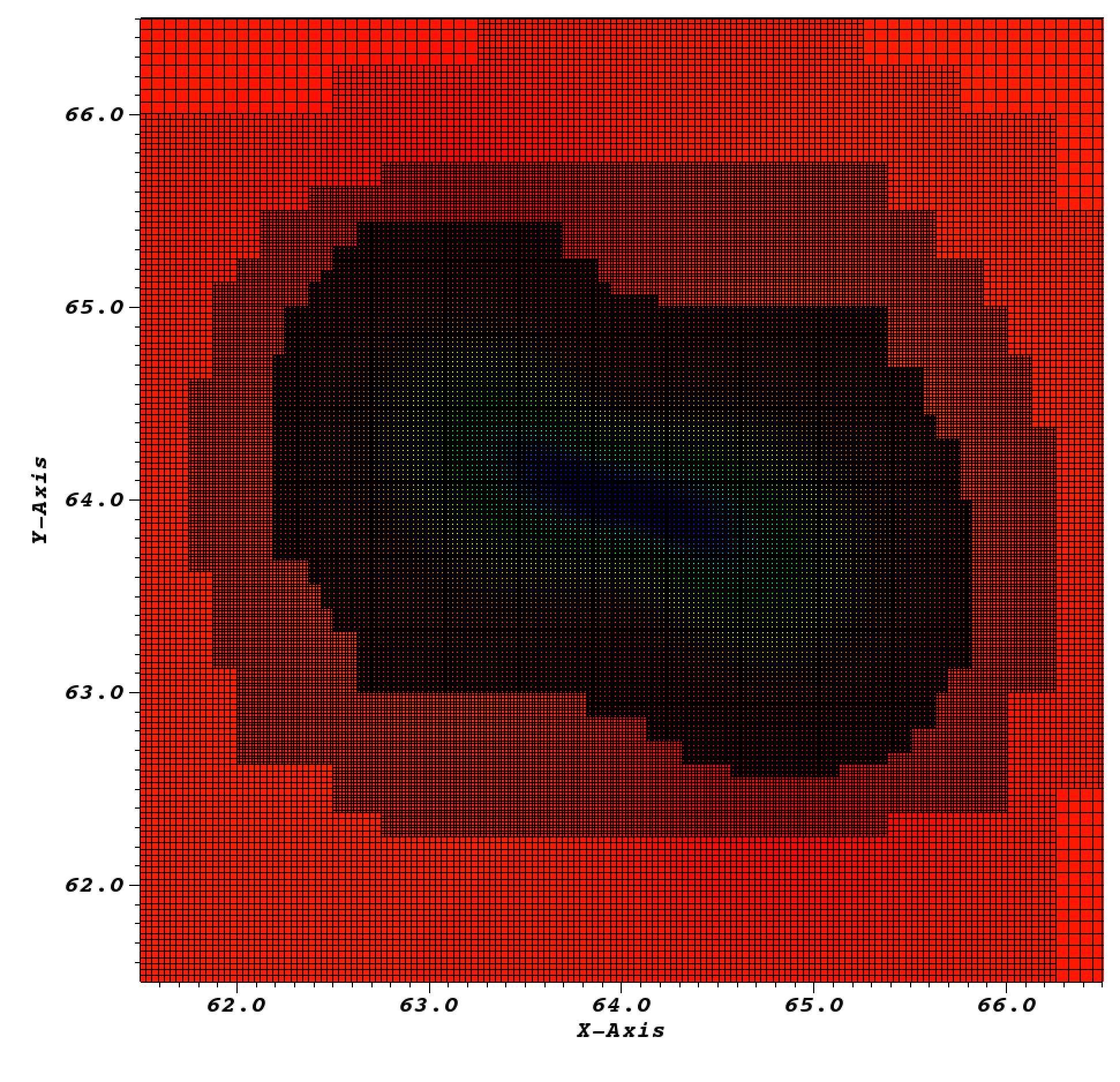}
}
\caption{Snapshots of the rotation plane of dynamically generated meshes by \texttt{GRChombo} during the evolution of the instabilities of a black ring (\textit{left}) and of a rapidly spinning spherical black hole (\textit{right}). AMR ensures that the mesh is adapted to the non-trivial topology of the horizon and new levels are added where new structure appears. This is essential in order to have enough resolution where it is needed while keeping the computational cost of the simulation under control.
\label{fig-Ring}}
\end{center}
\end{figure}

Some of the aforementioned instabilities are currently being investigated using \texttt{GRChombo} and will be discussed elsewhere. In Fig. \ref{fig-Ring} we display snapshots of the meshes that are dynamically generated by \texttt{GRChombo} during evolution for the case of an unstable black ring (\textit{left}) and a higher dimensional black hole (\textit{right}).  \texttt{GRChombo} not only can adapt the mesh to the non-trivial topology of the horizon but it also adds new levels in regions where new structures appear during the evolution. This essential capability ensures that all relevant length scales are correctly resolved whilst keeping the computational cost of the simulation under control. For these simulations, moving-box mesh refinement would be prohibitively expensive and hence it is not a realistic option. Previous works on higher dimensional black hole physics in asymptotically flat spaces using moving boxes include \cite{Zilhao:2010sr,Yoshino:2009xp}. In AdS, the code of \cite{Bantilan:2012vu} is based on \texttt{PAMR/AMRD}, whilst \cite{Chesler:2013lia} uses a fixed domain decomposition and pseudospectral discretisation.

\subsection{$N$-body problems}
\label{sec-newBHs}

\begin{figure}
\begin{center}
\subfigure{
\includegraphics[width=.3\textwidth]{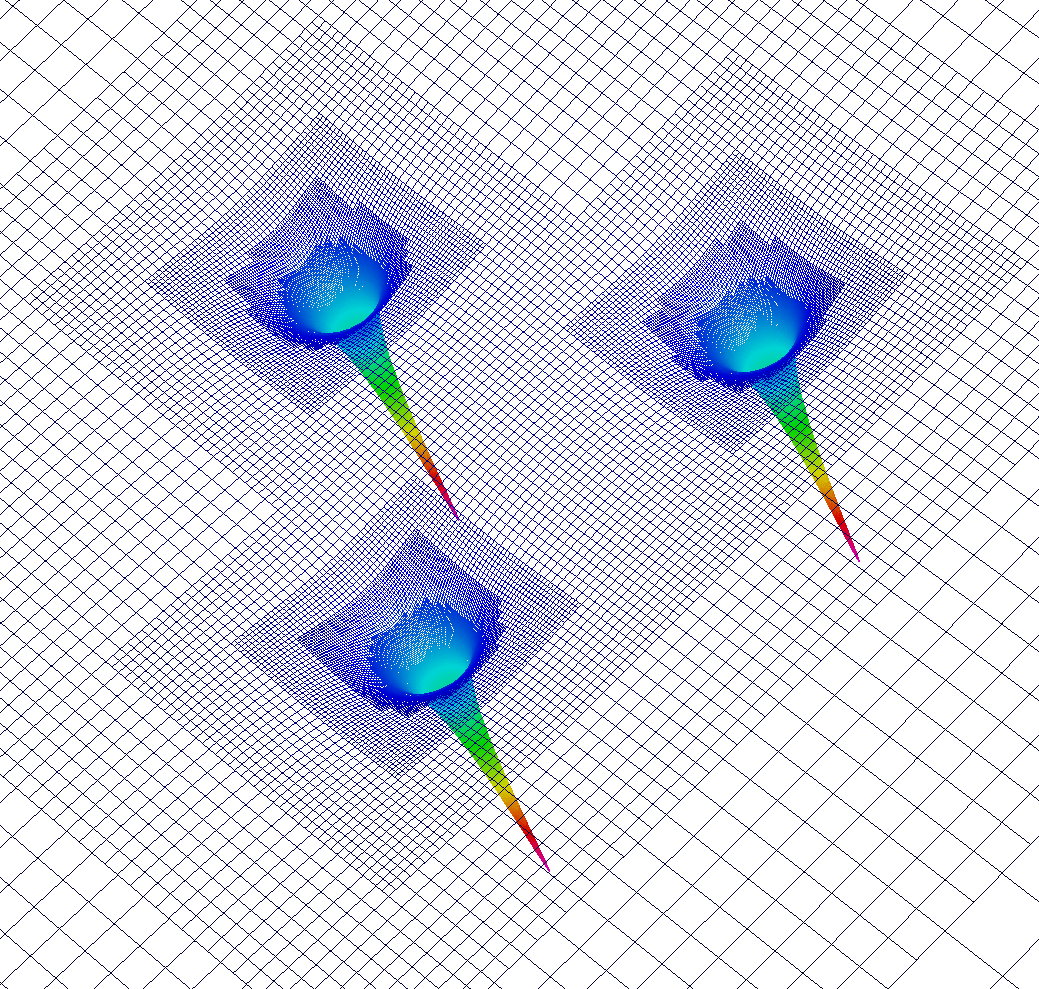}
}
\subfigure{
\includegraphics[width=.3\textwidth]{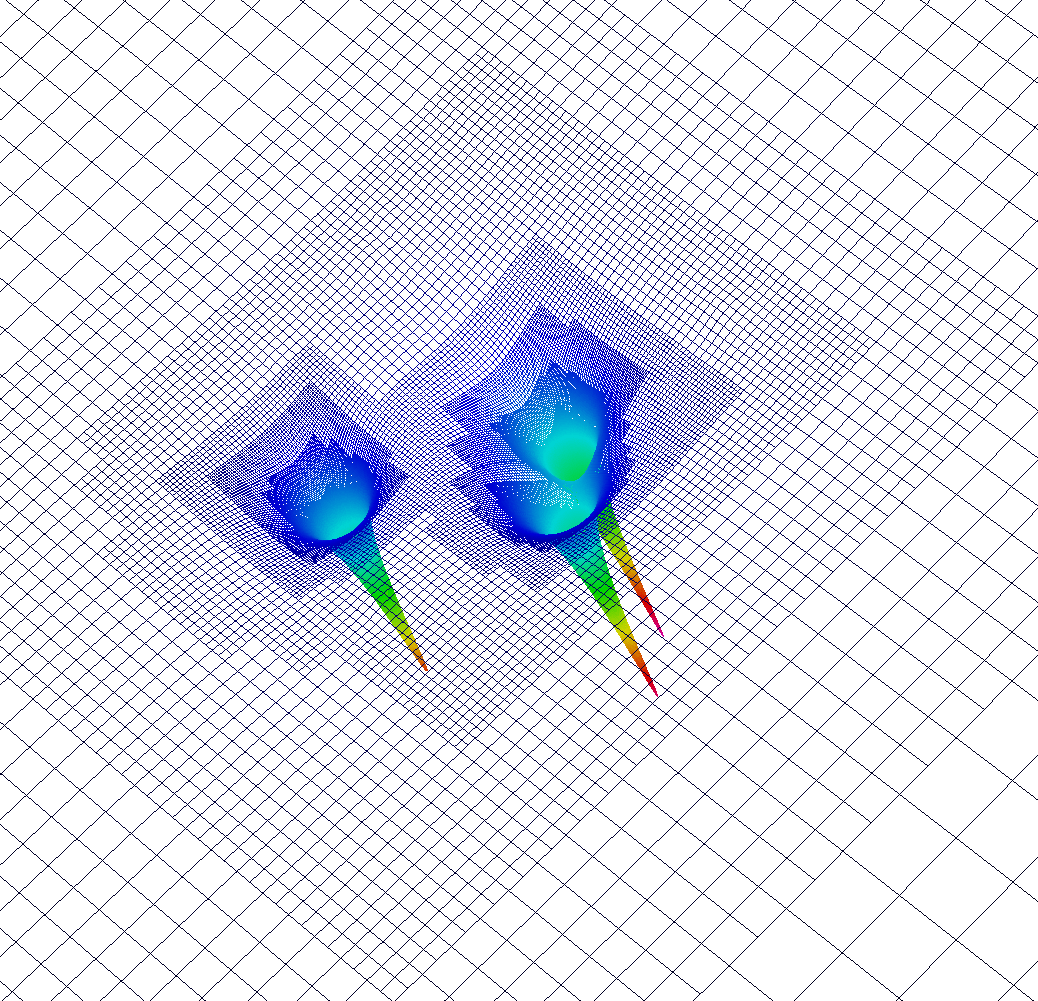}
}
\subfigure{
\includegraphics[width=.3\textwidth]{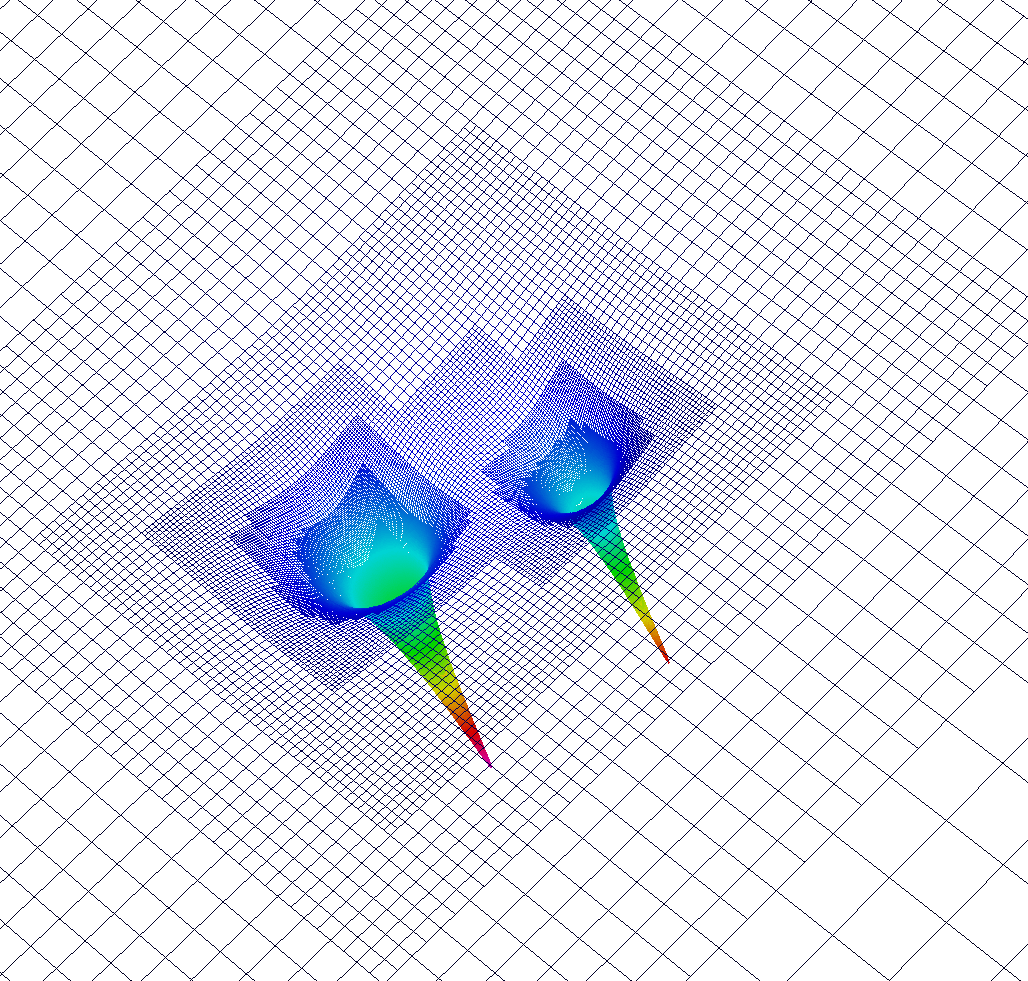}
}
\caption{Triple black hole merger: Three black holes are evolved with $\grchombo$. The mesh is shown, which has adapted to the local curvature in $\chi$, the variable plotted.
\label{fig-TriBH}}
\end{center}
\end{figure}

Simulations of single black holes and binary mergers often make use of symmetries in the problem or Newtonian approximations to predict the levels of resolution required at each point in space, for each time-step. In three (or more) body problems, the trajectories of the objects are generally not known ahead of time and must be calculated numerically. The shift vector can be used to predict the movement of black hole centres locally, but this tracking must be added in for each black hole and the boxes adjusted accordingly. For $\grchombo$, it is trivial to add multiple black holes to a spacetime, without actively tracking their central points, and so these many-body systems are as easy to set up as binary ones (although they clearly require greater computational resource to run). For example, the $\grchombo$ mesh can be seen adapting at each timestep in the triple black hole merger shown in Figure \ref{fig-TriBH}. 

\begin{figure}
\begin{center}
\subfigure{
\includegraphics[width=.3\textwidth]{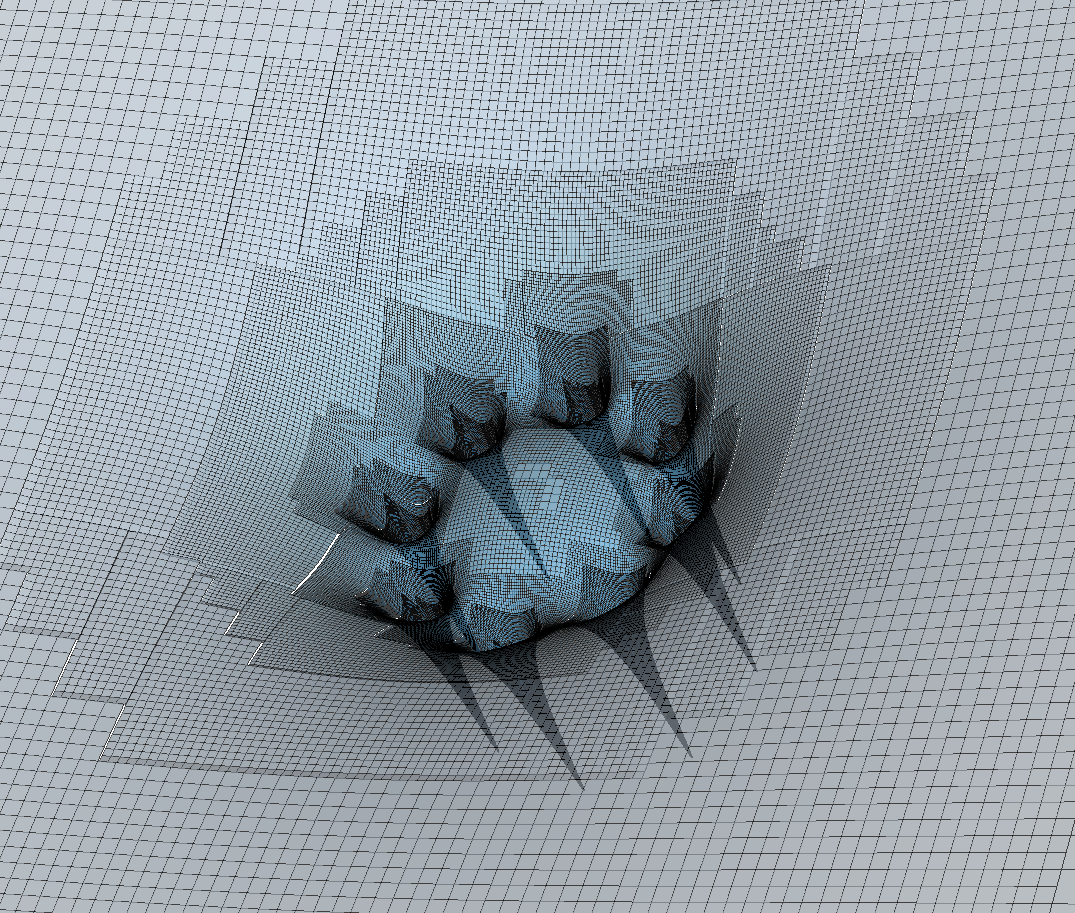}
}
\subfigure{
\includegraphics[width=.3\textwidth]{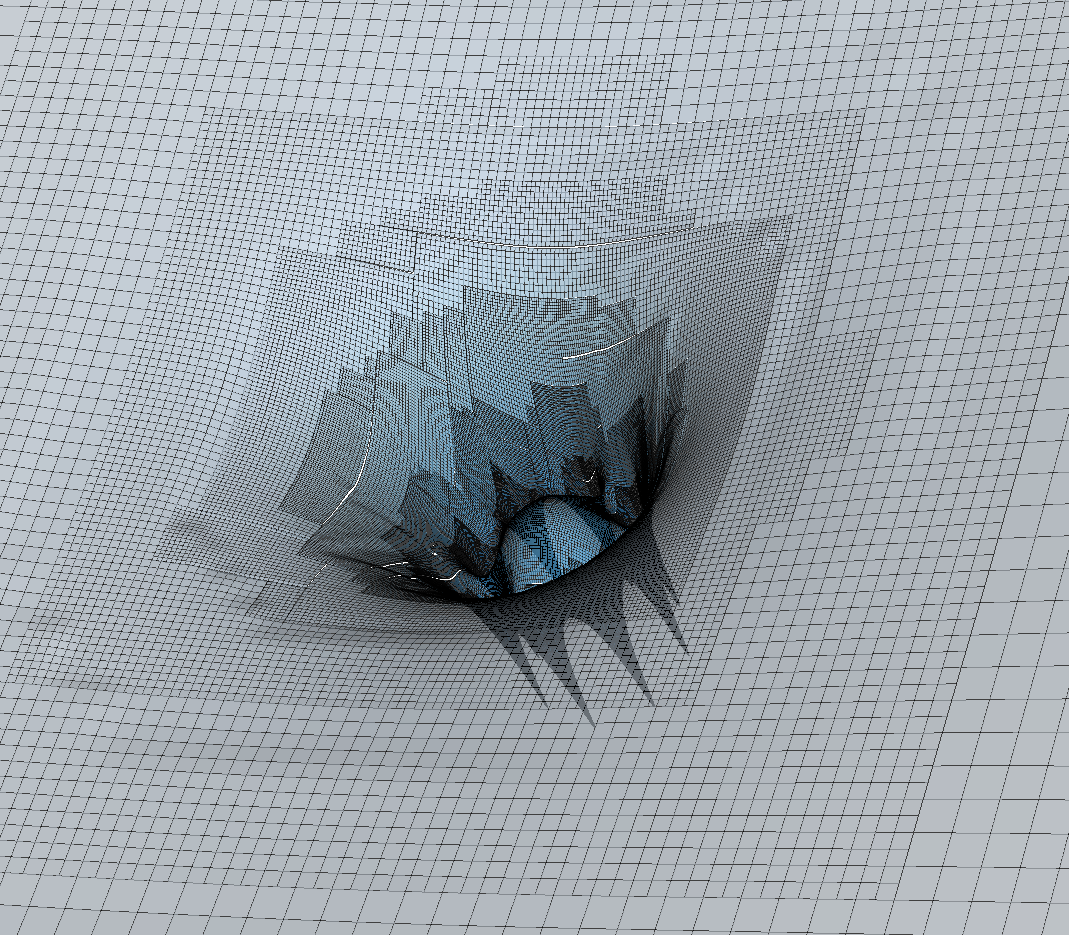}
}
\subfigure{
\includegraphics[width=.3\textwidth]{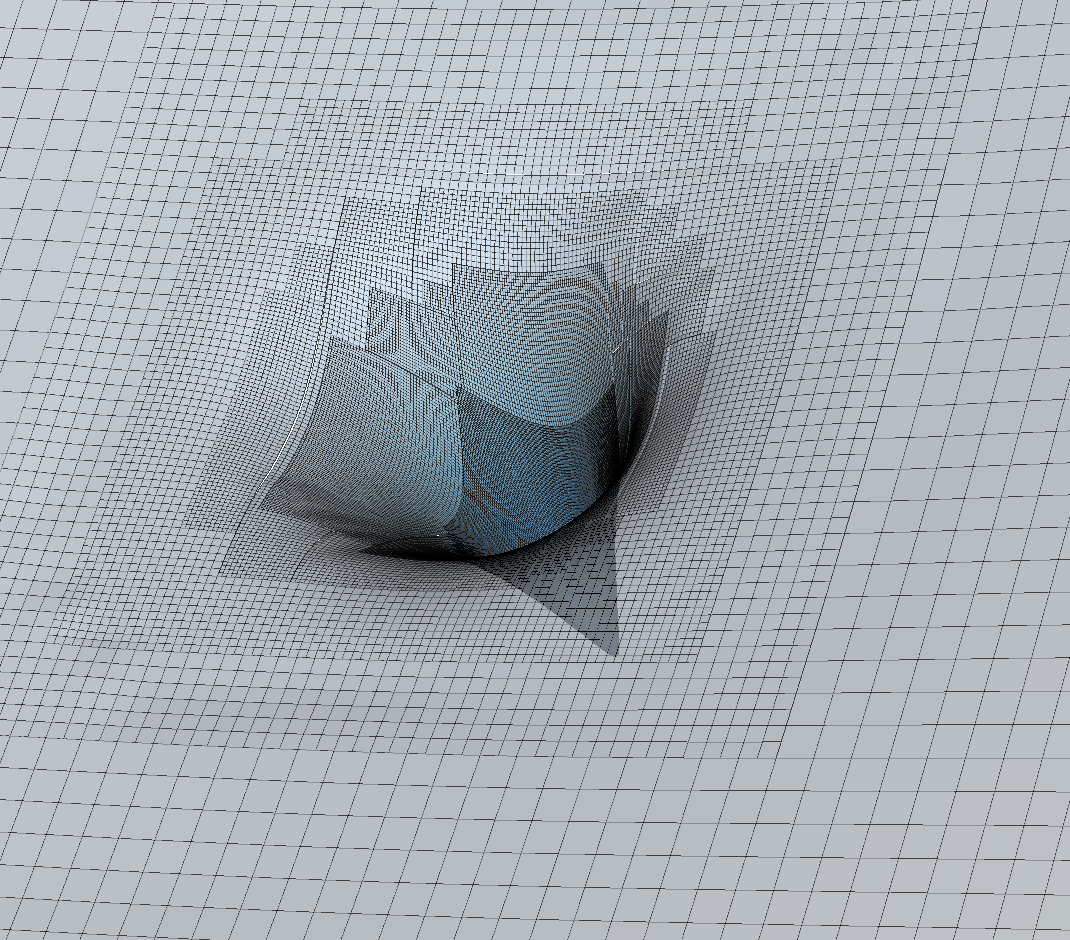}
}
\caption{Multiple Black Hole merger: A ring of black holes are evolved with $\grchombo$. The mesh is shown, which has adapted to the local curvature in $\chi$, the variable plotted.
\label{fig-RingBH}}
\end{center}
\end{figure}

Another (albeit rather contrived) example is the ring of black holes shown in Figure \ref{fig-RingBH}. The set up of this ring was no more difficult than that of the binary or triple black hole spacetime -- $\grchombo$ automatically remeshes the grid given a set of analytic initial conditions without further user intervention, and there is no need to try and predict the paths of the black holes individually. It is clear that other fields of research, such as magnetohydrodynamics, would benefit greatly from this ability to maintain a consistent level of resolution throughout a simulation, and make this resolution follow the inherently unpredictable movement in a many body system. 

\section{Testing $\grchombo$}
\label{sec-tests}

We detail the results of the standard Apples with Apples tests \cite{Babiuc:2007} in Sec. \ref{sec-apples} when turning off AMR and using fixed resolution grids. In Sec. \ref{sec-vacblack} we turn on the AMR abilities of the code and demonstrate that it can stably evolve spacetimes containing black-hole-type singularities. In \ref{sec-headon} we study convergence of our code in a head on collision of two black holes.  In Sec. \ref{sec-choptuik} we demonstrate the ability of the code to evolve matter content by considering scalar fields with gravity, by recreating the results of the sub-critical and critical cases of Choptuik scalar field collapse detailed in \cite{AlcubierreBook}. In Sec. \ref{sec-scaling} we discuss the weak scaling properties of the code on the Mira supercomputer.

The test figures referred to in this section can be found in Appendix \ref{sec-appendix}.

\subsection{Apples with Apples tests}
\label{sec-apples}

In this section we describe the results of applying the code to the standard Apples with Apples tests in \cite{Babiuc:2007}. Here we give a brief description of the key features of the tests, but the reader should refer to the paper for full specifications. Where we do not specify details, our treatment can be assumed to follow that of the standard tests. The AMR capabilities of the code are not utilised in these tests, which were designed for a uniform resolution, in order to make our results comparable to other codes. (We consider the effects of regridding on code performance in Section \ref{sec-vacblack}.)

\subsubsection{Robust stability test}
\label{subsubsec:rob}

The robust stability test introduces small amounts of random noise, scaled with the grid spacing, to all of the evolution variables, in order to test the code's robustness against numerical noise. The test was conducted at resolutions of $\rho = 4$, $\rho=2$ and $\rho=1$, corresponding to grid spacings of 0.005, 0.01 and 0.02 respectively. No dissipation was added in the test. 

As shown in Figure \ref{fig-robust}, the error growth in the evolution variables did not increase with increasing grid resolution, and the Hamiltonian constraint $H$ did not grow more for higher resolutions. Therefore, we conclude that the test is passed. 

\subsubsection{Linear wave test}

A wave of fixed amplitude is propagated across the grid in the $x$-direction with periodic boundary conditions. The amplitude is small enough that the non-linear terms are below numerical precision, such that the behaviour under the Einstein equation is approximately linear. The test measures the errors in magnitude and phase introduced by the code after 1000 crossing times. 

As can be seen from Figure \ref{fig-linear}, this error is 12 orders of magnitude smaller than the signal and therefore negligible.

\subsubsection{Gauge wave tests}

The BSSN formulation is known to produce unsatisfactory results for the gauge wave tests. $\grchombo$ is no different in this respect. As can be seen in Figure \ref{fig-gauge}, it becomes unstable after around 50 crossing times, with the Hamiltonian constraint increasing exponentially, even for a relatively small initial amplitude of the gauge wave of $A = 0.1$. 

As was shown in \cite{Alic:2011gg} stability can be achieved by adding in the CCZ4 constraint damping terms. $\grchombo$ shows exactly this behaviour (figure \ref{fig-gauge}).

\subsubsection{Gowdy wave test}

The Gowdy wave evolves a strongly curved spacetime: an expanding vacuum universe containing a plane polarised gravitational wave propagating around a 3-torus. 
In the expanding direction we use the analytic gauge, $\partial_t \alpha = -\partial_t \sqrt{\gamma_{zz}}$. The collapsing direction is evolved starting at $t=t_0$ with harmonic slicing for the lapse and zero shift. A Kreiss-Oliger dissipation coeffecient of $\sigma=0.05$ was used in both directions. 

The results for both the BSSN and CCZ4 codes in the collapsing direction are shown in Figure \ref{fig-gowdy-collapsing}, and in the expanding direction in Figure \ref{fig-gowdy-expanding}.

As is found in the Apples with Apples tests \cite{Babiuc:2007} for other simple BSSN codes, $\grchombo$ with BSSN and CCZ4 gives a less than satisfactory performance in this test in the expanding direction. The evolution is stable for approximately the first 30 crossing times, after which high frequency instabilities develop and cause code crash, due to the exponentially growing $\gamma_{zz}$ component. In \cite{Babiuc:2007} it was found that this behaviour of BSSN could be controlled with dissipation, but that long term accuracy was not achievable. 

In the contracting direction the evolution is stable for the full 1000 crossing times and we were able to confirm the convergence of our code. As shown in Figure \ref{fig-gowdy-convergence}, both BSSN and CCZ4 exhibit 4\textsuperscript{th} order convergence initially. While convergence is never lost, the order is reduced at later times. This is similar to the behaviour found in \cite{Babiuc:2007} and \cite{Cao:2011fu}.


\subsection{Vacuum black hole spacetimes}
\label{sec-vacblack}

In this subsection we show that our code can stably evolve spacetimes containing black holes. 

All the simulations presented here used the BSSN formulation of the Einstein equations, along with the gamma-driver and alpha-driver gauge conditions. Adding CCZ4 constraint damping gives better performance for the Hamiltonian constraint, as would be expected, but the results are broadly similar and so are not presented here. Unless otherwise stated, we perform the simulations with up to 8 levels of refinement and  we based our tagging/regridding criterion, \eqn{eqn:tagging}, on the value of $\chi$. We emphasise that the purpose of this subsection is to demonstrate that we can stably evolve black hole spacetimes, but we are not interested in extracting gravitational wave data or in studying convergence; this will be done in the next subsection. 

Where we refer to taking an $L^2$ norm of the Hamiltonian constraint $H$ in a test, this is calculated as follows (using the weighted variable sum function in VisIt):

\begin{equation}
|| H ||_{2} = \sqrt{\sum_i m_i H_i^2} , \label{eqn:L2norm}
\end{equation}
where $m_i$ is $V_i/V_{tot}$, the fraction of the total grid volume $V$ occupied by the $i$th box. Where the grid contains a black hole, we excise the interior by setting $H$ to zero within the region in which the lapse $\alpha$ is less than 0.3 (which is an approximate rule of thumb for the location of an event horizon for a black hole in the moving puncture gauge). The difference in the results is small, since the error norm is dominated by regridding errors at the boundaries between meshes. We also exclude the values on the outer boundaries of the grid, which can distort the results in cases where periodic boundaries are used. 

\subsubsection{Schwarzschild black hole}

First we evolve a standard Schwarzschild black hole in isotropic gauge, with a conformally flat metric, the lapse initially set to one everywhere, and the conformal factor $\chi$:

\begin{equation}
\chi = \left( 1+\frac{M}{2r} \right)^{-2} \label{eqn:SCchi}
\end{equation}

\begin{figure}
\begin{center}
\includegraphics[width=.6\textwidth]{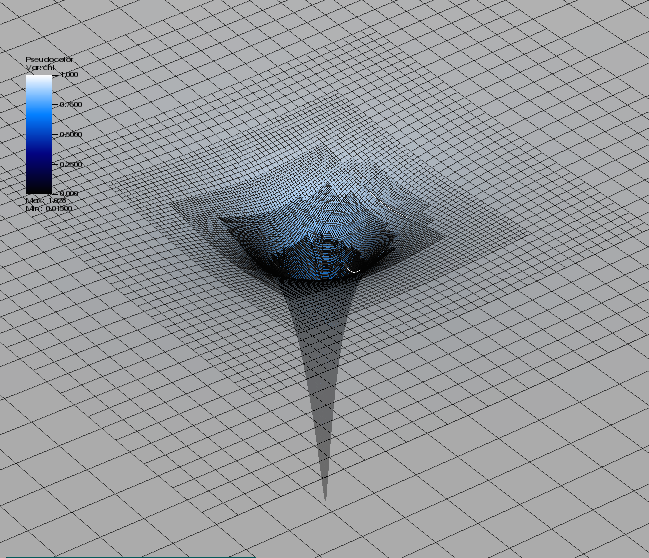}
\caption{The profile for $\chi$ through a slice perpendicular to the z axis is shown.
\label{fig-profile}}
\end{center}
\end{figure}

In this simulation, we chose the outer boundary of the domain to be at $600M$ and the spatial resolution in the coarsest mesh is $10M$.  We impose Sommerfeld boundary conditions. The initial value of $\chi$ through a slice is shown in Figure \ref{fig-profile}. We see the expected ``collapse of the lapse" at the singularity and the solution quickly stabilises into the ``trumpet" puncture solution described in \cite{Hannam:2008sg}. We find an apparent horizon and are able to evolve the black hole stably and without code crash for well over $t = 10000M$ time steps as shown in Figure \ref{fig-SC} (left). In this figure we show the $L^2$ norm of the Hamiltonian constraint across the whole grid, and it remains bounded throughout the evolution.

We monitor the ADM mass of the black hole by integrating over a surface near the asymptotically flat boundary, as seen in Figure \ref{fig-SC} (right). We also monitor the angular momentum and linear momentum of the black hole, and find that these remain zero as expected, as shown in Figure \ref{fig-SC} (right). These simple ADM measures rely on asymptotic flatness at the surface over which they are integrated, and so are sensitive to errors introduced by reflections at the boundaries, initial transients from approximate gauge choice or if the black hole is moving nearer the boundary (as in the boosted case). They are therefore less reliable as the simulation progresses, and we use them simply to confirm that we are evolving the correct spacetime initially.

\subsubsection{Kerr black hole}

In this sub-subsection we present the results of a simulation of the Kerr black hole spacetime in quasi isotropic gauge as in \cite{Brandt:1996si} with the angular momentum parameter $a = J/M$ set to 0.2. The domain size was chosen to be $(320M)^3$ and the grid spacing in the coarsest level was $4M$. We impose periodic boundary conditions for simplicity, which limits the duration of the simulation due to boundary effects. 

In Figure \ref{fig-KE} (left) we show the $L^2$ norm of the Hamiltonian constraint throughout the evolution. This plot shows that the amount of constraint violation remains stable during the simulation. In the right panel of Figure  \ref{fig-KE}  we display the ADM measures for the three components of the angular momenta and the mass. This Figure shows that these quantities remain (approximately) constant during the simulation.  

\subsubsection{Boosted black hole}

\begin{figure}
\begin{center}
\subfigure[Initial position]{
\includegraphics[width=.3\textwidth]{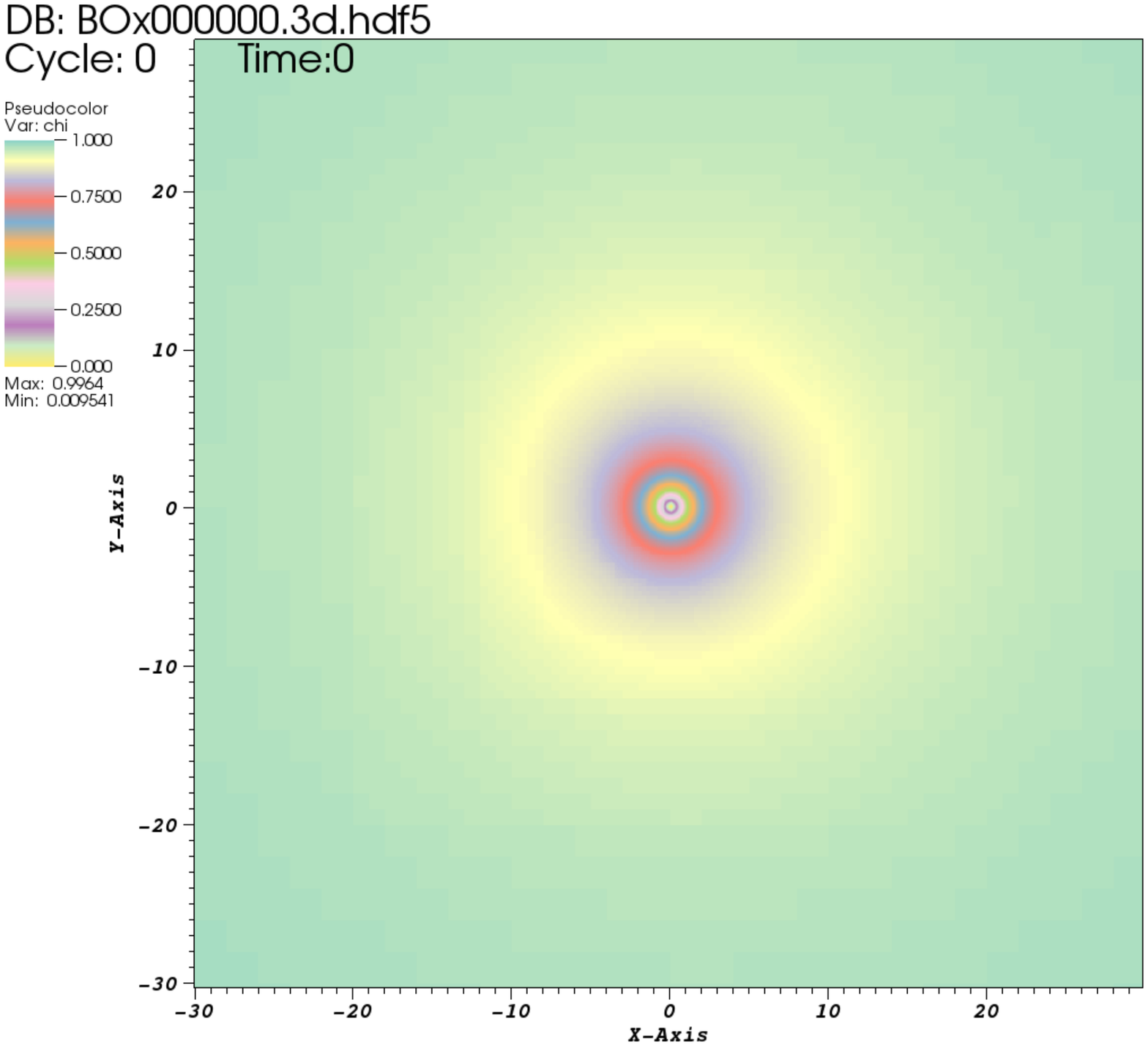}
}
\subfigure[Position at $t=100$]{
\includegraphics[width=.3\textwidth]{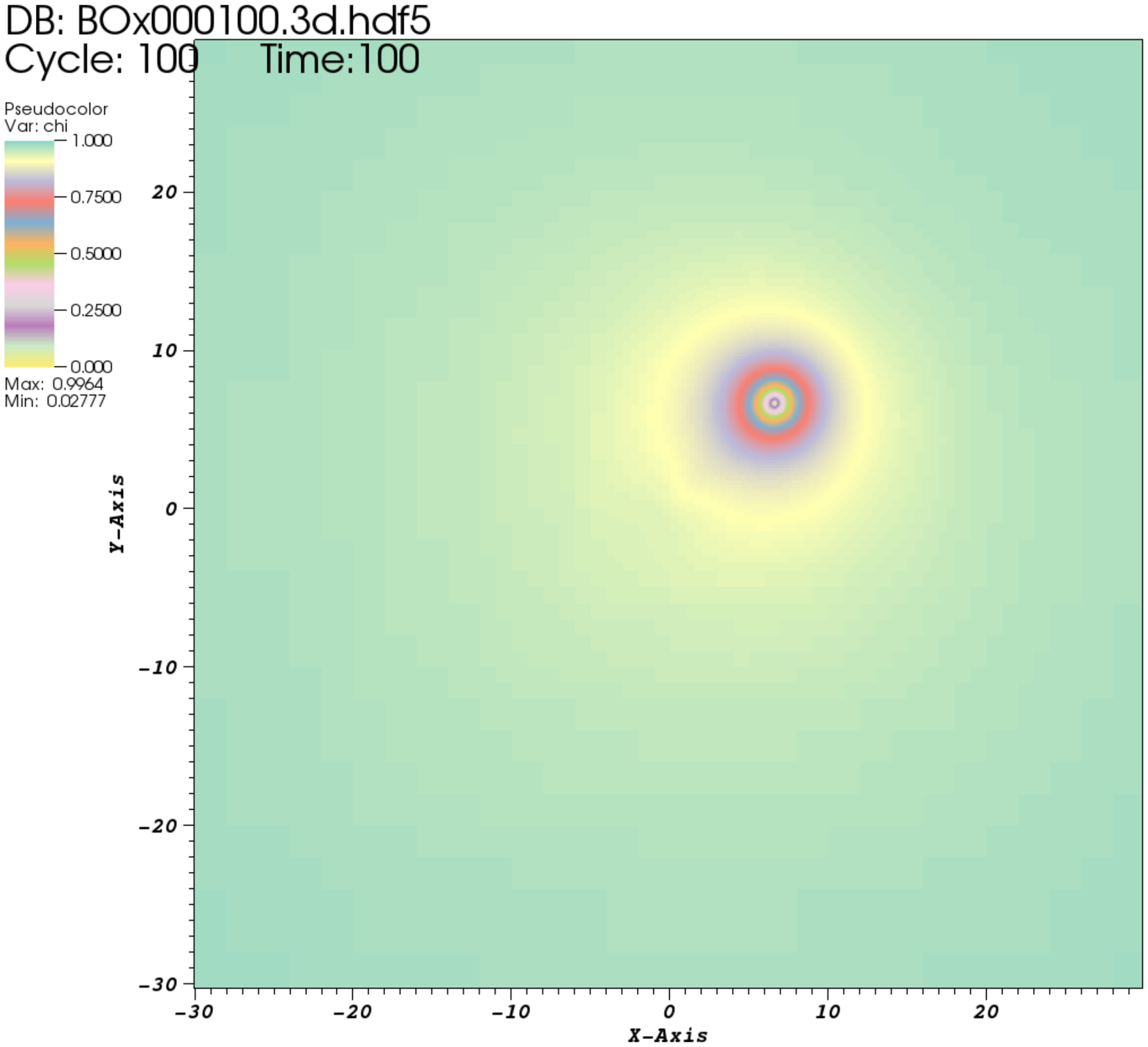}
}
\subfigure[Grid at $t=100$]{
\includegraphics[width=.3\textwidth]{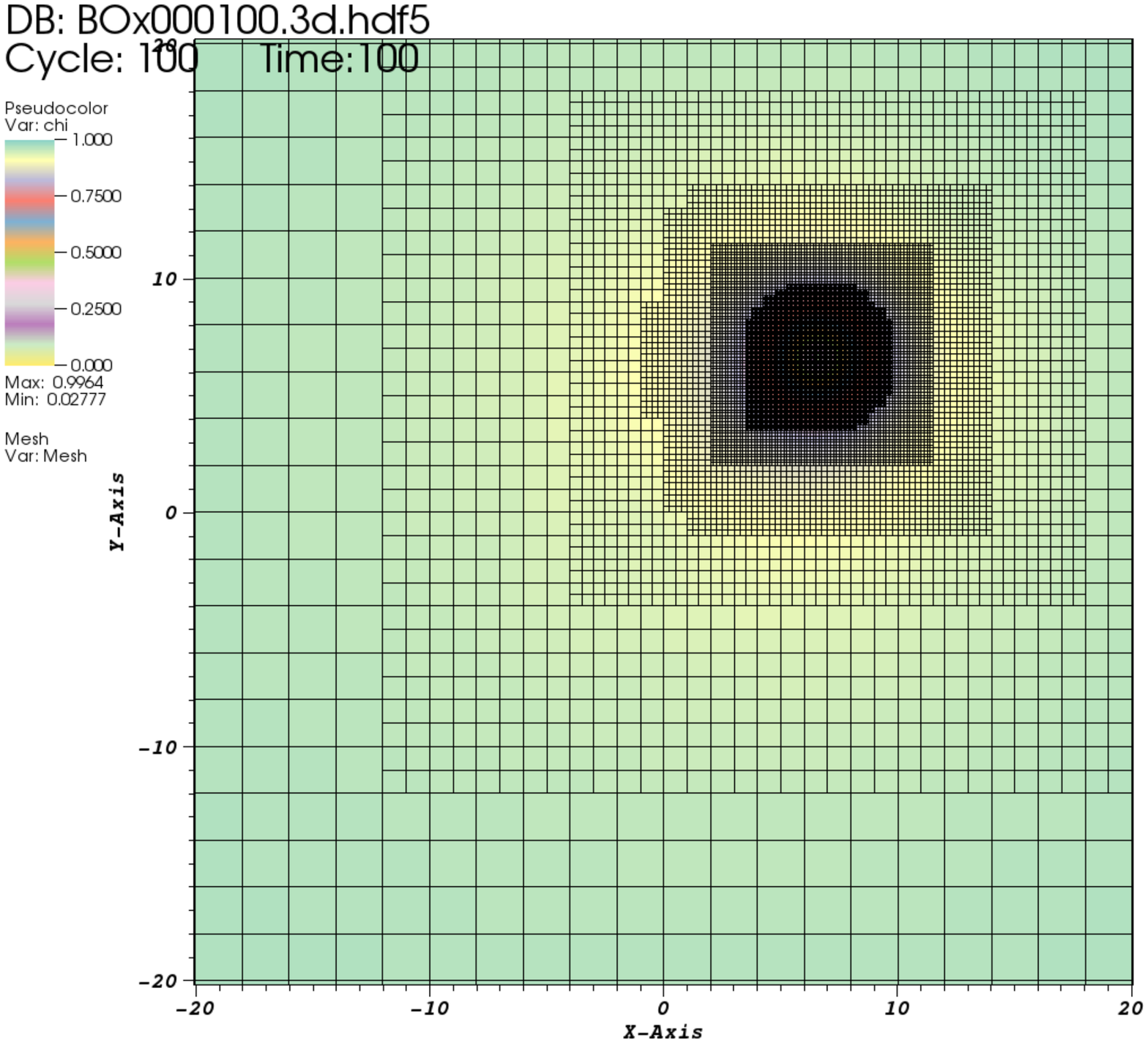}
}
\caption{Boosted black hole, movement: The boosted black hole moves across the grid diagonally with initial momenta of $P_x = 0.02$,  $P_y = 0.02$ and $P_z = 0.0$, as expected, and the grid adapts to this movement, with the high resolution grids following the movement.
\label{fig-Boostmove}}
\end{center}
\end{figure}

In this sub-subsection we evolve a boosted black hole using the perturbative approximation from \cite{ShapiroBook}, with initial momenta set to $P_x = 0.02$,  $P_y = 0.02$ and $P_z = 0.0$.  The domain size was chosen to be $(640)^3$, with spatial resolution in the coarsest grid of $4M$. We imposed periodic boundary conditions at the outer boundaries of the domain. The black hole moves across the grid diagonally as expected, as is seen in Figure \ref{fig-Boostmove}.

In the left panel of Figure \ref{fig-BoostBH} we show the $L^2$ norm of the Hamiltonian constraint across the domain as a function of time. This plot shows that the constraints remain bounded throughout the simulation. In the right panel of Figure \ref{fig-BoostBH} we display the components of the ADM linear momentum during the simulation. In the continuum limit they should be constant and in our simulation they are indeed approximately constant. 

\subsubsection{Binary inspiral}
\label{subsubsec:binary}

In this sub-subsection we superpose the initial perturbative solution for two boosted black holes in \cite{ShapiroBook}, sufficiently separated, to simulate a binary inspiral merger. The domain size was $(200M)^3$ with a grid spacing in the coarsest level of $5M$. As in some of the previous tests, for simplicity we imposed periodic boundary conditions at the outer boundaries of the domain. 

\begin{figure}
\begin{center}
\subfigure{
\includegraphics[width=.4\textwidth]{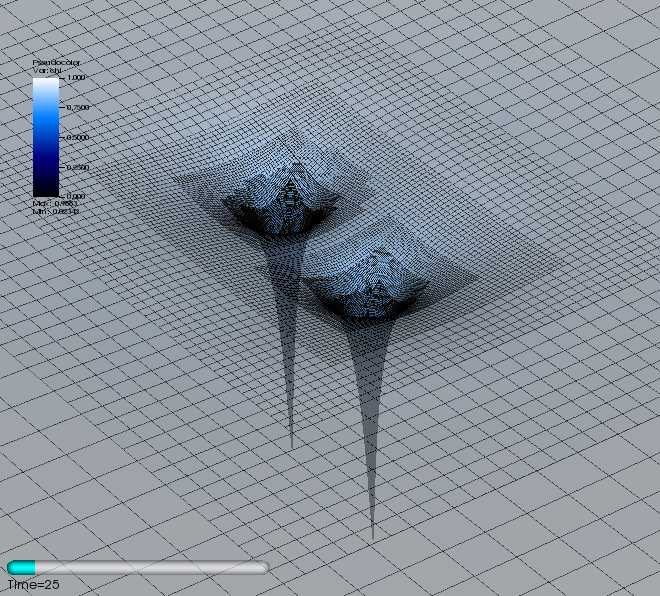}
}
\subfigure{
\includegraphics[width=.4\textwidth]{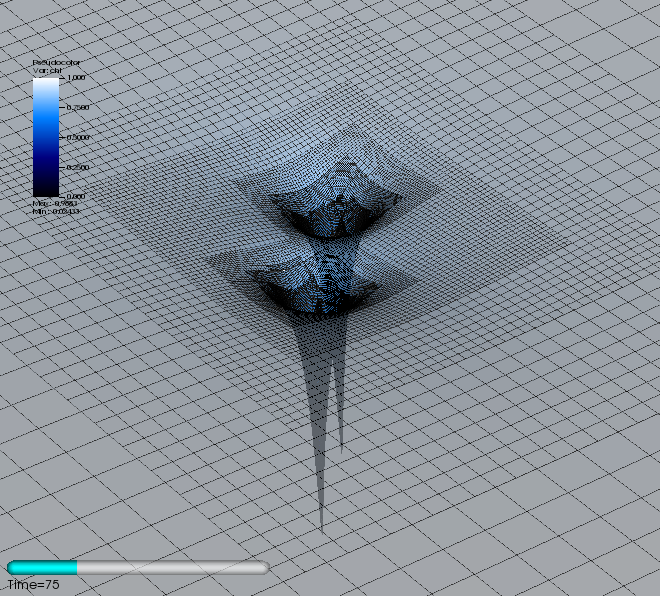}
}
\subfigure{
\includegraphics[width=.4\textwidth]{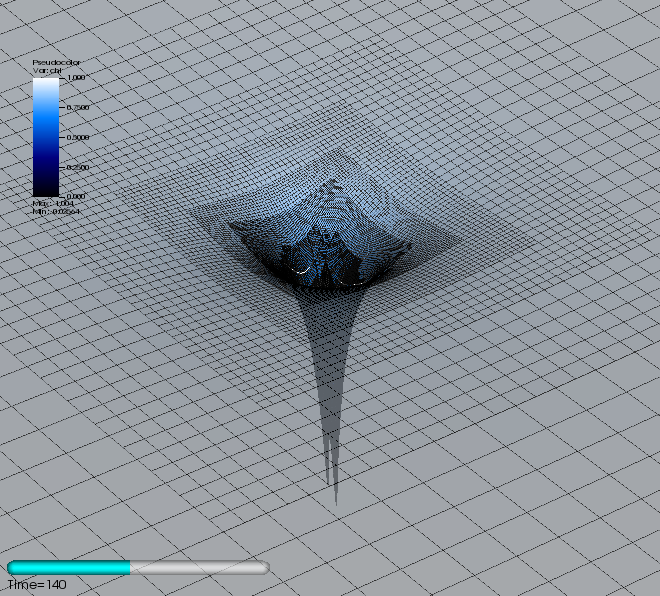}
}
\subfigure{
\includegraphics[width=.4\textwidth]{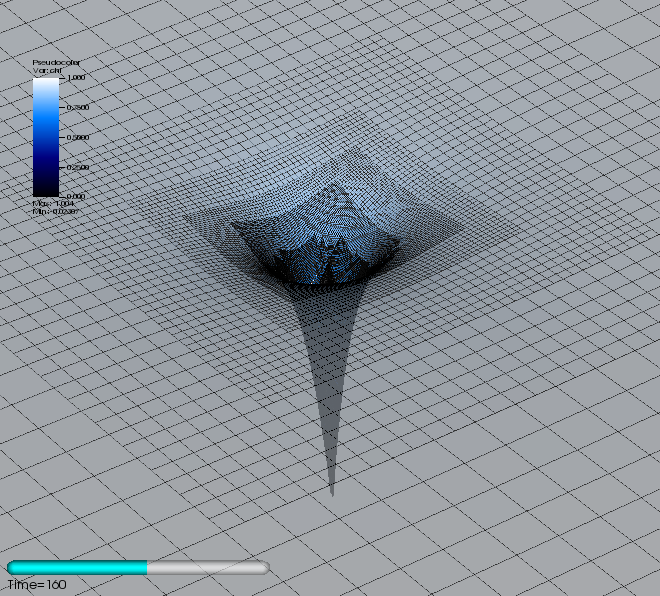}
}
\caption{Binary Black Hole merger: Two black holes are evolved with $\grchombo$. The final stages of the merger are shown. 
\label{fig-BiBH}}
\end{center}
\end{figure}

We are able to evolve the merger stably such that the two black holes merge to form one with a mass approximately equal to the sum of the two. The progression of the merger is shown in Figure \ref{fig-BiBH}. The time evolution of the $L^2$ norm of the Hamiltonian constraint across the grid is shown in Figure \ref{fig-Binary}. Again this remains stable throughout the simulation.

\subsection{Convergence test: head on collision of two black holes}
\label{sec-headon}

In this subsection we simulate the head on collision of two black holes and analyse the convergence of the code. We set up Brill-Lindquist initial data consisting of two static black holes of mass $0.5M$ with a separation of $10M$, located at the centre of the computational domain. We extract the gravitational wave signal (see below). An initial burst of radiation is seen, which is a property of the superimposed initial data, prior to the main signal. Even though this set up could be simulated in axisymmetry, we have evolved the system without imposing any symmetry assumptions. So the results below correspond to a full $3+1$ simulation with \texttt{GRChombo}.

We performed runs at three different resolutions with 7 levels of refinement, each level having half the grid spacing as
the previous one. The grid spacings were \linebreak $0.03125M/4M$ for the low resolution run, $0.02083M/2.66667M$ for the medium resolution run and $0.01563M/2M$ for the high resolution run. Here the numbers refer to the resolution on the finest/coarsest grids respectively. The outer boundary of the domain is located at $200M$ and we impose periodic boundary conditions for simplicity. This puts an upper bound on the time up to which we can evolve the system before boundary effects influence physical observables. 

\begin{figure}
\begin{center}
\subfigure{
\includegraphics[width=0.8\textwidth]{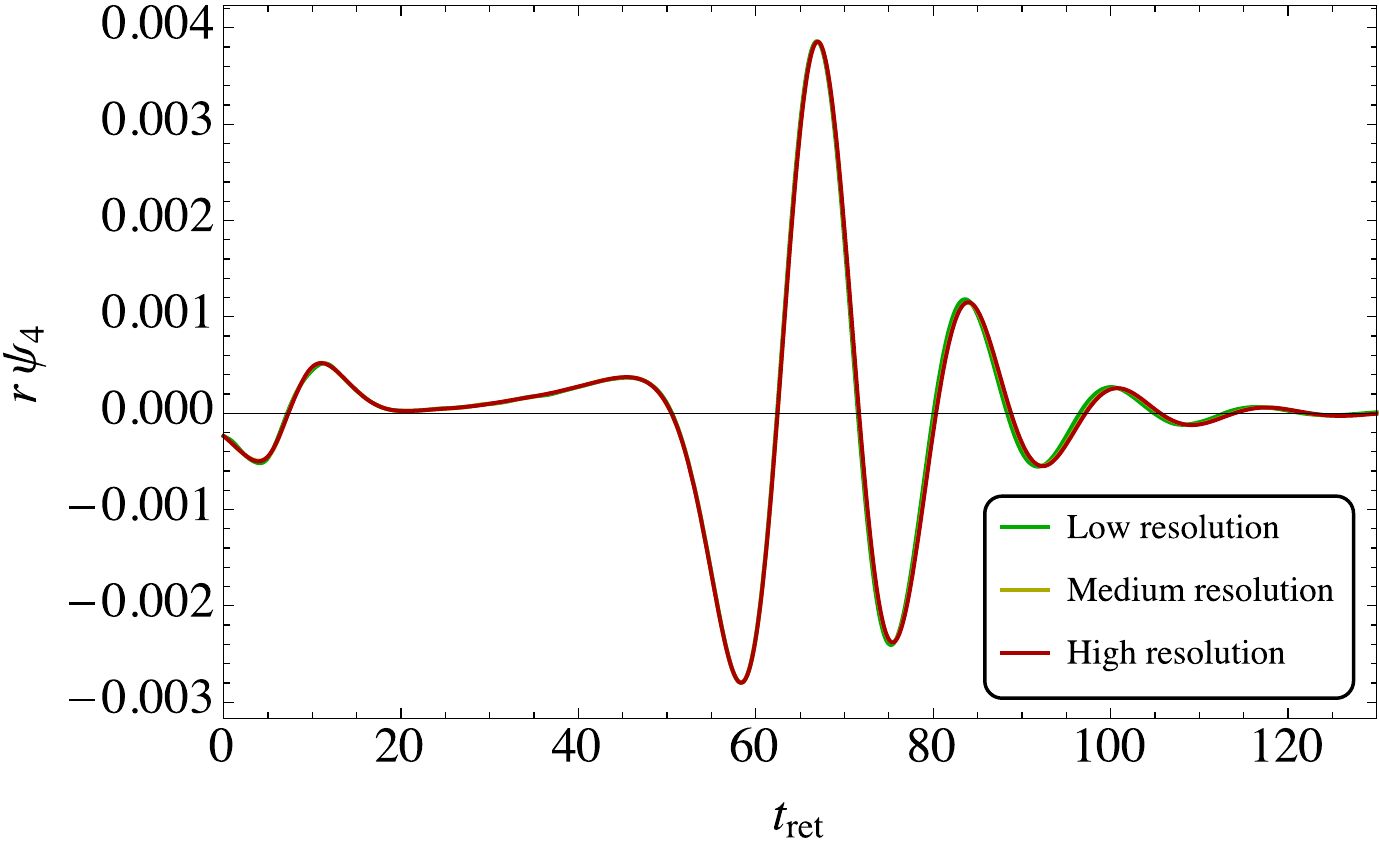}
}
\subfigure{
\includegraphics[width=0.75\textwidth]{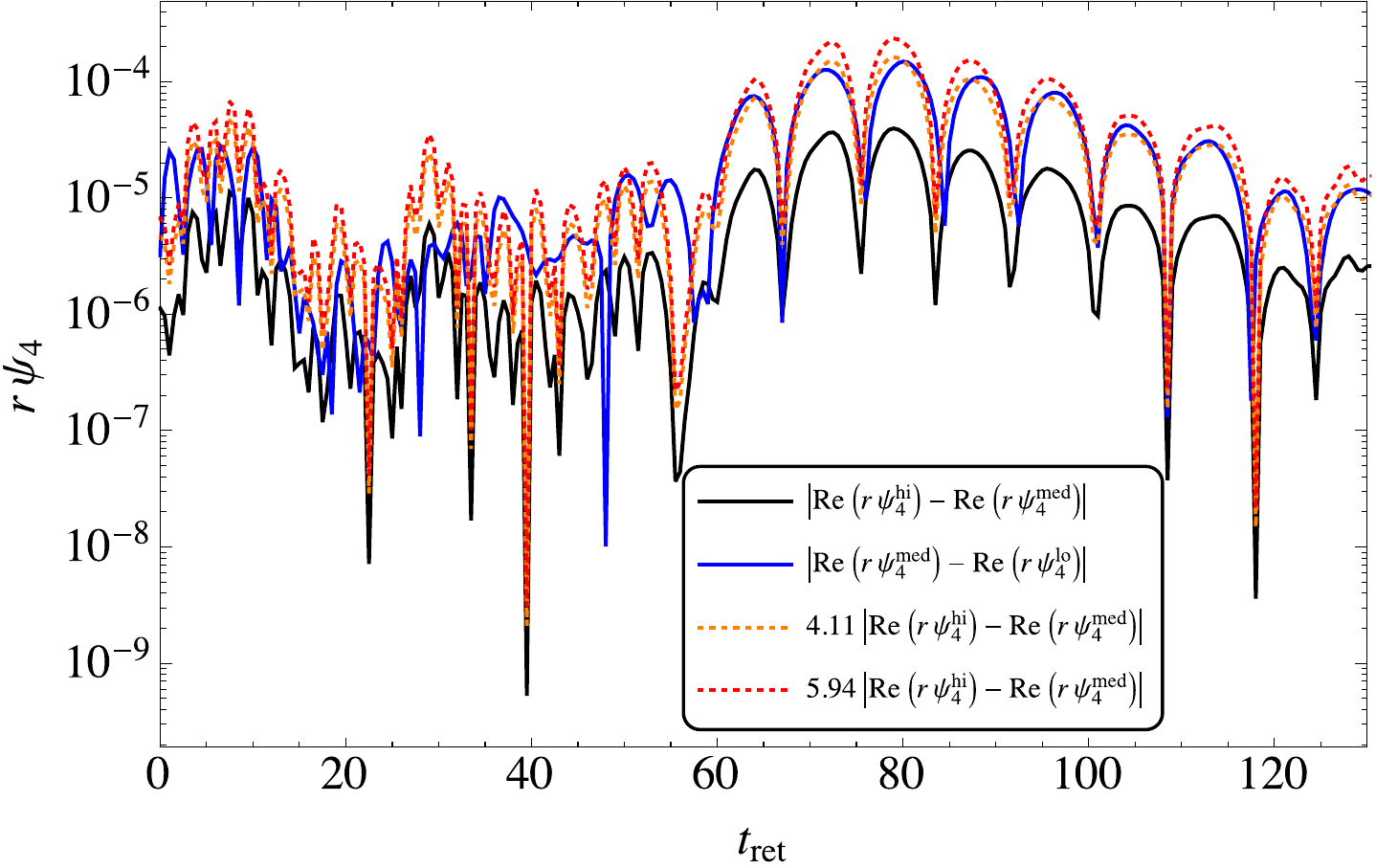}
}
\caption{Convergence test of head on collision. \textit{Top}: The real part of the $\ell=2$, $m=0$ mode of $r\Psi_4$ on the sphere of radius $R=60M$. \textit{Bottom}: Differences in the real part of the $\ell=2$, $m=0$ mode of $r\Psi_4$ between three different resolutions. We also show the data rescaled by a factor consistent with either third ($\times 4.11$) or fourth ($\times 5.64$) order convergence.
\label{fig-psi4}}
\end{center}
\end{figure}

In Figure \ref{fig-psi4} (top) we display the real part of the $\ell=2$, $m=0$ mode of $r\Psi_4$ extracted on a sphere of radius  $R=60M$ using 4th order interpolation. We use $320$ grid points in both the polar and azimuthal directions on the extraction sphere.  Following \cite{Loffler:2011ay}, we test convergence by comparing a physical quantity $\Psi$ at different resolutions. The convergence is of order $Q$ if for a set of grid spacings $h_1$, $h_2$, $h_3$, the differences between the numerically computed physical quantity $\Psi$ at successive resolutions satisfy
\begin{equation}
\frac{\Psi_{h_1} - \Psi_{h_2}}{\Psi_{h_2} - \Psi_{h_3}} = \frac{h_1^Q - h_2^Q}{h_2^Q - h_3^Q}\,.
\end{equation}
With the resolutions used in these runs, assuming $4^\textrm{th}$ order convergence the above factor is $\approx 5.953$, whilst assuming $3^\textrm{rd}$ order convergence the factor is $\approx 4.115$.

The gravitational wave content of the superimposed initial data is reflected in the non-zero initial signal. The collision of the two black holes takes place at $t\sim50$ on this plot, so the signal before this collision time should be regarded as unphysical. As can be seen in this plot, the results for the two higher resolutions are indistinguishable on the scale employed here, whilst the lowest resolution shows a very slight drift towards later times, but is still in very good agreement. The bottom plot in Figure \ref{fig-psi4} shows the absolute value of the difference between $r\Psi_4$ computed low and medium resolution (solid blue), medium and high resolution (solid black), and this latter curve scaled up by the convergence factor assuming $3^\textrm{rd}$ (dotted orange) and $4^\textrm{th}$ (dotted red) order convergence. This plot shows that in the highly dynamical stages of the evolution, when there is a lot of regridding and the boxes move around the domain, the convergence is closer to $3^\textrm{rd}$ order. On the other hand, when the system has nearly settled, and hence the boxes do not move much, the convergence order is closer to $4$. We can explain this loss of convergence due to regridding because in the interpolation used in \texttt{GRChombo} only the values of the functions are matched across levels, but not their derivatives. We hope to improve this aspect of the code in the future.


\subsection{Choptuik scalar field collapse}
\label{sec-choptuik}

\begin{figure}
\begin{center}
\subfigure{
\includegraphics[width=.3\textwidth]{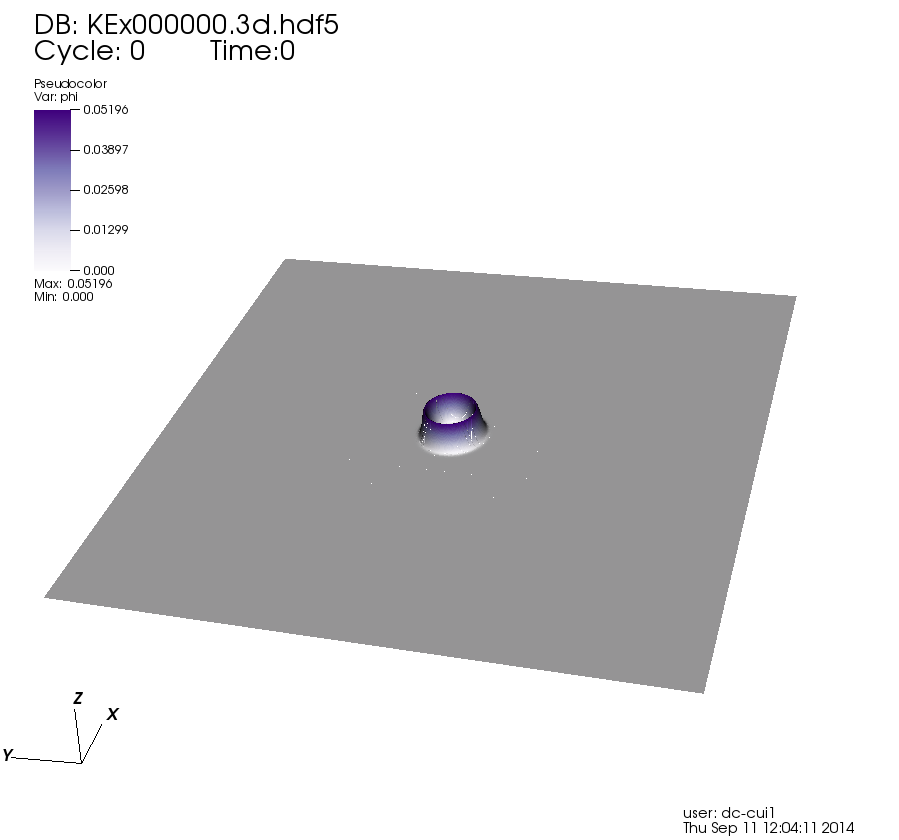}
}
\subfigure{
\includegraphics[width=.3\textwidth]{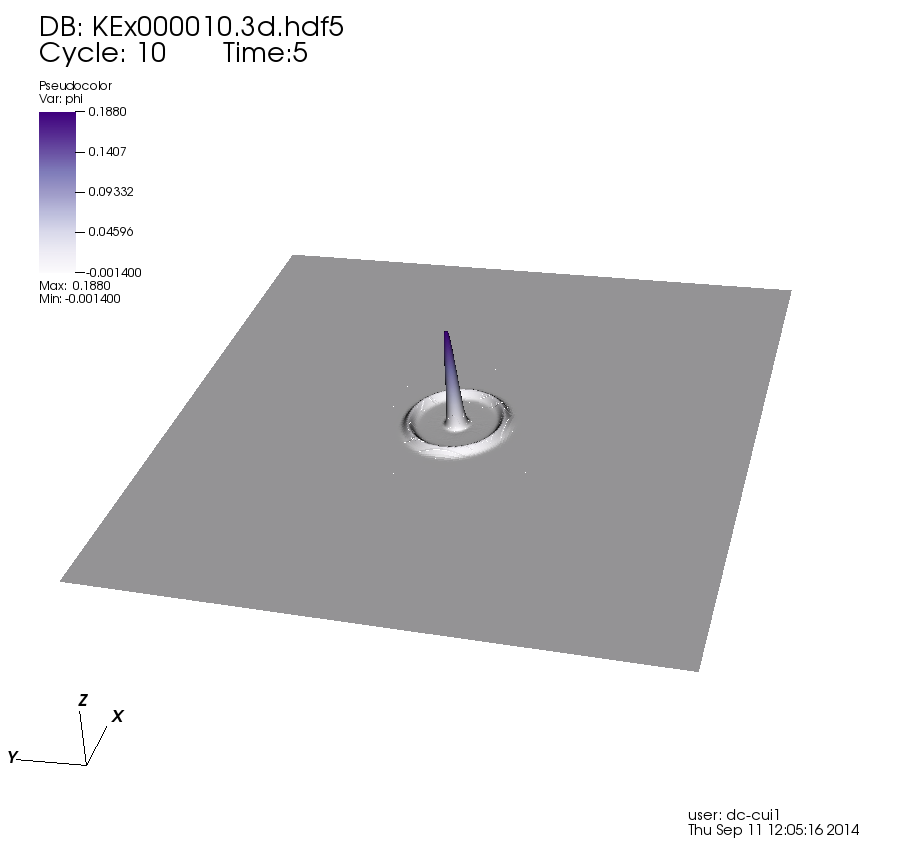}
}
\subfigure{
\includegraphics[width=.3\textwidth]{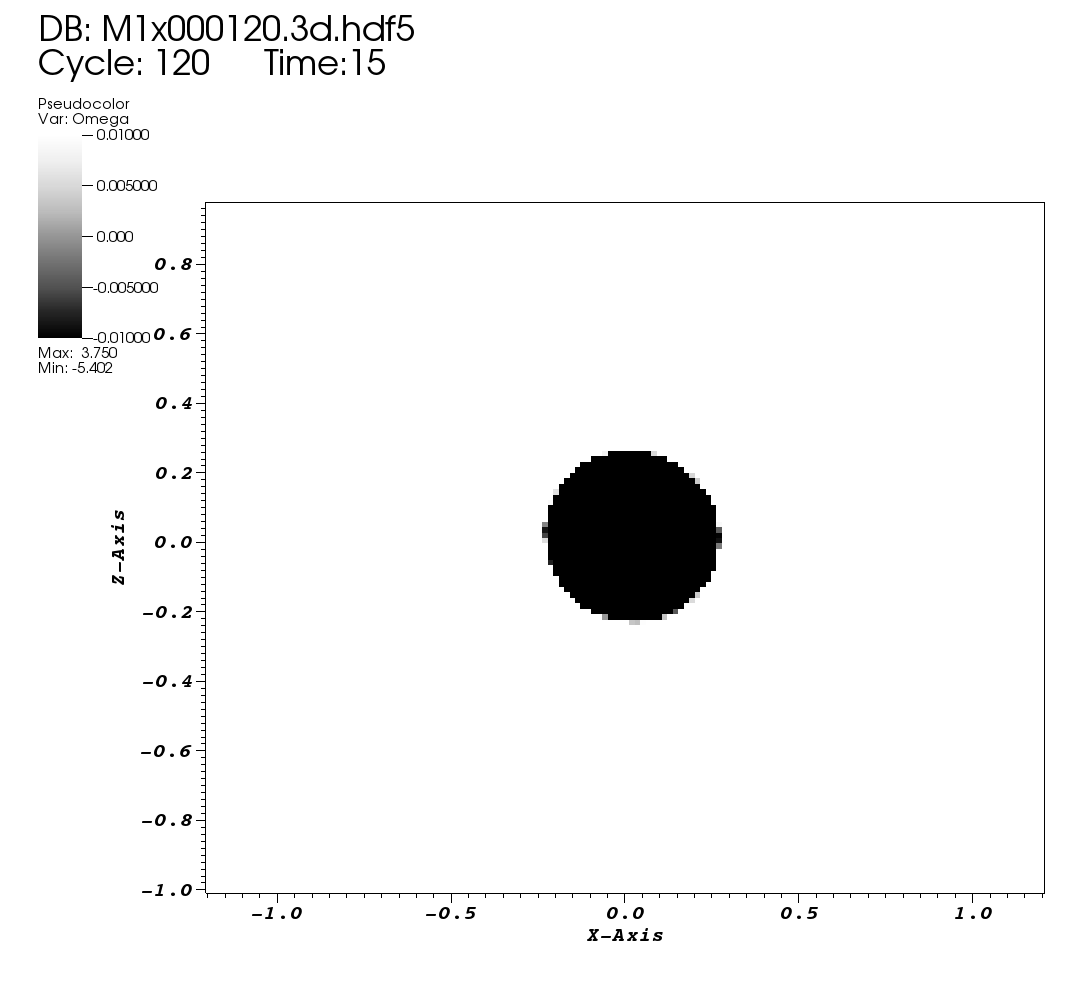}
}
\caption{In Choptuik scalar field collapse, the initial specially symmetric configuration in the first figure (which shows the values on a slice perpendicular to the z axis) collapses, splitting into an ingoing and an outgoing wave as seen in the second image. If the amplitude of the initial perturbation is greater than a certain critical value, the ingoing wave will result in the formation of a black hole, as seen from the output of the apparent horizon finder in the third figure, which shows that an apparent horizon with a mass of about 0.25 has formed by $t=15$.
\label{fig-chop3D}}
\end{center}
\end{figure}

We now test the scalar field part of the code, by simulating the Choptuik scalar field collapse as described in \cite{AlcubierreBook} and illustrated in Figure \ref{fig-chop3D}. The referenced description is for a 1+1 simulation which is evolved using a constrained evolution, such that the lapse $\alpha$ and the single degree of freedom for the metric, $A$, are both solved for on each slice using ODEs obtained from the constraint equations. The only degrees of freedom which are truly evolved are those of the field, $\phi$, $\Psi$ and $\Pi$.

Our evolution is carried out using the full $3+1$ BSSN equations, without assuming or adapting coordinates to spherical symmetry. We are able to replicate the results obtained in \cite{AlcubierreBook}, subject to some minor differences due to the fact that we evolve with the puncture gauge rather than according to the maximal slicing constraint equation, see Figures \ref{fig-chopsnaps} and \ref{fig-chopalp}, which can be found in Appendix \ref{sec-appendix}. Videos of the results can be viewed via our website at http://grchombo.github.io.

We see that $\grchombo$ can accurately evolve the field profile in the presence of gravity, and copes with the collapse of the supercritical case into a singularity, without code crash. For the subcritical cases we see that the field disperses as expected. 

\subsection{MPI Scaling properties}
\label{sec-scaling}

We now turn to the performance aspects of $\grchombo$. Here we perform a number of scaling tests to show that our code can exploit the parallelism offered in modern supercomputers to a reasonable extent. Whilst {\tt Chombo} does have the capability to partially utilise threads through hybrid OpenMP routines, we will limit our attention to pure MPI mode in these tests, as we have found that this gives significantly smaller run-to-run performance variations.

\begin{figure}[htp]
\begin{center}
\includegraphics[height=8cm]{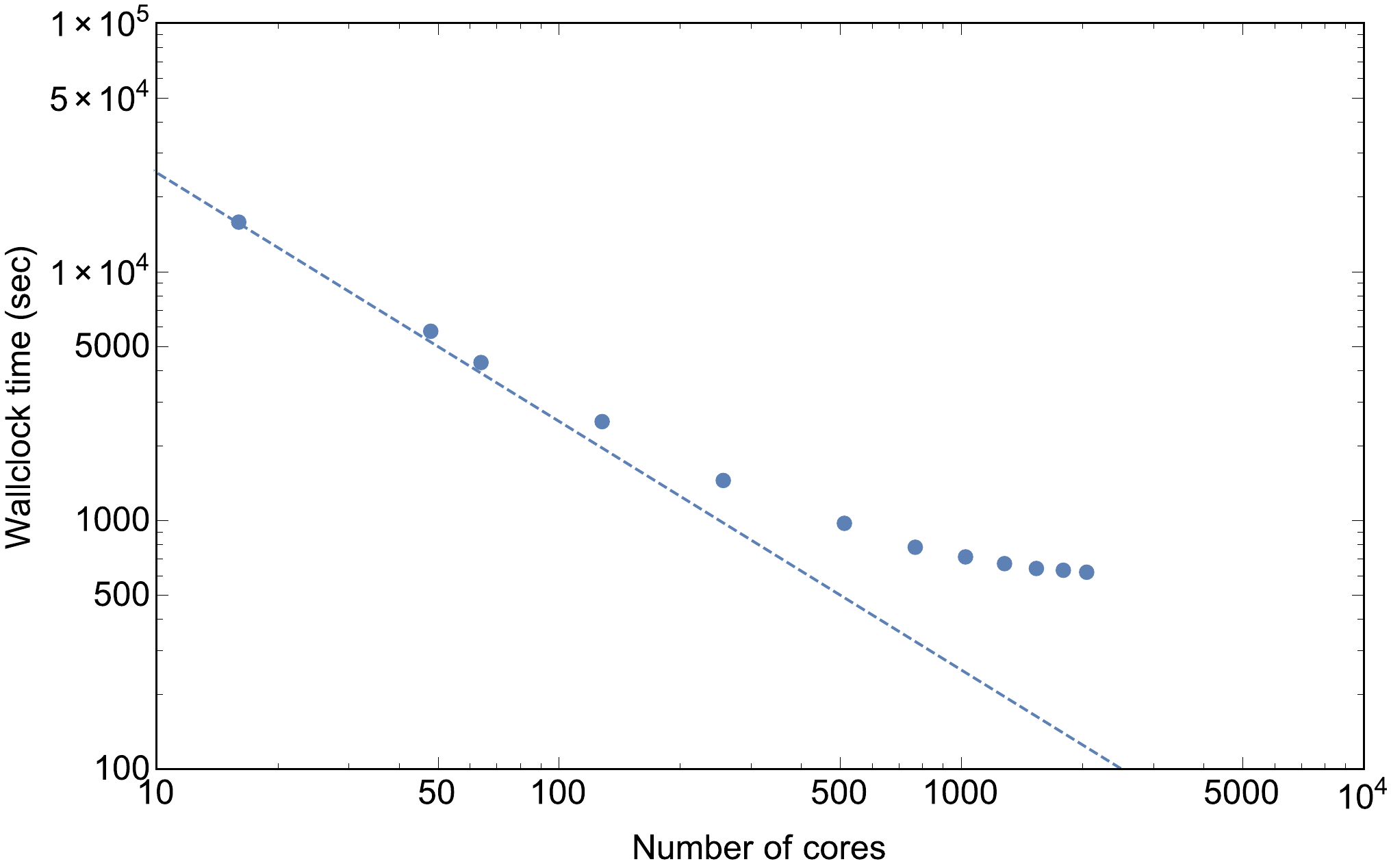}
\caption{
Strong scaling behaviour of GRChombo on the SuperMike-II cluster at the Louisiana State University. The code achieves excellent strong scaling up to 200 cores, and a useful scaling up to around 1000 cores.
\label{fig-StrongScaling}}
\end{center}
\end{figure}

Our strong scaling test is performed using a head on binary black hole system. We set up Brill-Lindquist initial data for two static black holes of mass $0.5M$, with a separation of $6M$. Our overall computational domain is a box of size $160M$, and at the coarsest level, we fix the total number of grid points to $320$ in each direction, giving a grid spacing of $0.5M$. The centre of mass of the system is at the centre of the domain. For the mesh refinement, we fix the total number of levels to six. The simulation is allowed to run up to the time of $2M$. The bulk of this test was performed on the SuperMike-II cluster at the Louisiana State University. Each compute node consists of two 2.6GHz 8-core Sandy Bridge Xeon processors, connected via a InfiniBand QDR fabric. We fix the computational load across all jobs and vary the number of core count from 16 to 2048. Our data in Figure \ref{fig-StrongScaling} shows excellent strong scaling up to 200 cores on this cluster. We continue to see a reasonable speedup up to around 1000 cores for this particular problem.

\begin{figure}[htp]
\begin{center}
\includegraphics[height=8cm]{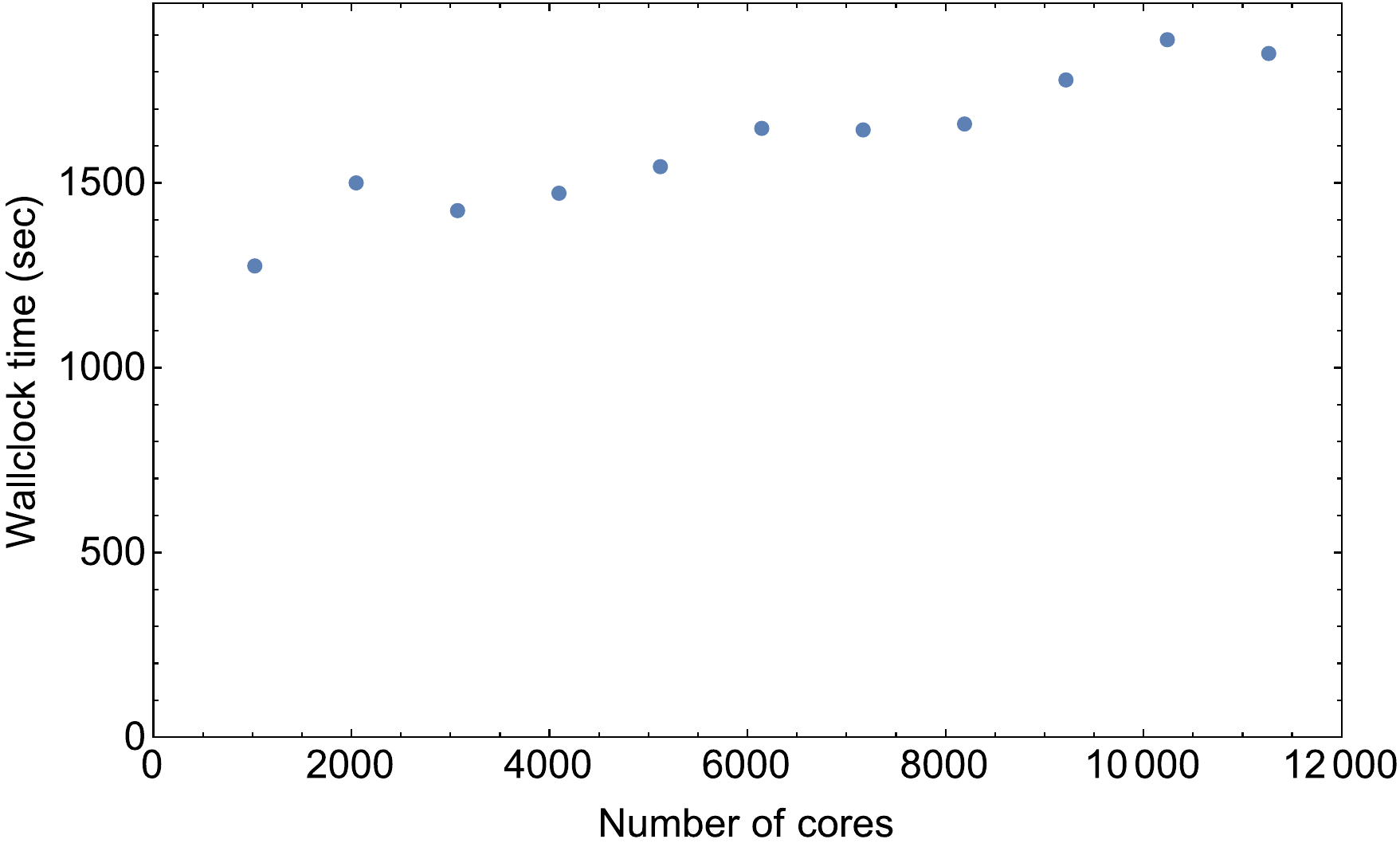}
\caption{
Weak scaling behaviour of GRChombo on the Mira Blue Gene/Q cluster at the Argonne National Laboratory over a 10x increase in core count.
\label{fig-WeakScaling}}
\end{center}
\end{figure}

Of course, in a production environment, it is often desirable to use additional cores to be able to run a larger simulation, rather than to speed up a problem of fixed size. In this scenario, weak scaling behaviour is of interest. We begin at 1024 cores with an identical setup to that in the strong scaling test. We then scale up the number of grid points at the coarsest level proportional to the increase in core count up to 10240, whilst adjusting the tagging threshold in order to maintain the shape and size of the refined regions. We also adjusted the time step size (i.e. the Courant factor) so that each simulation would reach the target stop time in the same number of steps. We use the Mira Blue Gene/Q cluster at the Argonne National Laboratory for this due to the larger number of cores available. Figure \ref{fig-WeakScaling} shows a less-than-perfect scaling behaviour in this setup, with the main bottleneck appearing in the regridding and box generation stages. We are working together with the developers of {\tt Chombo} to improve this aspect of the code performance. It is worth noting, however, that even in its current state the code still shows a useful level of scalability: the wallclock time increases by less than 2x over the 10x increase in core count.

\subsection{Performance comparison}
\label{sec-performance}
Lastly, we demonstrate that $\grchombo$'s performance on standard $3+1$ black hole problems is comparable to that of an existing numerical relativity code.

Our comparison target is the $\Lean$ code~\cite{Sperhake:2006cy,Zilhao:2010sr}, a $3+1$ numerical relativity code designed to evolve four and higher dimensional vacuum spacetimes. $\Lean$  is based on the {\texttt{Cactus}} computational toolkit~\cite{Cactuscode:web} and realises moving-box mesh refinement via the \texttt{Carpet} package~\cite{Schnetter:2003rb,CarpetCode:web}, both of which are part of the open-source {\texttt{Einstein Toolkit}}~\cite{Loffler:2011ay,EinsteinToolkit:web}. Initial data is constructed either analytically or numerically by employing the {\texttt{TwoPunctures}} spectral solver~\cite{Ansorg:2004ds}. In order to track apparent horizons, $\Lean$ makes use of~\texttt{AHFinderDirect}~\cite{Thornburg:2003sf,Thornburg:1995cp}.

The $\grchombo$ setup is identical to that in the strong scaling test as detailed in Sec. \ref{sec-scaling}. The $\Lean$ code is subject to the limitation of Carpet, where successive levels may only occur a collection of nested-box hierarchies, whose sizes are typically related by a power of two. In this case, we first fix boxes of side lengths $160$, $80$, $40$ and $20M$ at the centre of the domain, encompassing both black holes, then fix further boxes of side lengths 5 and $2.5M$ centred at each of the black holes. During the evolution, $\Lean$ has the capability to track the black holes and move or merge the finer boxes as appropriate, however the shape and size of the boxes remain unchanged. The $\grchombo$ code is not subject to this box structure limitation, and therefore we simply tune the regridding threshold so that the size of the finest level matches that of the $\Lean$ setup. We make no attempt to match the sizes of the intermediate levels as this would defeat the spirit of fully-flexible AMR.

\begin{figure}[htp]
\begin{center}
\includegraphics[height=8cm]{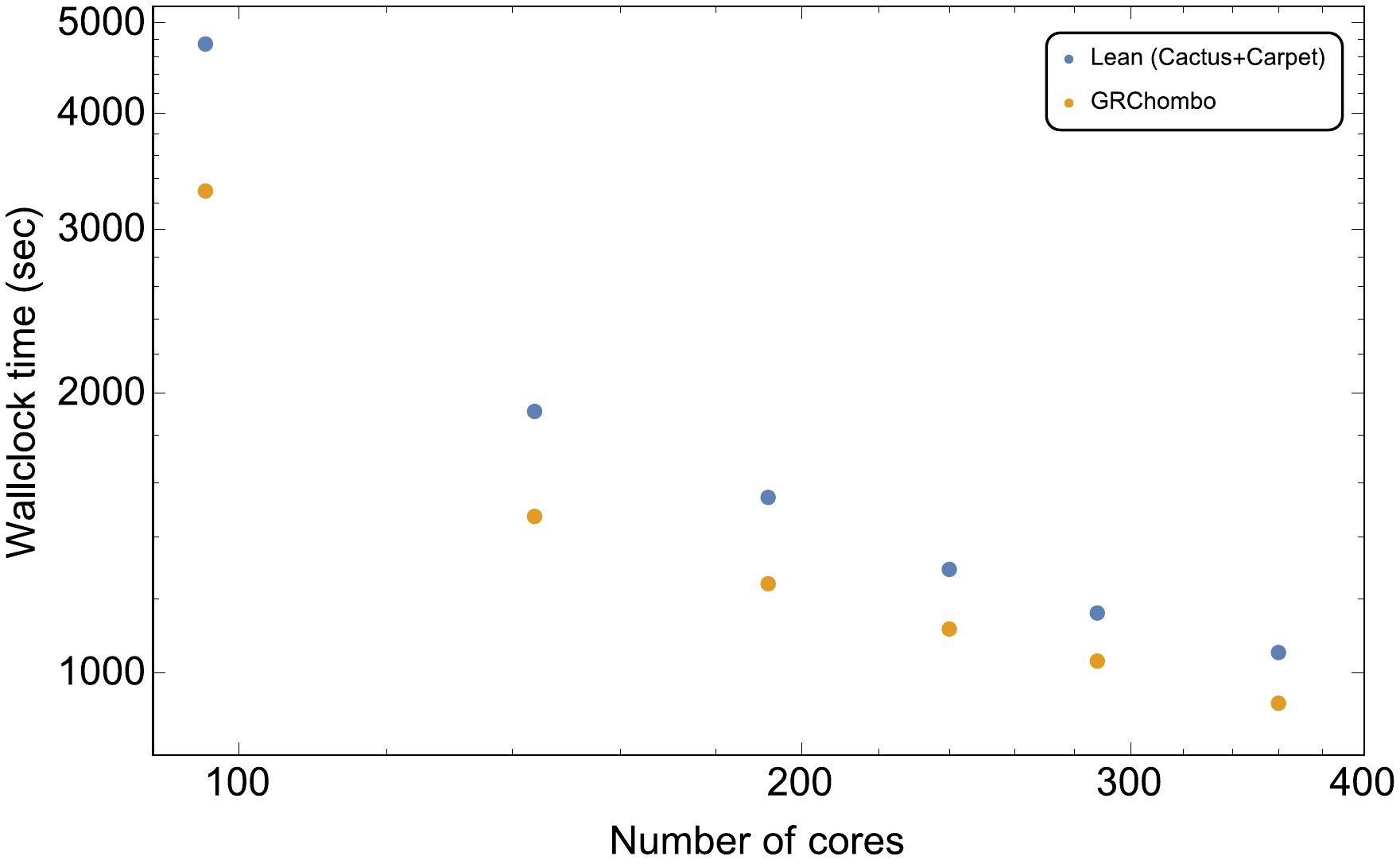}
\caption{
Runtime and scaling comparison between $\grchombo$ (orange) and $\Lean$ (blue). The leftmost data points show disproportionately large wallclock times as the machine becomes memory-limited at this core count.
\label{fig-LeanComparison}}
\end{center}
\end{figure}

Our comparison tests were performed on the COSMOS VIII shared memory facility. Both codes were executed on the same SGI UV1000 machine, utilising up to 60 Nehalem EX 2.67GHz CPUs with 6 cores per CPU, giving up to 360 cores in total. In all of these runs, we pin one MPI rank to each core and disabled all checkpointing activity since we wish to exclude I/O bottleneck. We allowed the simulation to run up to coordinate time $t = 2$, and measured the wall-clock time taken to execute the time evolution portion of the code (i.e. we excluded the time spent during initial setup).

Within the range of 150-360 cores, both $\grchombo$ and $\Lean$ exhibit similar performance and strong scaling characteristics (figure \ref{fig-LeanComparison}). Below 150 cores, we cannot meaningfully test the strong scaling behaviour as the machine becomes memory-limited. We have not performed this comparison on a larger cluster due to the lack of resource availability, but we have no reason to expect any significant difference provided that the problem size is also scaled up appropriately. Having said this, we believe that a framework like Cactus probably remains the better choice when it comes to these standard problems, owing to the wealth of existing tools and resources and a more mature community of users. Instead, we intend for $\grchombo$ to be complementary to existing numerical relativity codes in order to open up new avenues of research by enabling a wider range of problems to be tackled at a feasible level of resources (see section \ref{sec-newphys} for examples of such problems).

\section{Discussion}
\label{sec-dis}

In this paper, we introduced and described $\grchombo$, a new multi-purpose numerical relativity code built using the {\tt Chombo} framework. It is a $3+1$D finite difference code based on the BSSN/CCZ4 evolution scheme. It supports Berger-Collela type AMR evolution with Berger-Rigoutsos block structured grid generation, and is fully parallelized via the Message Passing Interface, and time evolution is via standard 4-th order Runge-Kutta time-stepping.

We illustrated some areas of physics that can potentially benefit from this code, such as multiple black hole mergers and scalar field collapse. Such fields require a code which adapts to changes in the range and location of scales at different points in space and time in the simulation. We emphasise that setting the initial conditions for these mergers are trivial -- $\grchombo$ automatically remeshes the grid given a set of analytic initial conditions without further user intervention.

We showed that $\grchombo$ successfully passes the standard ``Apples with Apples'' tests\footnote{We note that we perform as well as any other BSSN code in the Gowdy wave test as we expected in a pre-determined gauge. Although, \cite{Cao:2011fu} managed to achieve long-term evolution by considering different gauge conditions.}. In addition to these tests, we evolved standard single black hole spacetimes (Schwarzschild and Kerr) and showed that it is stable to more than $T =10000M$. Using the moving puncture gauge, we also show that $\grchombo$ stably evolves the merger of two and three black holes in inspiral and head-on collisions. We simulated the supercritical collapse of a scalar field configuration, and found that it forms a black hole as expected, to show that the code supports non vacuum spacetimes. Finally we tested the MPI scaling properties of the code, both strongly and weakly, and compared this with an alternative numerical relativity code based on the popular Cactus framework.

Nevertheless, despite its power, the AMR capability of $\grchombo$ has to be treated with care. As we mentioned earlier, coarse-fine boundaries could be a significant source of inaccuracy, even though the Hamiltonian constraint may still be kept under control. A way to reduce coarse-fine boundary errors is to introduce conservative refluxing during interlevel operations. Although refluxing requires significant overhead, we intend to implement it in the next iteration of $\grchombo$. The litmus test for accuracy of $\grchombo$ is its ability to make accurate predictions of outgoing gravitational wave-forms. We leave this, and the introduction of a set of ``best practices'' for the use of AMR in general relativistic systems, for a follow-up work.

\section*{Acknowledgements}

We would first like to thank the $\Lean$ collaboration for allowing us to use their code as a basis for comparison, and especially Helvi Witek for helping with the setting up and running of the $\Lean$ simulation. We would like to thank Erik Schnetter, Ulrich Sperhake, Helvi Witek, Luis Lehner, Carlos Palenzuela and Tom Giblin for many useful conversations, and members of the $\mathtt{Chombo}$ collaboration, Daniel Martin and Brian Van Straalen. We would especially like to thank Juha J{\"a}ykk{\"a} and James Briggs for their amazing technical support.  This work was undertaken on the COSMOS Shared Memory system at DAMTP, University of Cambridge operated on behalf of the STFC DiRAC HPC Facility. This equipment is funded by BIS National E-infrastructure capital grant ST/J005673/1 and STFC grants ST/H008586/1, ST/K00333X/1. EAL acknowledges support from an STFC AGP grant ST/L000717/1. PF and ST are supported by the European Research Council grant ERC-2011-StG279363HiDGR. PF is also supported by the Stephen Hawking Advanced Research Fellowship from the Centre for Theoretical Cosmology, University of Cambridge. MK is supported by an STFC studentship. He started his work on this project as a summer student funded by the Bridgwater Summer Undergraduate Research Programme at the Centre for Mathematical Sciences, University of Cambridge, and by King's College, Cambridge. HF is supported by the US Department of Energy (DOE), and this research used resources of the Argonne Leadership Computing Facility, which is a DOE Office of Science User Facility, both supported under Contract DE-AC02-06CH11357. Part of the performance test for this work was performed on Louisiana State University's High Performance Computing facility.

\bibliography{all.bib}
\bibliographystyle{utcaps}

\newpage

\appendix

\section{Results from code tests}
\label{sec-appendix}

In this appendix we collect the figures of the code tests described in Sec. \ref{sec-tests}.

\begin{figure}[h]
\begin{center}
\subfigure[Hamiltonian constraint]{
\includegraphics[height=6cm]{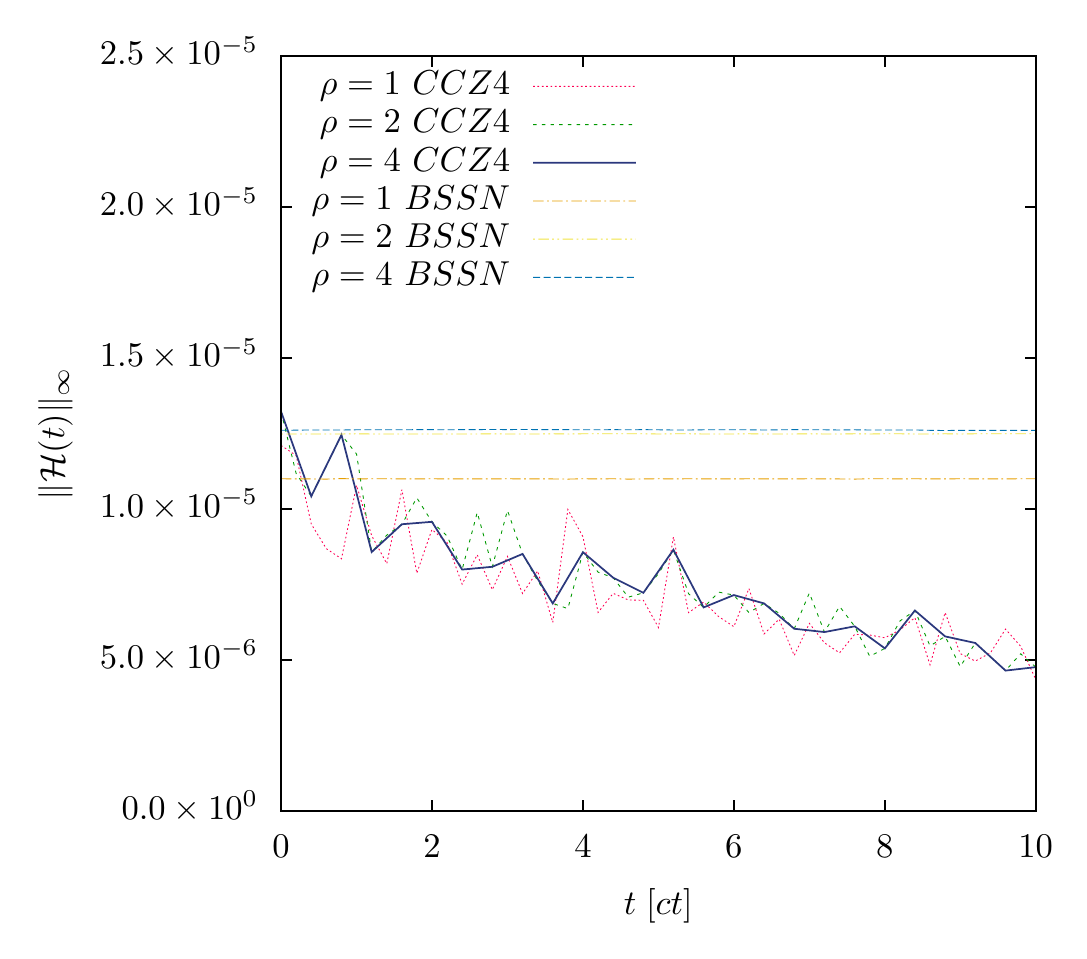}
}
\hspace{0.75cm}
\subfigure[Deviation in $\tilde \gamma_{xx}$]{
\includegraphics[height=6cm]{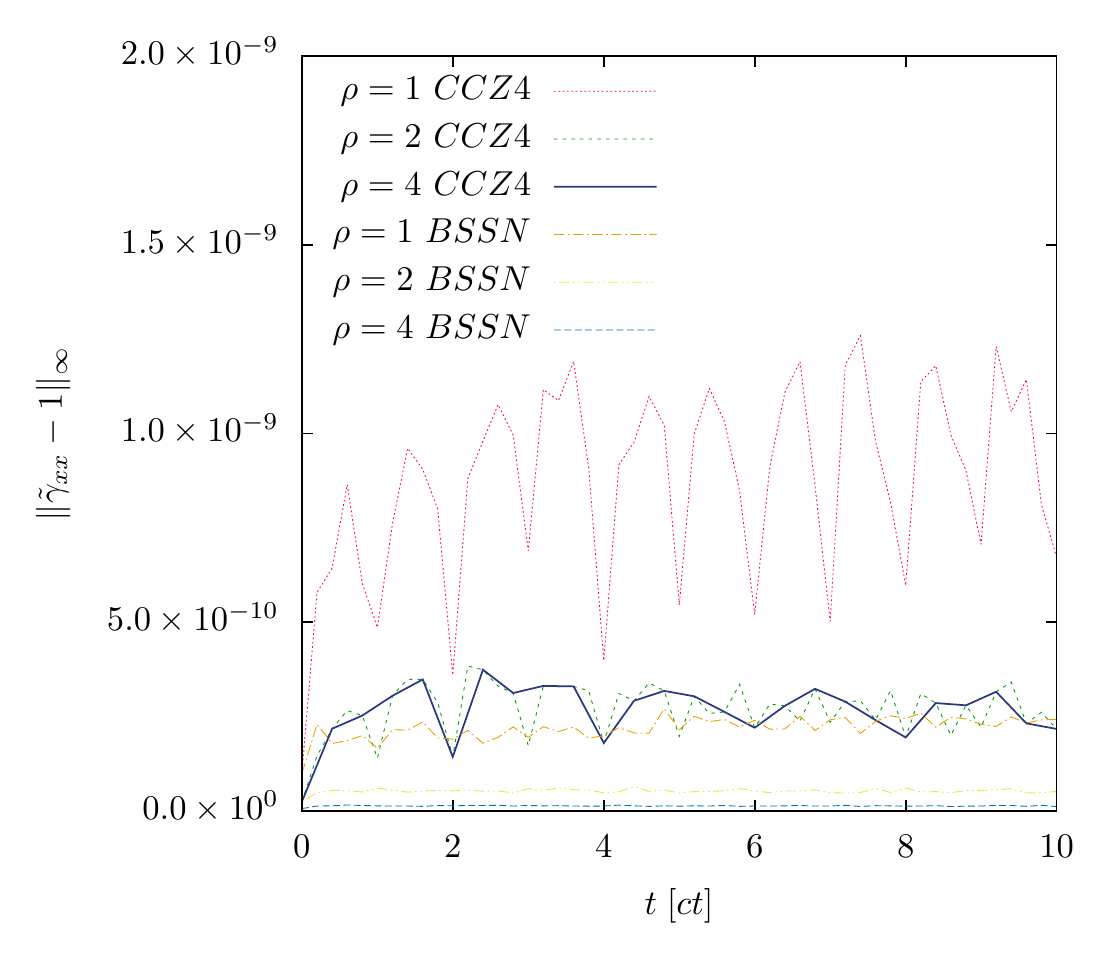}
}
\caption{
Robust stability test for both the BSSN and the CCZ4 codes, with resolutions $\rho=2,4$ respectively. \textit{Left}: time evolution of the $L_{\infty}$ norm of the Hamiltonian constraint. \textit{Right}: deviation of $\tilde \gamma_{xx}$ from 1. Neither norm grows at an increasing rate with increasing resolution, and so the test is passed.
\label{fig-robust}}
\end{center}
\end{figure}

\begin{figure}[H]
\begin{center}
\subfigure[$g_{yy}$ at $T=1000$]{
\includegraphics[height=6cm]{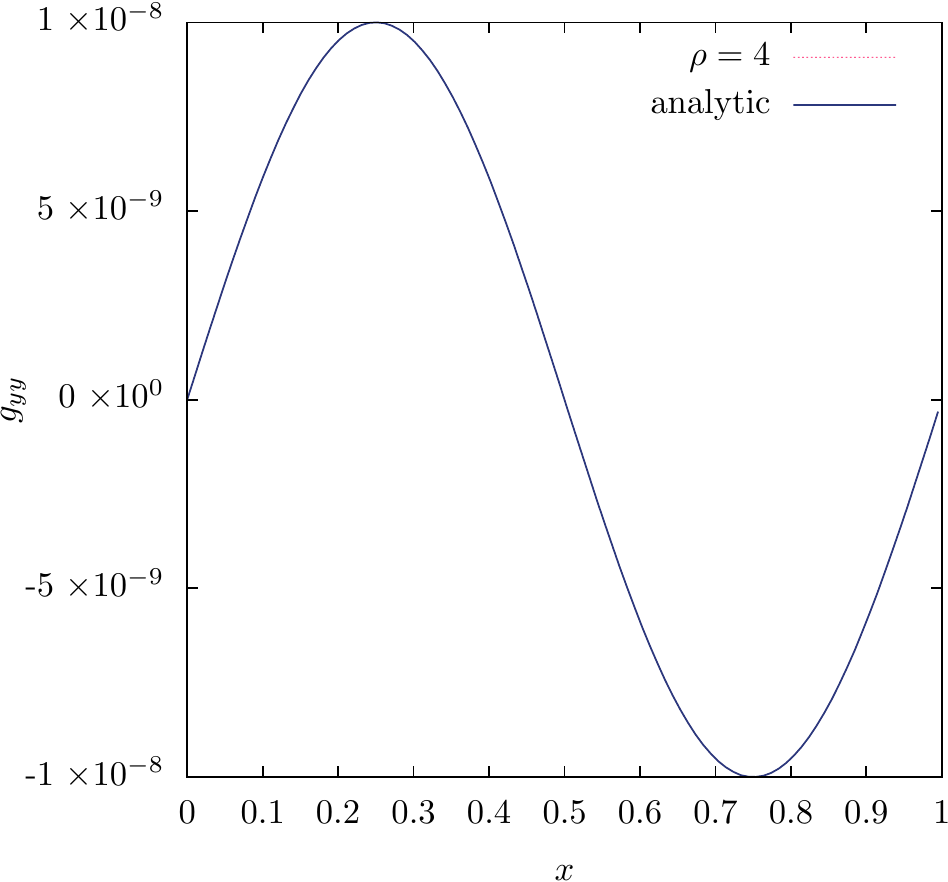}
}
\hspace{0.75cm}
\subfigure[Error in $g_{yy}$ at $T=1000$]{
\includegraphics[height=6cm]{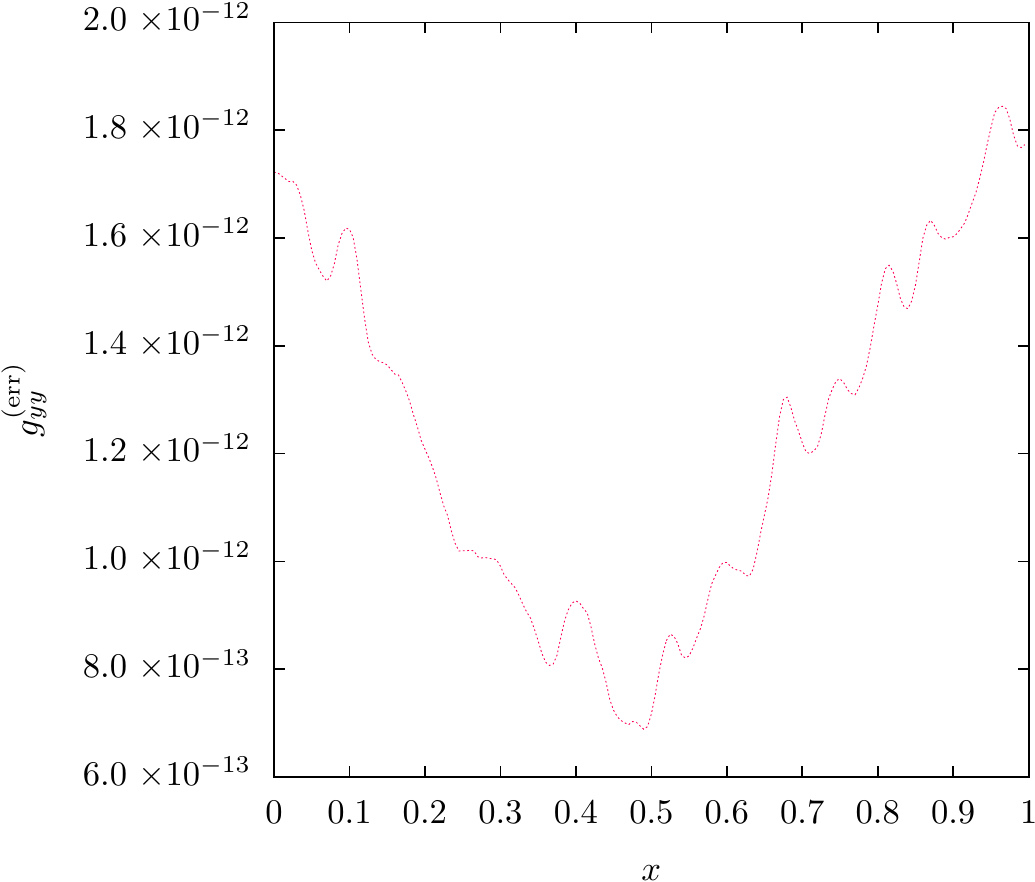}
}
\caption{
Linear wave test. \textit{Left}:  analytical solution and the evolved $g_{yy}$ component of the metric at $T=1000$ at resolution $\rho=4$, but the two are indistinguishable. \textit{Right}: absolute value of the error across the grid at $T=1000$, from which we can see more easily that some small errors in the magnitude and phase have been introduced.
\label{fig-linear}}
\end{center}
\end{figure}

\begin{figure}
\begin{center}
\includegraphics[height=10cm]{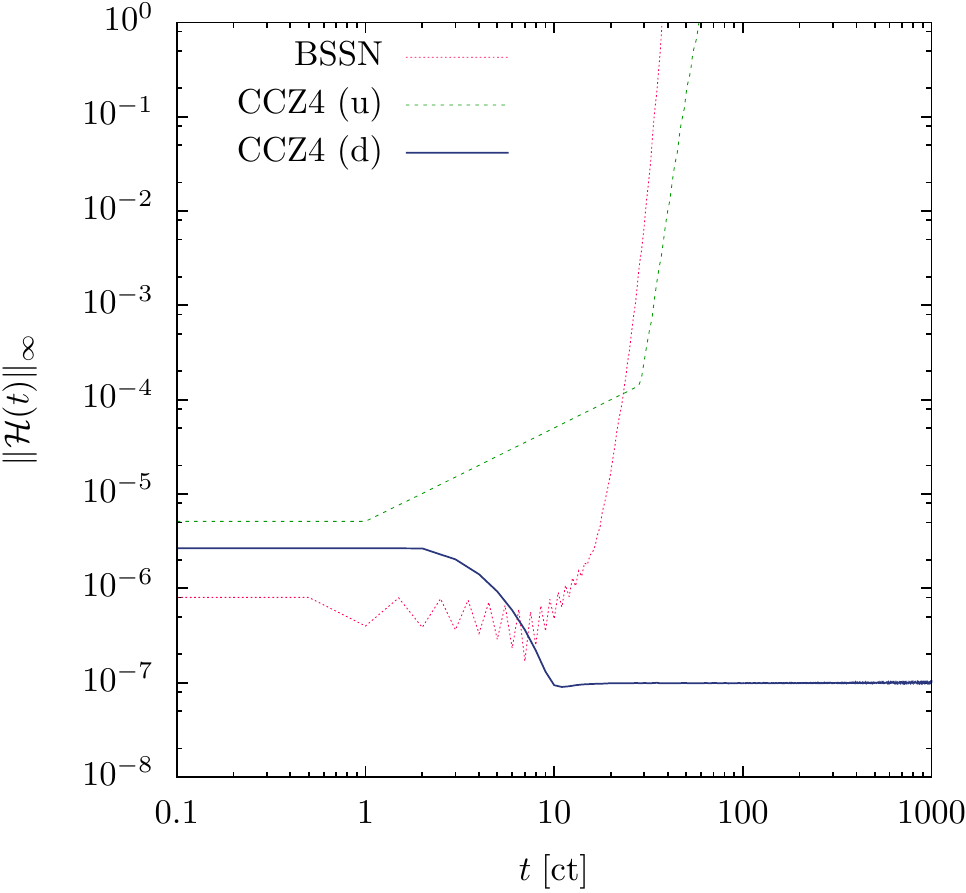}
\caption{Gauge wave test. The increase in the $L_\infty$ norm of the Hamiltonian constraint means that the BSSN code only remains stable for less than 50 timesteps. Undamped (u) CCZ4, i.e. CCZ4 with $\kappa_1 = 0$, performs similarly. Damped (d) CCZ4 with $\kappa_1 = 1$ is stable for the full 1000 timesteps. The test was performed with initial amplitude of $A=0.1$, Kreiss-Oliger dissipation coeffecient of $\sigma=0.1$ and a resolution of $\rho=4$. 
\label{fig-gauge}}
\end{center}
\end{figure}

\begin{figure}
\begin{center}
\subfigure[BSSN Lapse $\alpha$ (collapsing)]{
\includegraphics[height=6cm]{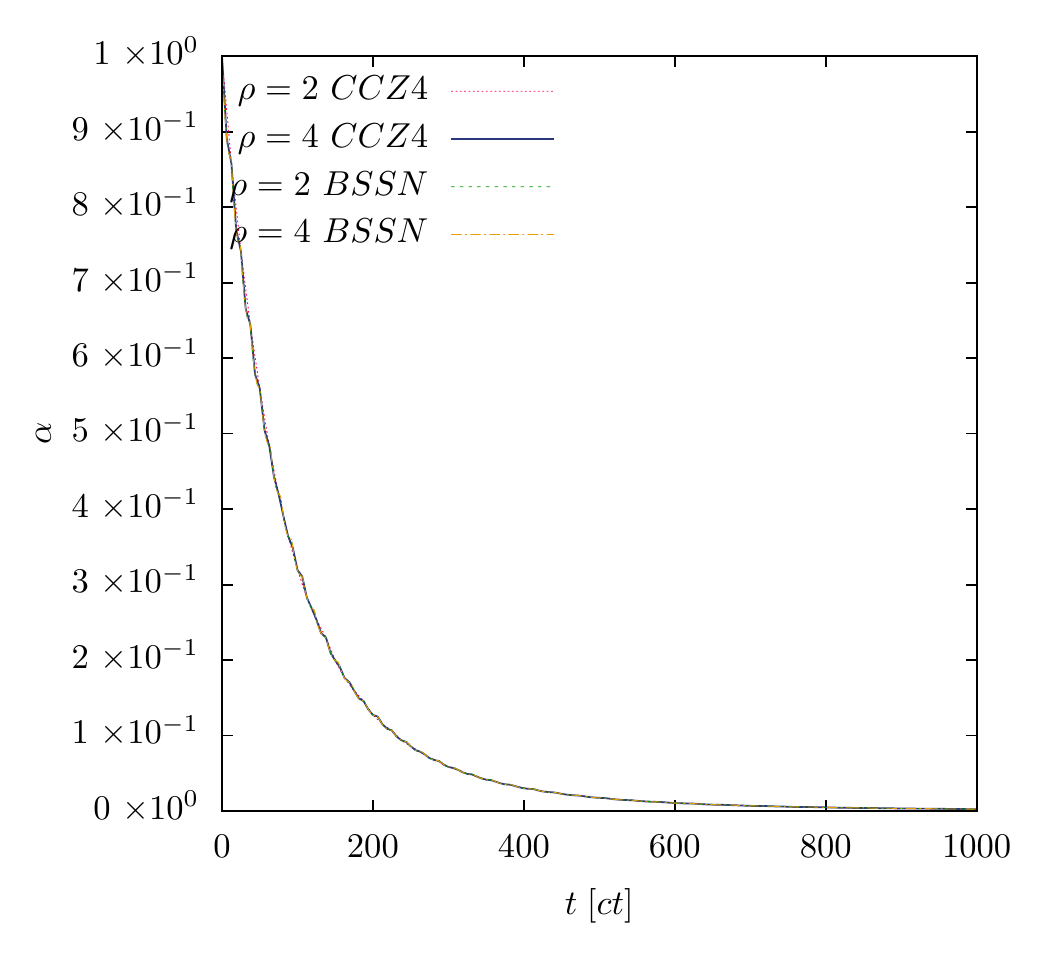}
}
\hspace{0.75cm}
\subfigure[Hamiltonian constraint (collapsing)]{
\includegraphics[height=6cm]{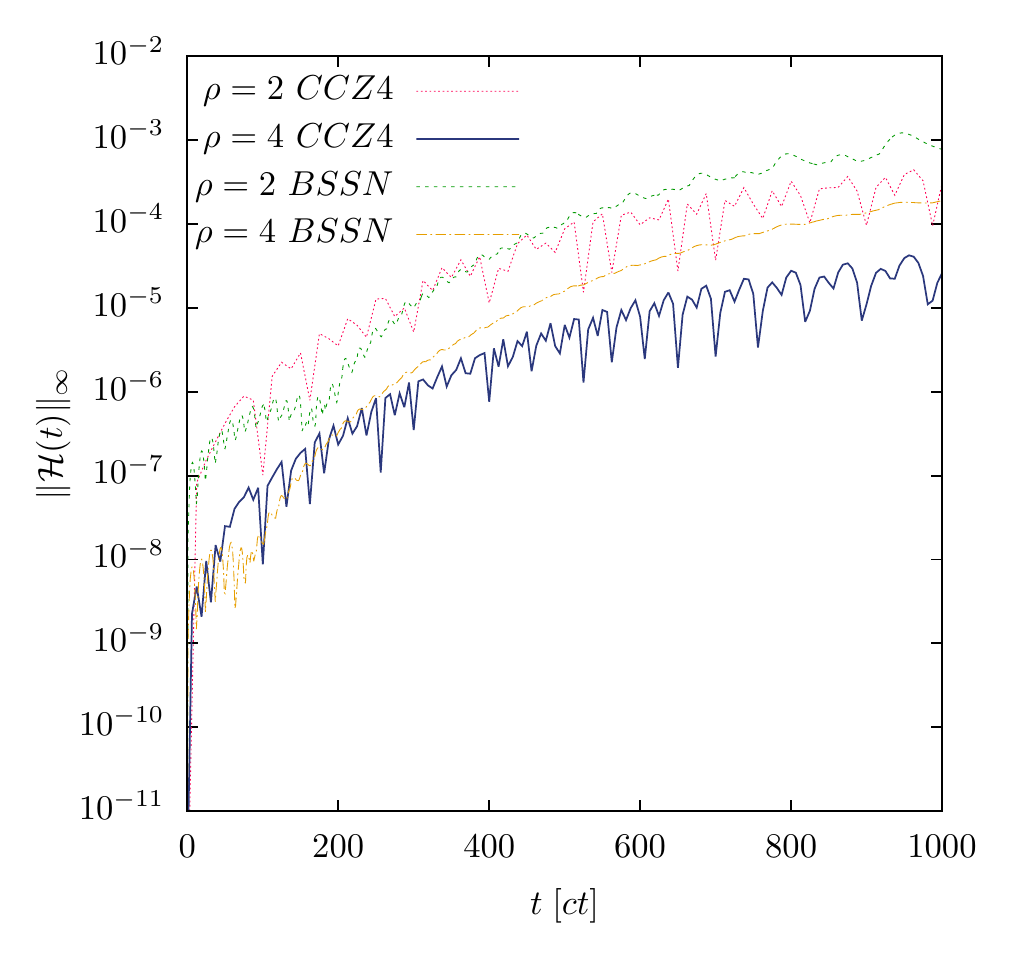}
}
\caption{Gowdy wave test, collapsing. \textit{Left}: minimum value of the lapse $\alpha$ across the grid as the spacetime is evolved towards the singularity. As expected, the harmonic gauge causes the evolution to ``slow down'' as the singularity is approached. \textit{Right}: evolution of the  $L_\infty$ norm of the Hamiltonian constraint for two resolutions. The test reaches $T=1000$ crossing times without crashing.
\label{fig-gowdy-collapsing}}
\end{center}
\end{figure}

\begin{figure}
\begin{center}
\subfigure[$K$ (expanding)]{
\includegraphics[height=6cm]{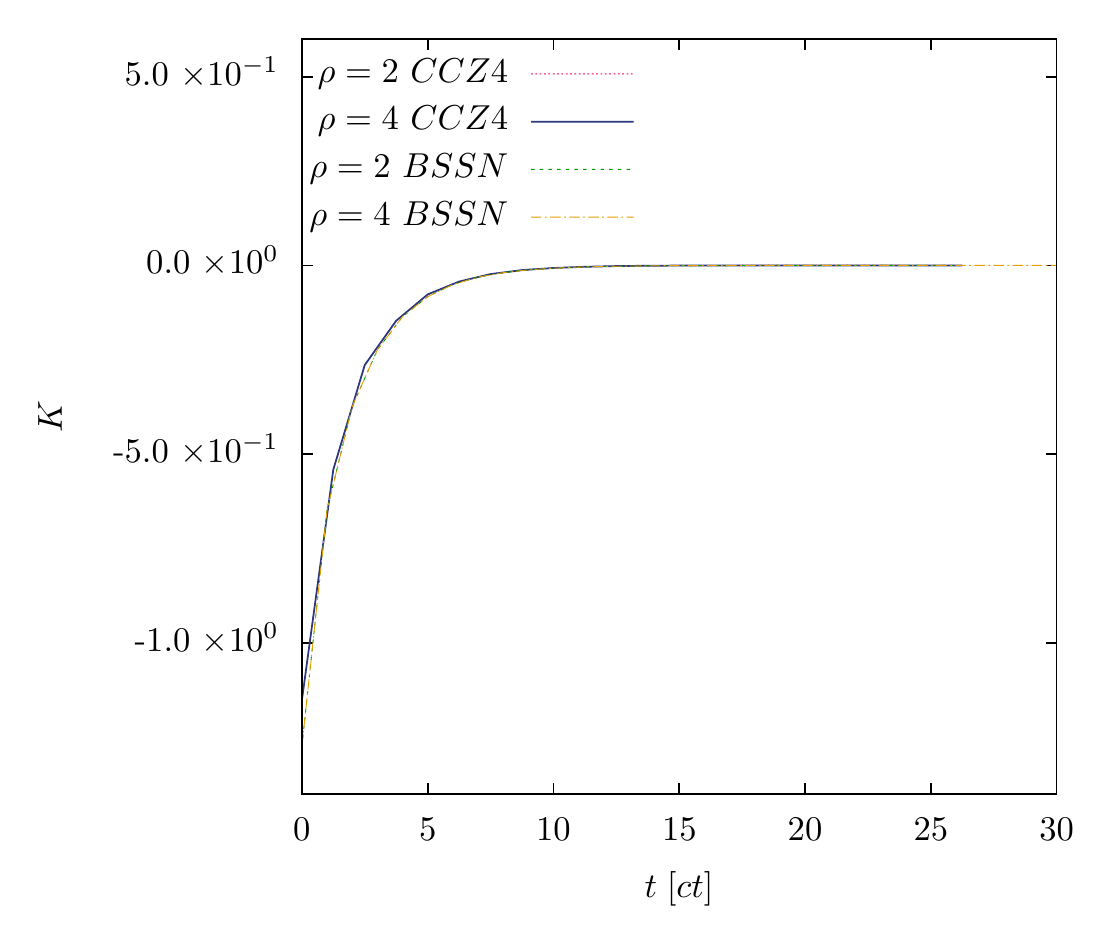}
}
\hspace{0.75cm}
\subfigure[Hamiltonian constraint (expanding)]{
\includegraphics[height=6cm]{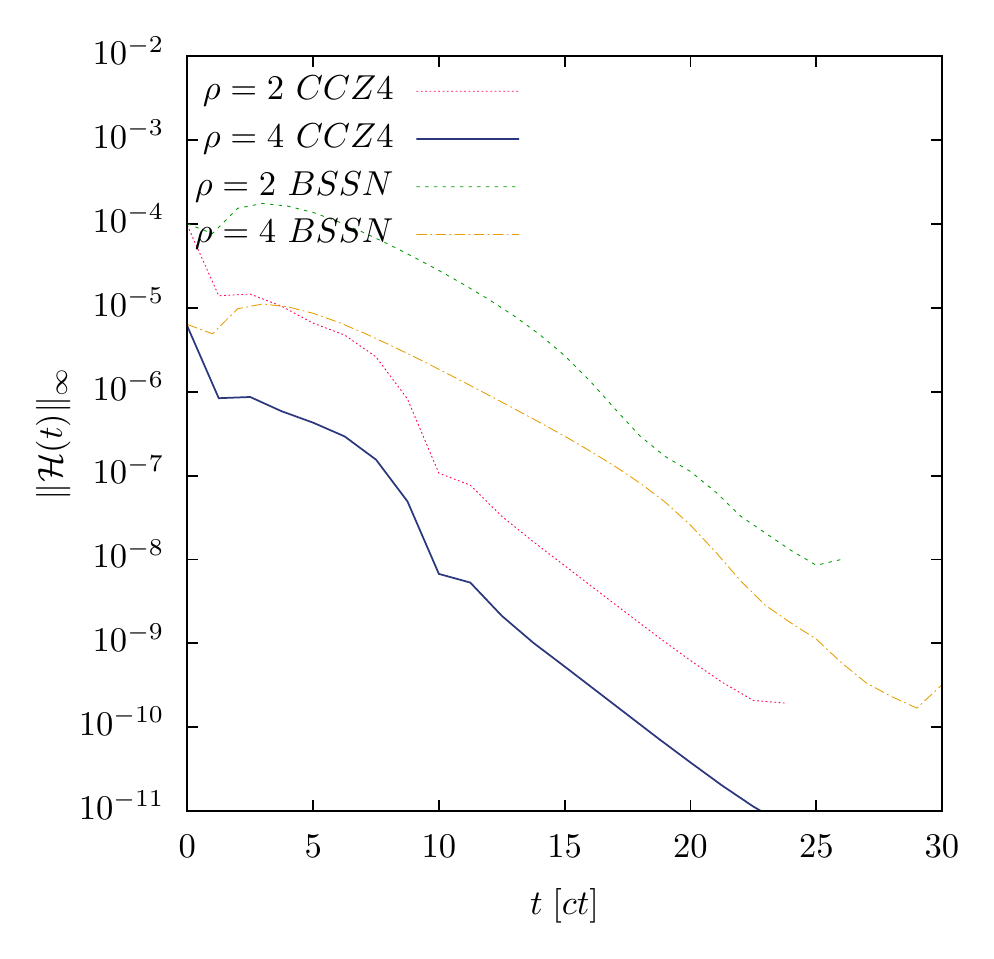}
}
\caption{Gowdy wave test, expanding. \textit{Left}: trace of the extrinsic curvature $K$ as the Gowdy wave spacetime is evolved in the collapsing direction. This correctly asymptotes to zero as the spacetime expands, but becomes unstable at around $t=30$. \textit{Right}: evolution of the  $L_\infty$ norm of the Hamiltonian constraint for two different resolutions.
\label{fig-gowdy-expanding}}
\end{center}
\end{figure}

\begin{figure}
\begin{center}
\subfigure[Convergence (expanding)]{
\includegraphics[height=6cm]{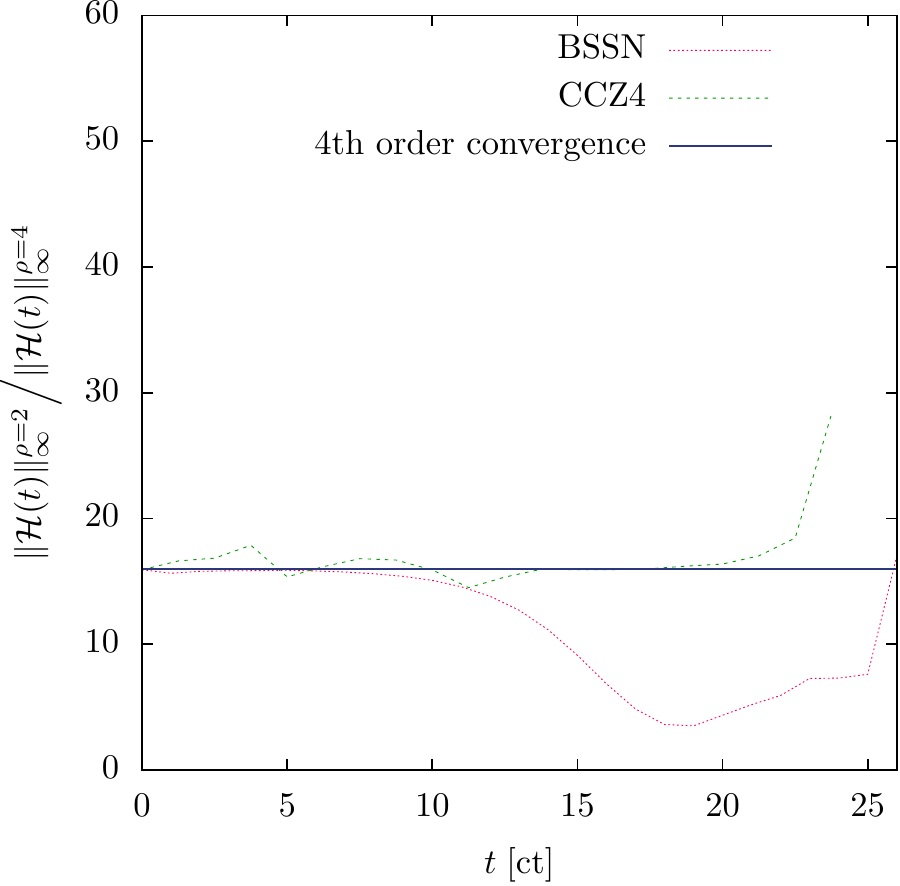}
}
\hspace{0.75cm}
\subfigure[Convergence (collapsing)]{
\includegraphics[height=6cm]{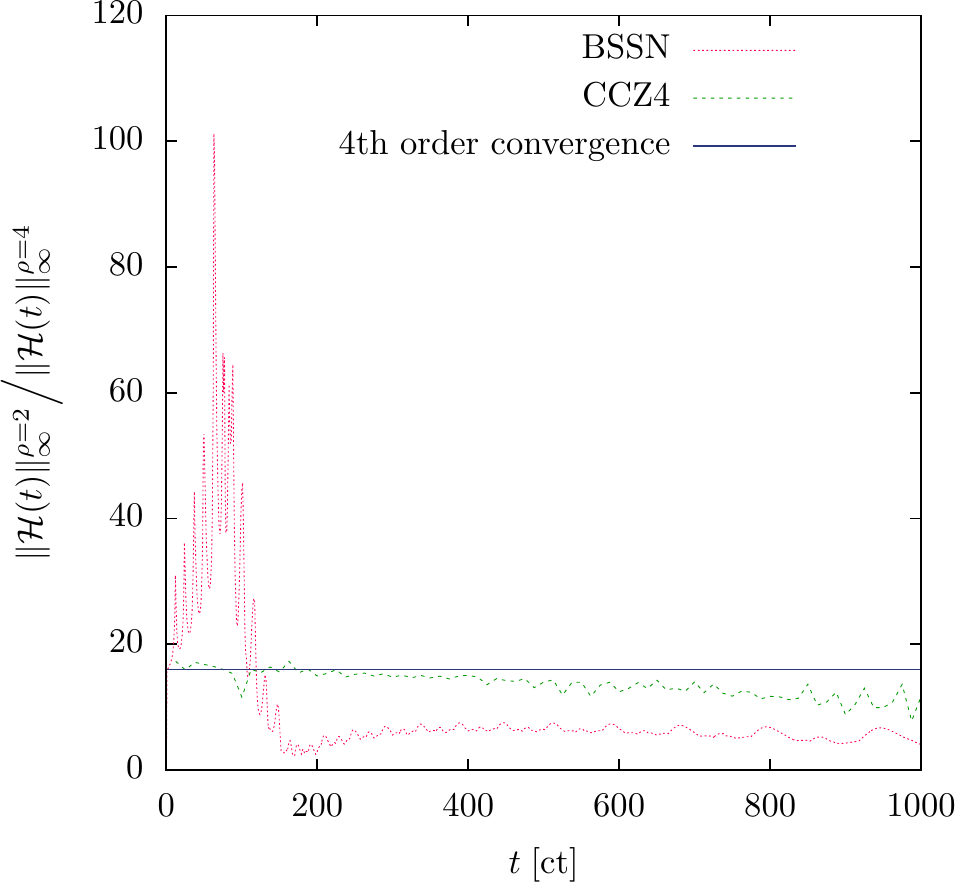}
}
\caption{Gowdy wave test, convergence. The ratio of the $L_\infty$ norm of the Hamiltonian constraint for the resolutions $\rho=4$ and $\rho=2$ is shown, for the expanding and collapsing directions for the BSSN and CCZ4 codes. A value of 16 indicates 4th order convergence, which is demonstrated by the codes initially, although lost at later times by BSSN. 
\label{fig-gowdy-convergence}}
\end{center}
\end{figure}

\begin{figure}
\begin{center}
\subfigure[Hamiltonian constraint]{
\includegraphics[height=6cm]{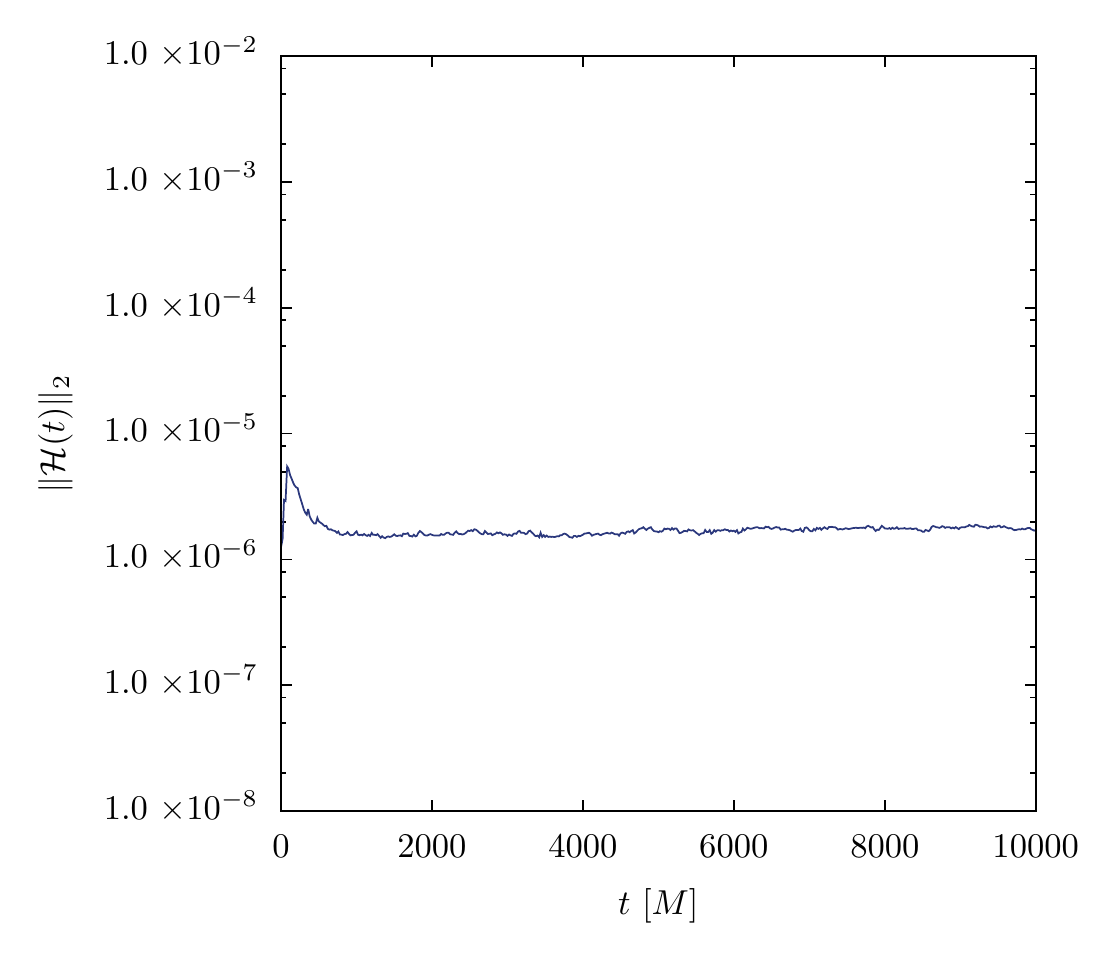}
}
\hspace{0.75cm}
\subfigure[ADM quantities]{
\includegraphics[height=6cm]{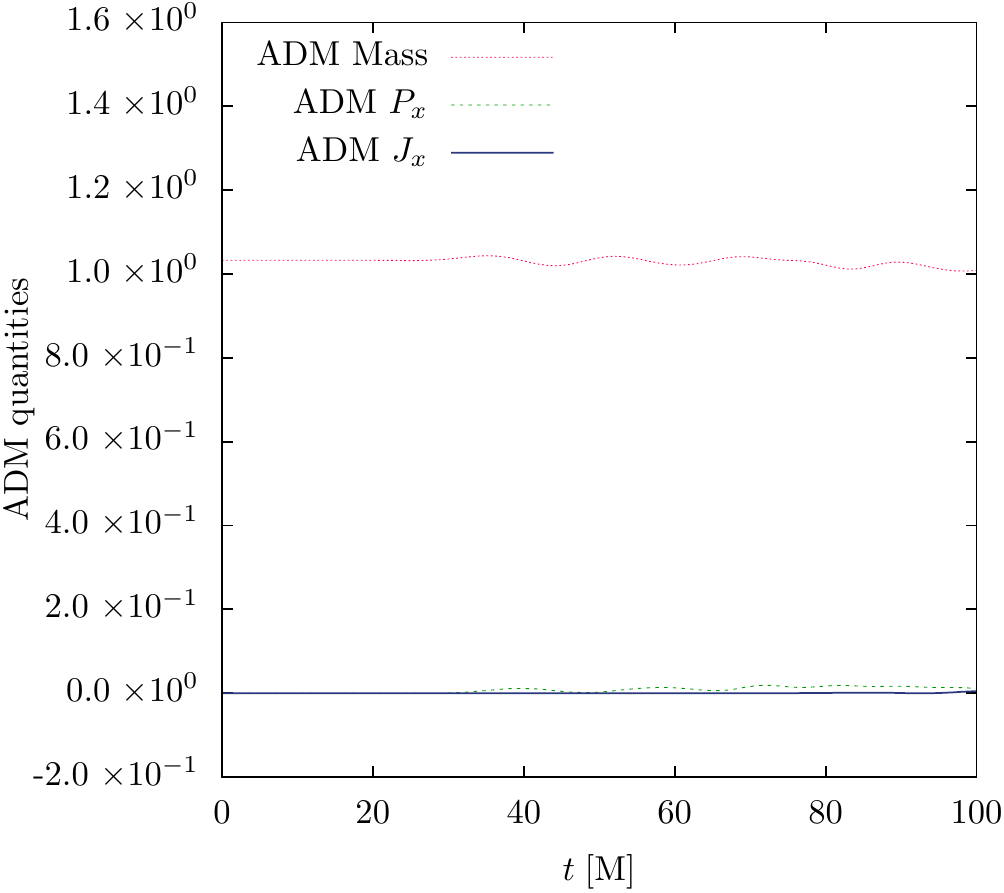}
}
\caption{Schwarzschild black hole simulations. \textit{Left:} Evolution of the $L^2$ norm of the Hamiltonian constraint up to $t=10000M$, showing long term stability.  \textit{Right:} ADM Mass, angular momentum and linear momentum (in the $x$ direction) during the initial stages of the evolution. These quantities remain approximately constant. 
\label{fig-SC}}
\end{center}
\end{figure}

\begin{figure}
\begin{center}
\subfigure[Hamiltonian constraint]{
\includegraphics[height=6cm]{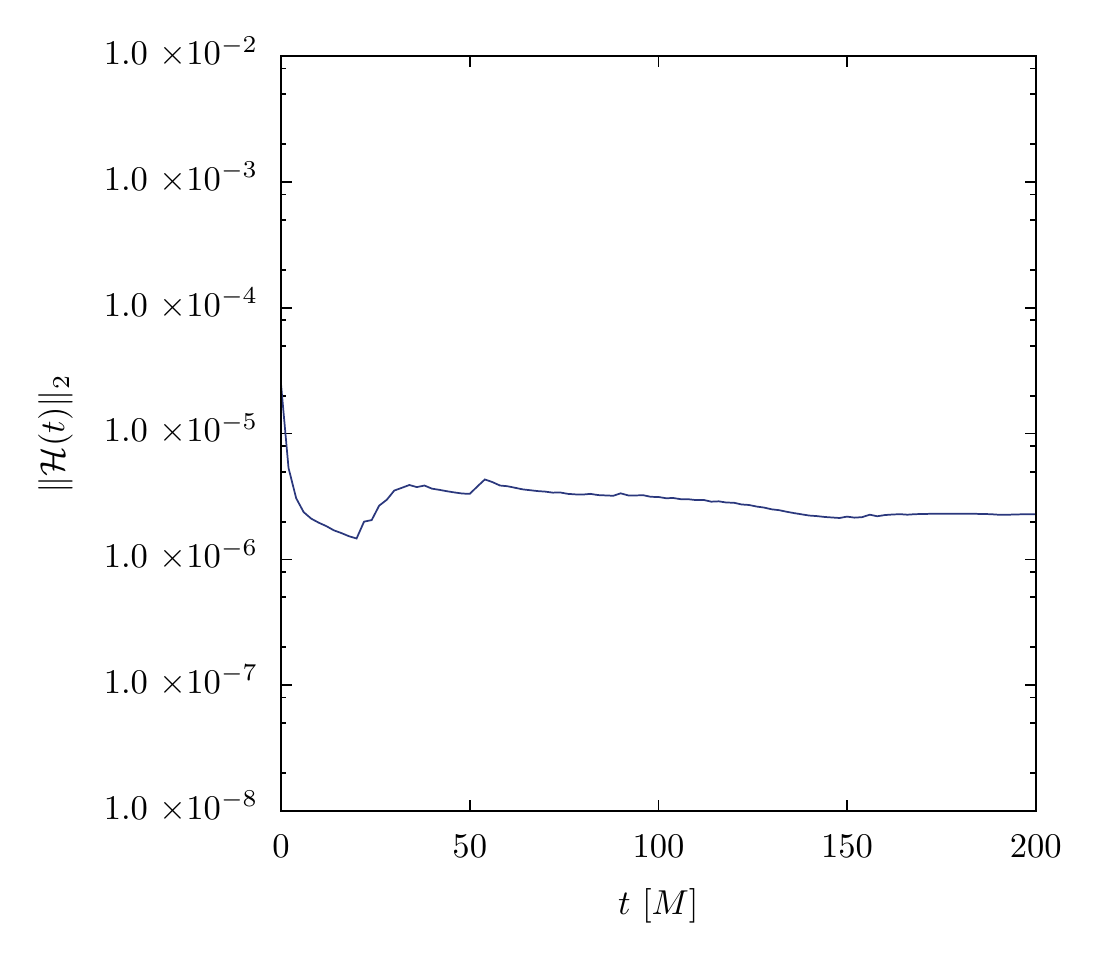}
}
\hspace{0.75cm}
\subfigure[ADM quantities]{
\includegraphics[height=6cm]{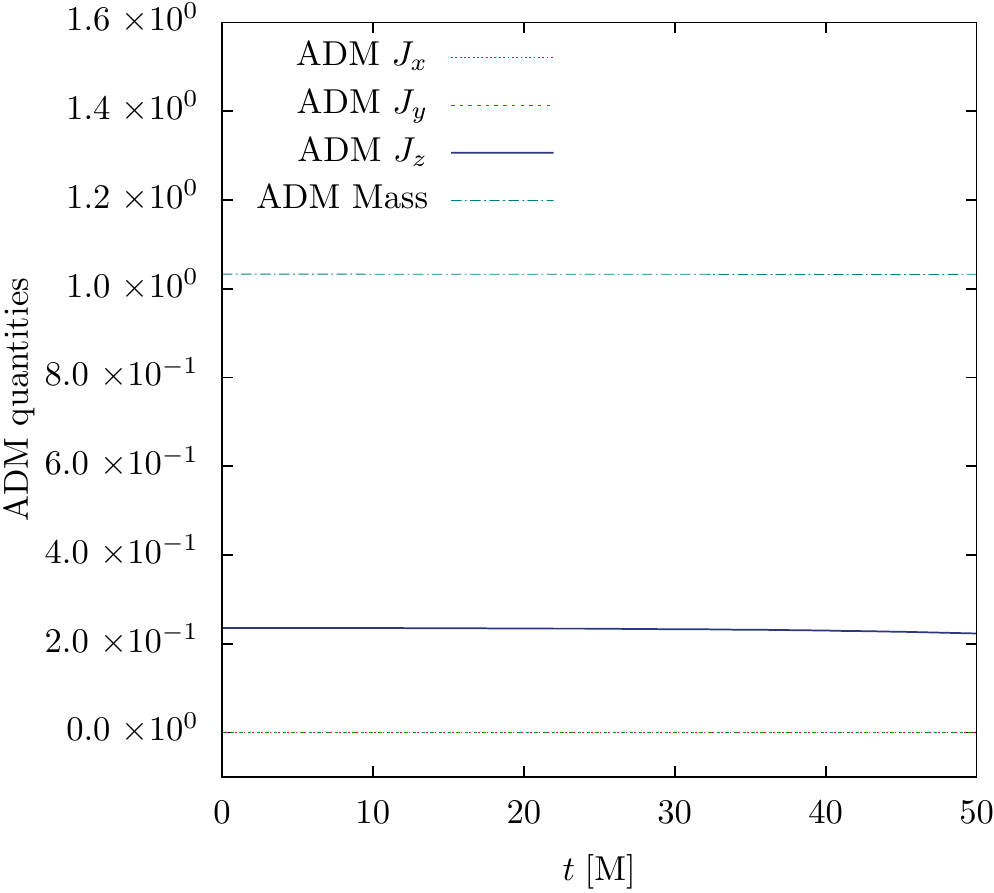}
}
\caption{Kerr black hole simulations. \textit{Left}:  Evolution of the $L^2$ norm of the Hamiltonian constraint. \textit{Right}: Components of the angular momentum and mass of the Kerr black hole during the evolution. The ADM quantities remain constant. 
\label{fig-KE}}
\end{center}
\end{figure}

\begin{figure}
\begin{center}
\subfigure[Hamiltonian constraint]{
\includegraphics[height=6cm]{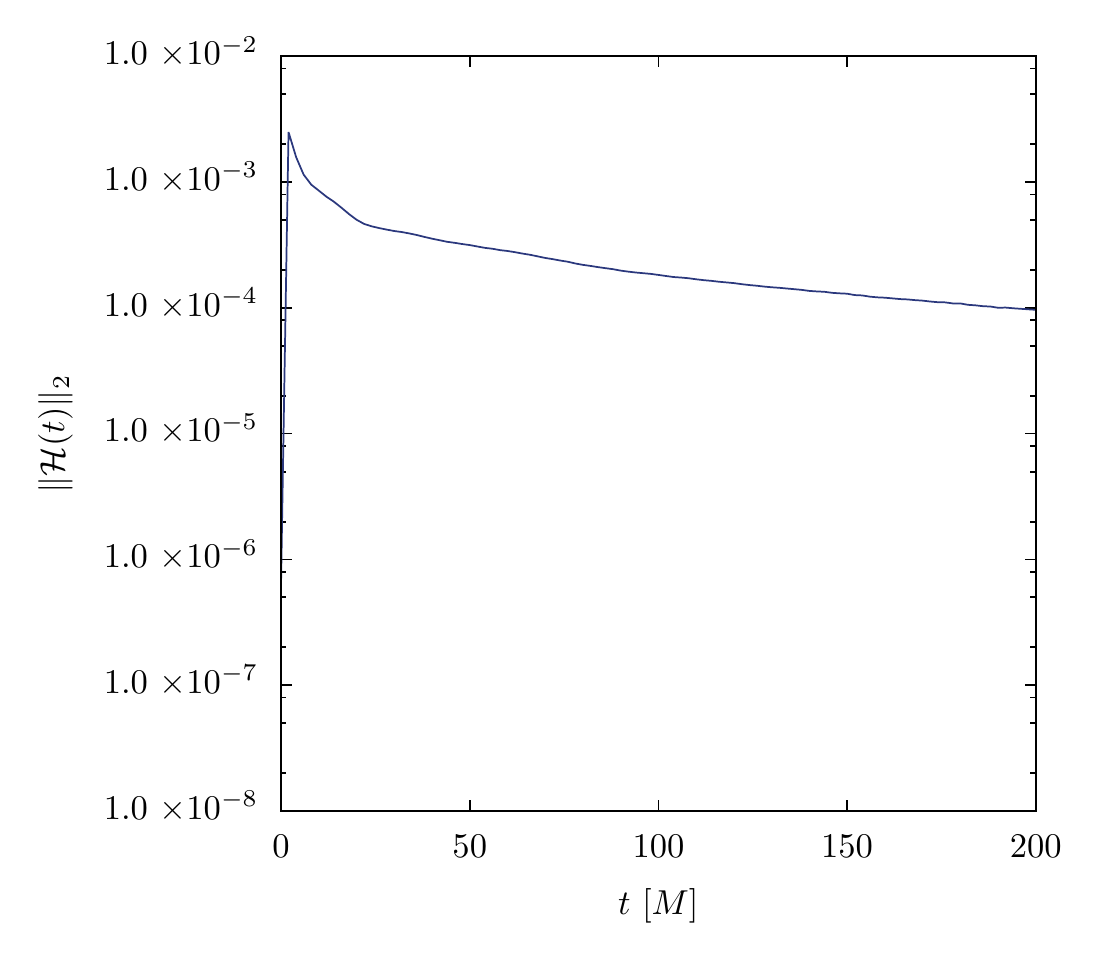}
}
\hspace{0.75cm}
\subfigure[ADM linear momentum]{
\includegraphics[height=6cm]{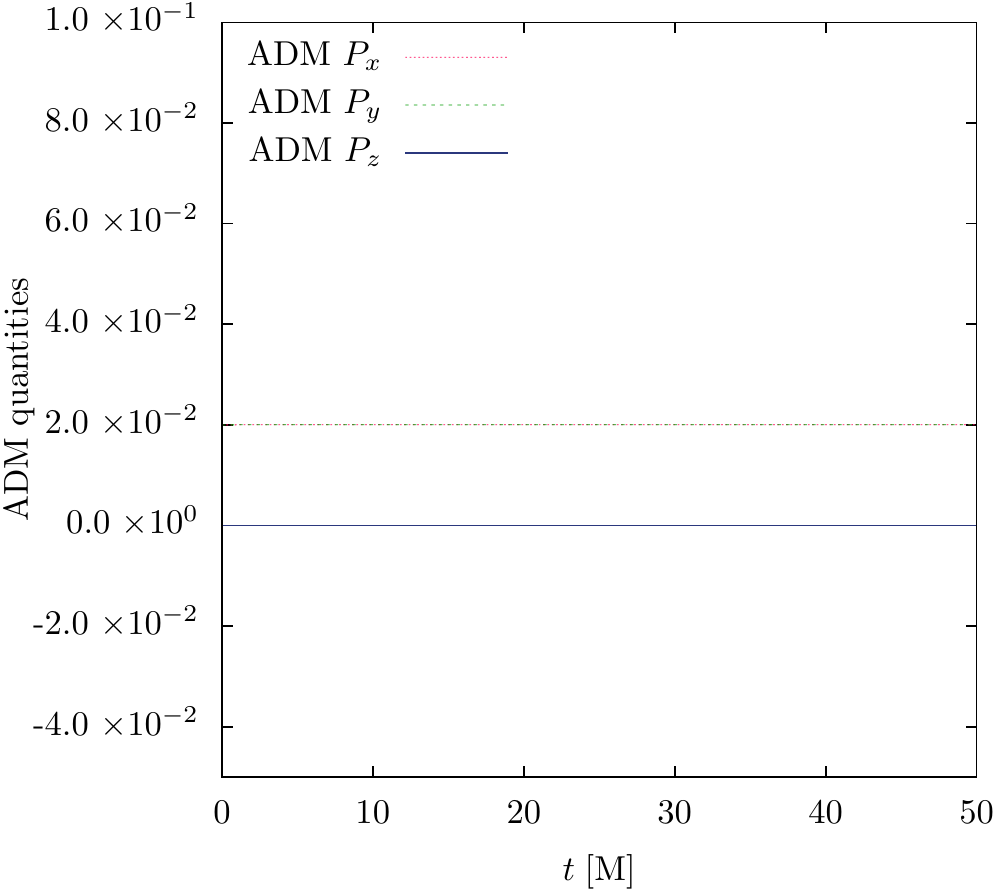}
}
\caption{Boosted black hole simulations.  \textit{Left}: Evolution of the $L^2$ norm of the Hamiltonian constraint. \textit{Right}: Components of the ADM linear momentum during the evolution. They remain constant.
\label{fig-BoostBH}}
\end{center}
\end{figure}

\begin{figure}
\begin{center}
\includegraphics[height=6cm]{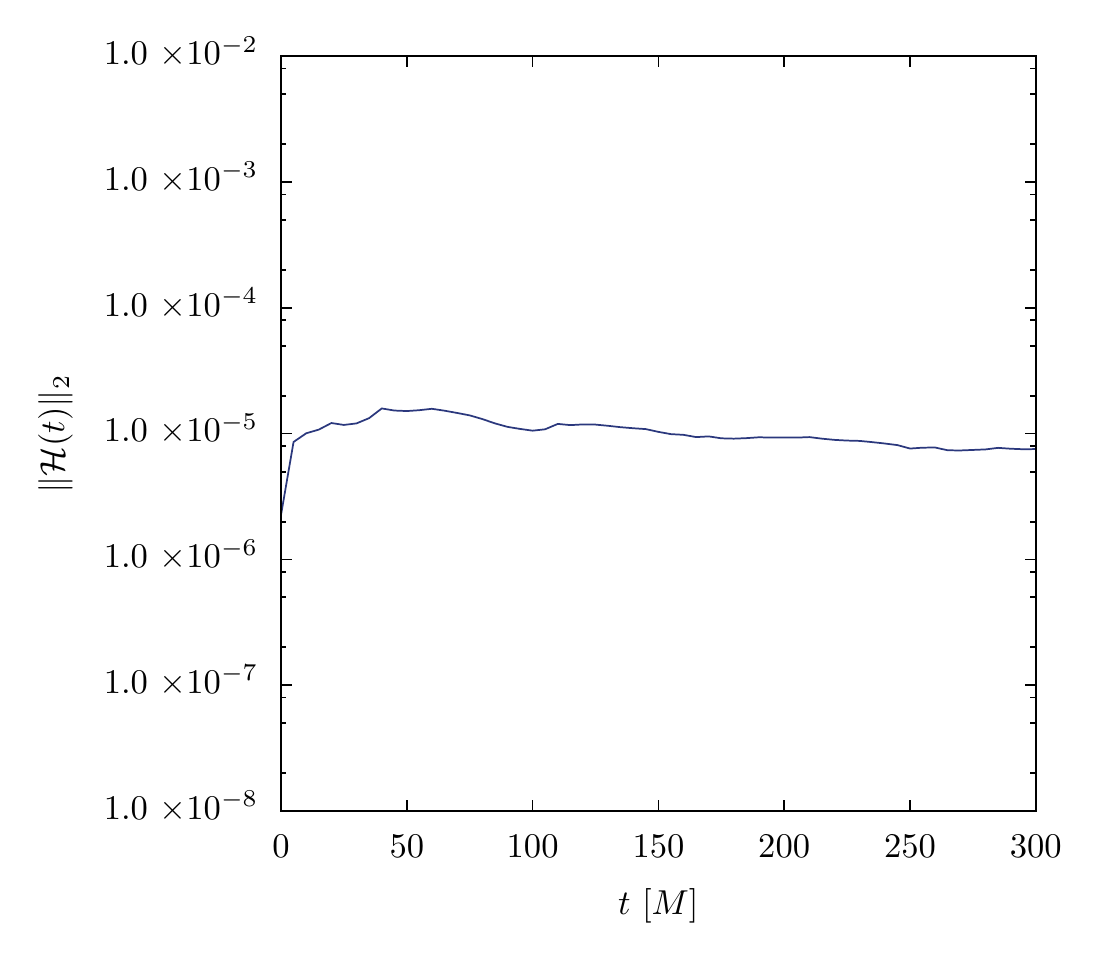}
\caption{$L^2$ norm of the Hamiltonian constraint across the whole domain for the binary black merger. The constraint violation remains bounded throughout the evolution, which includes the merger and ring-down phases.
\label{fig-Binary}}
\end{center}
\end{figure}

\begin{figure}
\begin{center}
\subfigure[Subcritical profiles of $\phi$]{
\includegraphics[height=6cm]{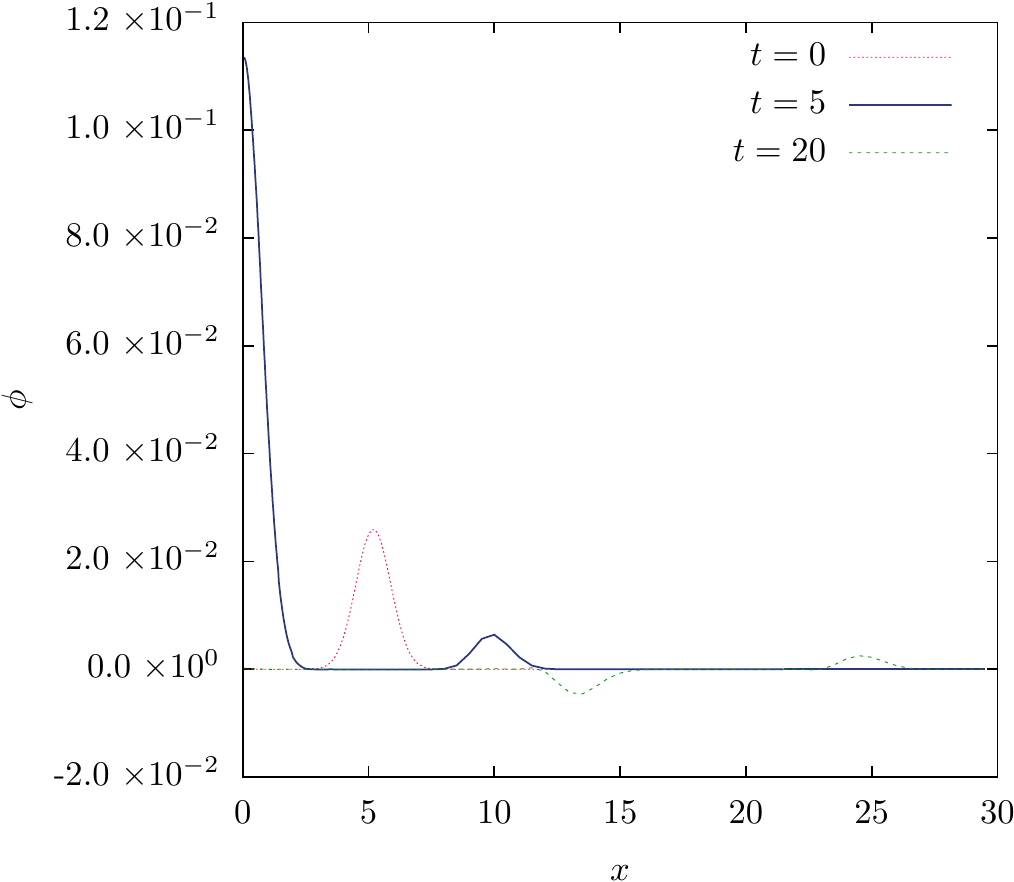}
}
\hspace{0.75cm}
\subfigure[Supercritical profiles of $\phi$]{
\includegraphics[height=6cm]{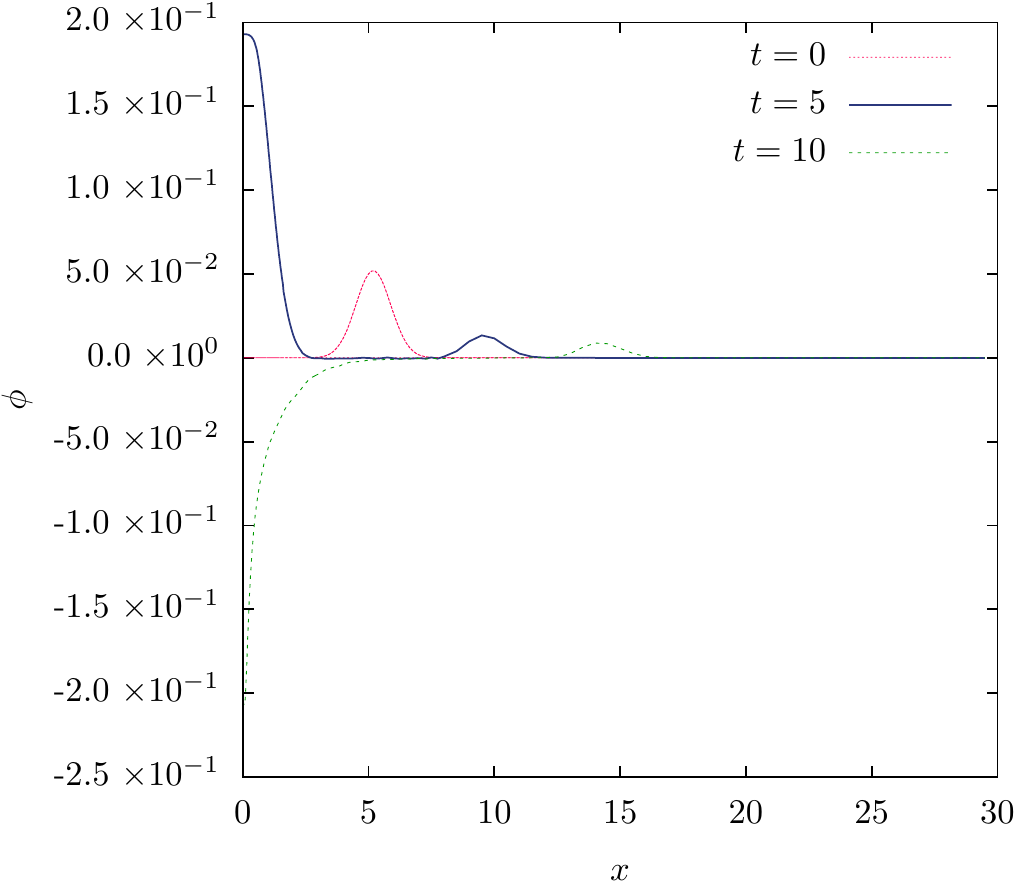}
}
\caption{Choptuik scalar field collapse. The profiles shown for the fields at $t=$ 0, 5 and 20 differ from those in \cite{AlcubierreBook} due to the different gauge conditions used. In the supercritical case we show the snapshot at $t=$ 10 rather than 20 as this is the point at which the evolution is frozen in the gauge choice in \cite{AlcubierreBook}. In the puncture gauge the evolution of the region within the event horizon continues and the result is that the large spike in the field effectively falls into the puncture, resulting in a zero field value at the centre of the coordinate grid. 
\label{fig-chopsnaps}}
\end{center}
\end{figure}

\begin{figure}
\begin{center}
\subfigure[Subcritical lapse]{
\includegraphics[height=6cm]{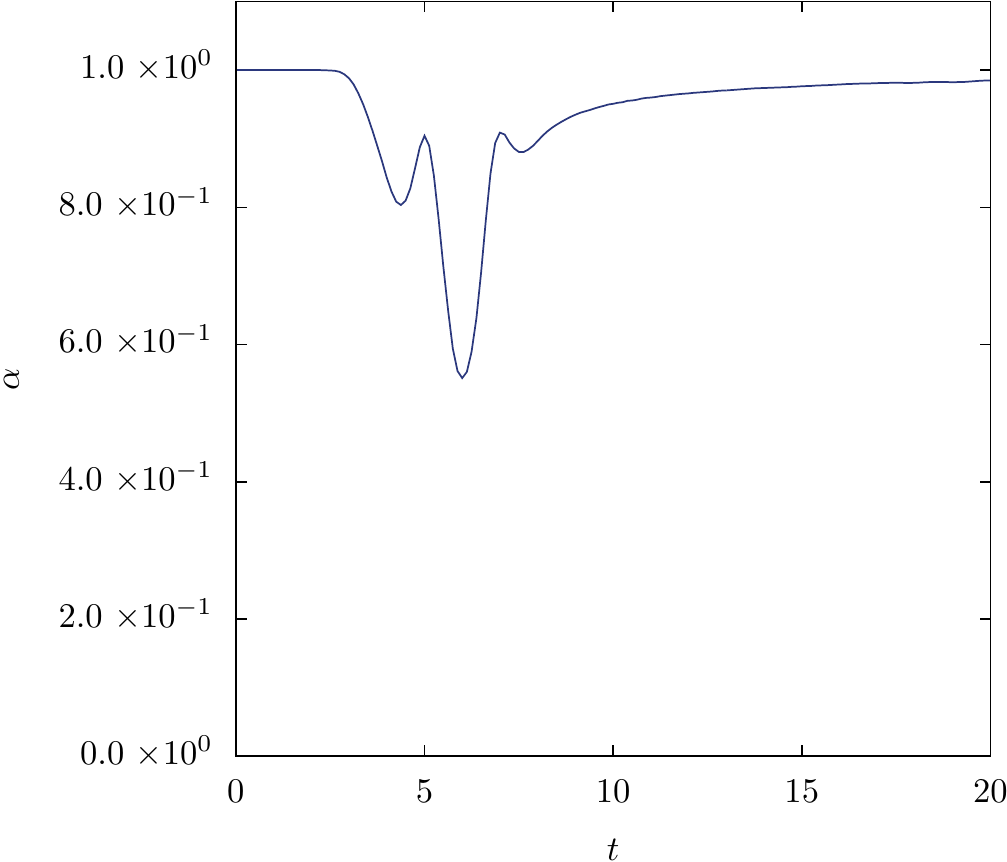}
}
\hspace{0.5cm}
\subfigure[Supercritical lapse]{
\includegraphics[height=6cm]{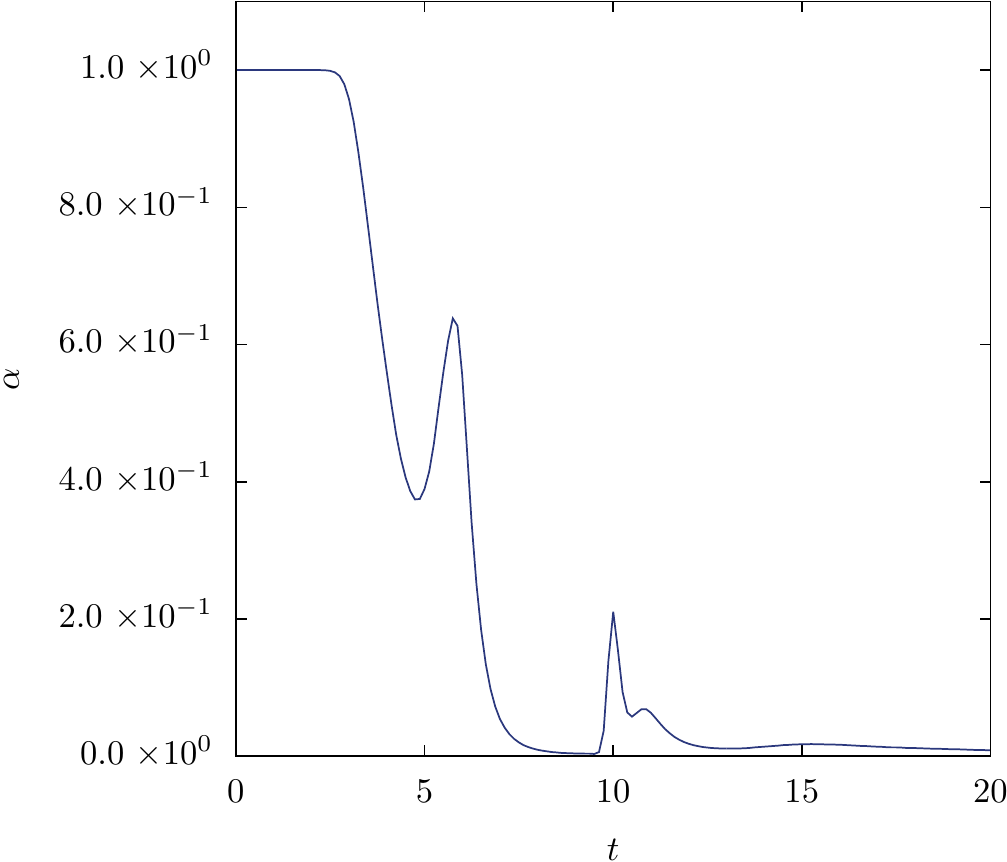}
}
\caption{Choptuik scalar field collapse. The values of the lapse at the centre of the grid are given. It can be seen the the profiles are very similar to those obtained by Alcubierre in \cite{AlcubierreBook}, and that the one for the supercritical case shows the characteristic collapse of the lapse which is symptomatic of black hole formation.
\label{fig-chopalp}}
\end{center}
\end{figure}

\end{document}